\documentclass[12pt]{article}

\pdfoutput=1

\usepackage[top=80pt,bottom=85pt,left=85pt,right=85pt]{geometry}
\usepackage{amssymb}
\usepackage{amsmath}
\usepackage{graphicx,color,subfigure}
\usepackage{float}
\usepackage{sectsty}
\usepackage{cite}
\usepackage[debug,pageanchor=false]{hyperref}
\definecolor{link}{rgb}{.8,.15,.1}
\hypersetup{colorlinks=true,linkcolor=link,citecolor=link,urlcolor=link,linktocpage}

\usepackage{bbm}

\newcommand{\bbone}{\ensuremath{\mathbbm{1}}}
\newcommand{\R}{\mathbb{R}}

\newcommand{\de}{\partial} 
\newcommand{\del}{\partial} 

\DeclareMathOperator{\e}{e}
\DeclareMathOperator{\g}{\gamma}
\DeclareMathOperator{\eps}{\epsilon}
\DeclareMathOperator{\dd}{\text{d}}
\newenvironment{sistema}%
{\left\lbrace\begin{array}{@{}l@{}}}%
	{\end{array}\right.}

\def\be{\begin{equation}}
\def\ee{\end{equation}}

\def\d {{\rm d}}

\def\cale         {{\cal E}}

\def\calg         {{\cal G}}

\def\calh         {{\cal H}}
\def\cali         {{\cal I}}

\def\calk         {{\cal K}}
\def\call         {{\cal L}}

\def\calp         {{\cal P}}

\def\calt         {{\cal T}}

\def\ii         {{\rm i}}

\makeatletter
\@addtoreset{equation}{section}
\makeatother

\begin{document}

\begin{titlepage}

\begin{center}

\vskip .5in 
\noindent

{\Large \bf{Timelike structures\\ of ten-dimensional supersymmetry}}

\bigskip\medskip

Andrea Legramandi,$^1$ Luca Martucci$^2$ and Alessandro Tomasiello$^1$\\

\bigskip\medskip
{\small 

$^1$ Dipartimento di Fisica, Universit\`a di Milano--Bicocca, \\ 
\& INFN, sezione di Milano--Bicocca,  Piazza della Scienza 3, I-20126 Milano, Italy 
\\	
\vspace{.3cm}
$^2$ 
Dipartimento di Fisica ed Astronomia ``Galileo Galilei'', Universit\`a di Padova\\
\& INFN, Sezione di Padova, Via Marzolo 8, I-35131 Padova, Italy	
}

\vskip .5cm 
{\small \tt andrea.legramandi@unimib.it, luca.martucci@pd.infn.it  alessandro.tomasiello@unimib.it}
\vskip .9cm 
     	{\bf Abstract }
\vskip .1in
\end{center}

\noindent

In several contexts, supersymmetry can be reformulated in terms of calibrations, namely forms whose integrals measure minimal energies. It has been conjectured that this should be possible in general. For type II supergravity, we present a new system of equations which realizes this expectation. Besides the customary D-brane calibrations, it also includes NS5-brane and KK5-monopole calibrations.
It is equivalent to supersymmetry under the assumption that the Killing vector associated to supersymmetry is timelike. No assumption is made on a factorization of spacetime. We also obtain a version of the system which is manifestly S-invariant and we present an application to near-horizon backgrounds. Using calibration, a definition of central charges in purely gravitational terms is given.

\noindent

\vfill
\eject

\end{titlepage}

\tableofcontents


\section{Introduction} 
\label{sec:intro}

In the study of supersymmetry, one often finds that more transparent interpretations are obtained by encoding the fermionic degrees of freedom in non-spinorial objects. For example, the problem of finding BPS solutions is usually simplified by mapping the spinorial transformation parameters to a set of equivalent tensors, such as forms. 

In supergravity, various techniques have been deployed for this purpose: $G$-structures (starting from \cite{strominger,gauntlett-pakis}), generalized (complex) geometry \cite{hitchin-gcy,gualtieri,gmpt,gmpt2} and spinorial geometry (see \cite{gran-gutowski-papadopoulos} for a recent review). This has worked rather well: not only are the equations obtained in this fashion easier to solve than the original spinorial ones, but in some cases they also have an elegant physical interpretation. 

For example, the problem of finding AdS$_4$ or Minkowski$_4$ compactifications of type II supergravity reduces with generalized complex geometry to a set of ``pure spinor equations'' \cite{gmpt2}. In this formulation the metric only appears indirectly: the usual notions of Riemannian geometry are replaced by natural operations involving forms, namely wedge products and exterior differentials, which makes the equations much easier to solve. Moreover, the pure spinor equations can be interpreted in terms of \emph{calibration conditions} \cite{martucci-smyth,koerber-martucci-ads}. In differential geometry, a calibration is a closed form that measures if a submanifold minimizes its volume \cite{harvey-lawson}. In supergravity we have a similar concept \cite{gutowski-papadopoulos,gutowski-papadopoulos-townsend}, sometimes called \emph{generalized} calibration, dealing with the various branes of the theory: in this case a calibrated brane minimizes its energy and the calibration condition is equivalent to imposing that the brane preserves part of the background supersymmetry (though generalizations to non-supersymmetric settings are also possible \cite{lust-marchesano-martucci-tsimpis}).

These successes have fueled speculations that a reformulation of the supersymmetric equations in terms of calibration conditions might exist in supergravity even without assuming a factorization into an external spacetime and an internal manifold. This is sometimes called ``supersymmetry--calibrations correspondence''. For compactifications to six dimensions, there is evidence \cite{lust-patalong-tsimpis} that such a correspondence still holds. However, for two dimensions  \cite{prins-tsimpis,rosa} found equations that are rather elegant, but that so far don't appear to have a straightforward calibration interpretation. Thus, a general answer to this question has been elusive so far.                                                                               

In \cite{10d}, the general problem of BPS solutions in type II  was considered without assuming any factorization.
Building on \cite{hackettjones-smith,figueroaofarrill-hackettjones-moutsopoulos,koerber-martucci-ads}, a system of equations in terms of forms was found; it is equivalent to supersymmetry, and it reproduces the pure spinor equations when specialized to four-dimensional compactifications. As in previous less general settings, part of the BPS system of  \cite{10d} can be interpreted in terms of calibrations \cite{martucci-electrified}. 
However, two of the equations in \cite{10d} are rather clumsy and  have no clear physical interpretation. 

In this paper, we point out a new, alternative reformulation of the ten-dimensional BPS equations in type II supergravity. The equations are all written using just forms and exterior algebra. The two main ones have a clear physical interpretation in terms of calibrations for D-branes and for NS5-branes. Another equation seems to be related to a similar concept for Kaluza-Klein (KK) monopoles. Moreover, we have found a way to supplement them with two reasonably elegant equations that make the full system equivalent to the BPS system \emph{for timelike solutions}, which means that the Killing vector $K$ naturally associated to the spinorial parameters is timelike. We recall that $K^2\le 0$, so in the space of all solutions the subset $K^2<0$ is actually the generic case, while $K^2=0$ has measure zero. Let us stress that fluxes are not expected to be completely determined in terms of BPS conditions and indeed we will discuss which equations of motion we must impose in order to get a proper supergravity solution.
(This is similar to what happens for example in \cite{gauntlett-pakis,gran-gutowski-papadopoulos-3}, where some components of the flux are shown to be undetermined by supersymmetry.)

While we do not consider our system  the final say in the matter, it certainly points in the right direction. The new NS5 equation is rather natural: it implies rather straightforwardly the equation of motion for the NS three-form in the Killing direction, and we show that it behaves well under dualities. Moreover, we show in both IIA and IIB that it can be interpreted as the existence of a generalized calibration, very similarly to the interpretation given in \cite{martucci-smyth} to the pure spinor equations in terms of D-branes. NS5-branes do not have an effective world-volume description from open strings, but we manage to demonstrate the interpretation using dualities: in IIA by using a reduction from the M5 calibration, in IIB by using S-duality with the D5 one. 

The KK equation is probably to be improved in the future, but it points to the possibility that the sought-after calibration reformulation of supergravity might so far not have worked because of the failure to consider various gravitational defects. 

Even if the timelike requirement can be seen as a limitation of our results, it is actually met in a lot of situations in which a complete classification is still missing. For example, one can use the BPS system to study some vacuum compactifications with extended supersymmetry but also stationary black-hole backgrounds. In particular, we take a first step in this direction by facing the problem of finding AdS$_2$ near-horizon solutions. We explicitly show how to specialize our system to that type of geometry.

In a slightly separate development, while studying how our various calibrations behave under dualities, we also managed to complete the supersymmetry system for IIB supergravity in such a way as to be manifestly covariant under $\mathrm{SL}(2,\mathbb{Z})$  transformations. Moreover, we extend this result also to the case of $\mathcal{N} = 1$ vacua, which were excluded from our system since they do not meet the timelike requirement; in particular, we focused on the four-dimensional ones.  
                                              
We begin in section \ref{sec:sys} by presenting our system, its derivation, and its interplay with the equations of motion.  
We discuss in section \ref{sec:dualities} its duality transformations, and use them to write a manifestly $\mathrm{SL}(2,\mathbb{Z})$-invariant version in type IIB. 
In \ref{sec:d-cal} we interpret one of the equations in our system as the calibration condition for an NS5-brane; using calibrations we manage in section \ref{sec:kk} to define central charges in purely gravitational terms, this reformulation can be generalized to KK-monopole charge and we then argue that another equation can be interpreted in terms of KK5-monopoles calibration. We then discuss some applications. In section \ref{sec:ads2} we show how to apply our system to AdS$_2\times M_8$ solutions, which is relevant for black hole horizons. In section \ref{sec:4d-sl2z} we show how to apply the manifestly $\mathrm{SL}(2,\mathbb{Z})$-invariant system to four-dimensional vacua (see also \cite{heidenreich-mcallister-torroba,heidenreich}).



\section{System} 
\label{sec:sys}

After some definitions and mathematical preliminaries in sections \ref{sub:geo} and \ref{sec:IIBspinorial}, we will give our system in section \ref{app:sys}. We will show that it is necessary for supersymmetry in section \ref{sub:der}.

\subsection{Some spinorial geometry} 
\label{sub:geo}

We start by reviewing quickly some aspects of the forms associated to the supersymmetry parameters of type II theories. More details can be found in \cite{10d} and App.~\ref{app:tech}.

In ten-dimensional space-time the irreducible spinorial representation is given by sixteen-dimensional Majorana--Weyl spinors. One can choose the gamma matrices $\gamma^M$ to be all real and we underline the indices that must be interpreted as flat when it is not evident from the context. In this basis, $\gamma^{\underline{0}}$ is the only antisymmetric matrix, while the other ones are all symmetric.  This can be summarized by
\begin{equation}
\g_M^{t} = \g_{\underline{0}} \g_M \g_{\underline{0}} \, .
\end{equation}

In order to extract from a spinor $\epsilon$ its geometrical content more transparently, it is often convenient to use its associated bispinor $\epsilon\otimes \overline{\epsilon} = \epsilon\otimes \epsilon^t \g_{\underline{0}} $. Since the antisymmetric products of $k$ gamma matrices $\gamma^{M_1\ldots M_k}$ are a basis for the space of bispinors, $\epsilon\otimes \overline{\epsilon}$ can be expanded on it using the Fierz identity: 
\begin{equation}
\label{eq:fierzeps}
\eps \otimes \, \overline{\eps} = \sum_{k = 0}^{10} \frac{1}{32 \, k!} \, (\overline{\eps}\g_{M_k \dots M_1}\eps) \, \g^{M_1 \dots M_k} \, .
\end{equation}
This bispinor can in turn be understood as a sum of forms of different degrees using the Clifford map
\begin{equation}\label{eq:cl}
C_k = \frac{1}{k !} C_{M_1 \dots M_k} \g^{M_1 \dots M_k} \longrightarrow C_k = \frac{1}{k !} C_{M_1 \dots M_k} \d x^{M_1} \wedge \dots \wedge \d x^{M_k} \, ,
\end{equation}
which is an isomorphism between the space of bispinors and the space of differential forms. In what follows, we will make no distinction between a differential form and a bispinor.

If $\epsilon$ is chiral, only forms of even degrees survive. If $\epsilon$ is also Majorana we see that
\begin{equation}
	\begin{split}
	\overline{\eps}\g_{M_k \dots M_1}\eps &= (\overline{\eps}\g_{M_k \dots M_1}\eps)^t = - \eps^t (\g_{M_k \dots M_1})^t \g_{\underline{0}} \eps \\
	&= - (-)^{k} (-)^{k(k-1)/2} \, \, \overline{\eps}\g_{M_k \dots M_1}\eps \, ,
	\end{split}	
\end{equation}
which sets to zero the degrees $k=0,3,4,7,8$ in (\ref{eq:fierzeps}), so that in fact only $k=1,5,9$ are present. Moreover, the chiral operator $\gamma=\gamma^{\underline{01\ldots 9}}$ can be translated in terms of form operations: 
\begin{equation}
\label{eq:gamma=}
\g C_k = * \lambda (C_k) \, ,
\end{equation}  
where $*$ indicates the Hodge dual and $\lambda$ acts on a $k$-form by $\lambda (C_k)\equiv (-1)^{k(k-1)/2} C_k$. Depending on the $\eps$ chirality we have $\g \eps \otimes \, \overline{\eps} = \pm \eps \otimes \, \overline{\eps}$; thus the nine-form is dual or anti-dual to the one-form, while the five-form is self-dual or anti-self-dual.
So in the end, if $\epsilon$ is Majorana--Weyl of chirality $\pm$, its bilinear can be written in terms of forms as 
\begin{equation}\label{eq:eeKOK}
	\eps \otimes \, \overline{\eps} = K + \Omega \pm * K \, \, ,
\end{equation}
where $K$ and $\Omega$ are a one- and five-form with components
\begin{equation}
K_M \equiv \frac{1}{32} \overline{\eps}\g_M \eps \qquad \text{and} \qquad \Omega_{M_1 \dots M_5} \equiv \frac{1}{32} \overline{\eps}\g_{M_1 \dots M_5}\eps \,.
\end{equation}
Notice that $*\Omega=\pm\Omega$ for  $\epsilon$ of chirality $\pm$.

These forms have notable algebraic properties. For starters, using
\begin{equation}
\label{eq:gammacontr}
\g_M C_k \g^M = (-)^k (10 - 2k) C_k \, \, 
\end{equation}
and (\ref{eq:gamma=}), we have
\begin{equation}
	K \eps = K_M \g^M \eps = \frac{1}{32} \g^M \eps \overline{\eps} \g_M \eps = - \frac{1}{4} (1 \pm \g) K \eps = - \frac{1}{2} K \eps\,,
\end{equation}
from which
\begin{equation}
\label{eq:keps=0}
K \eps = 0 \, \, .
\end{equation}
From equation \eqref{eq:keps=0} we can obtain that $K^M$ is a null vector:
\begin{equation}
K^M K_M = \frac{1}{(32)^2} \overline{\eps} \g^M \eps \overline{\eps} \g_M \eps = - \frac{1}{2 \cdot 32}  \overline{\eps} K \eps = 0 \, .
\end{equation}
Moreover, remembering (\ref{eq:gammadestrasinistra}),
\begin{equation}
	K \, \eps \overline{\eps} =  \eps \overline{\eps} \, K = 0 \quad \Longrightarrow \quad K \wedge  \eps \overline{\eps} = \iota_K  \eps \overline{\eps} = 0 \,.
\end{equation}
From (\ref{eq:eeKOK}) we then have
\begin{equation}
\label{eq:ikk^omega}
K \wedge  \Omega = \iota_K  \Omega = 0 \, \, .
\end{equation}
Therefore we can rewrite the 5-form as
\begin{equation}\label{eq:OKPsi}
\Omega = K \wedge \Psi
\end{equation}
for some four-form $\Psi$, which can be chosen to satisfy $\iota_K\Psi=0$. As we review in appendix \ref{app:s-group}, it is a $\mathrm{Spin}(7)$ form. Notice that, in particular, $K$ is determined by $\Omega$.

\subsection{Type II spinorial geometry}
\label{sec:IIBspinorial}

In type II theories, we have two spinorial parameters, $\epsilon_1$ and $\epsilon_2$, both Majorana--Weyl; in our conventions, they have both chirality $+$ in IIB, and chiralities $+$ and $-$ respectively in IIA. Their bilinears are obtained by applying (\ref{eq:eeKOK}) twice: 
\begin{equation}
\begin{split}
\eps_1 \otimes \, \overline{\eps}_1 \equiv& K_1 + \Omega_1 + * K_1 \\
\qquad \eps_2 \otimes \, \overline{\eps}_2 \equiv& K_2 + \Omega_2 \mp * K_2 \qquad \text{for} \qquad \overset{\text{\tiny{IIA}}}{\text{\tiny{IIB}}} \,,
\end{split}
\end{equation}
but this time we can also define the mixed bispinor
\begin{equation}
\eps_1 \otimes \, \overline{\eps}_2 \equiv \Phi \,,
\end{equation}
a collection of forms with the property that $* \lambda (\Phi) = \Phi$. If $\eps_1$ and $\eps_2$ have the same chirality then $\Phi$ will contain only forms with odd degree, otherwise it will contain only forms with even degree:
\begin{equation}
\Phi = 
\begin{sistema}
\Phi_0 + \Phi_2 + \Phi_4 + \Phi_6 + \Phi_8 + \Phi_{10} \quad \text{for IIA} \\
\Phi_1 + \Phi_3 + \Phi_5 + \Phi_7 + \Phi_9 \qquad \, \, \, \, \, \, \, \, \, \,  \text{for IIB} \\
\end{sistema} \quad  .
\end{equation}
The bispinor $\eps_2 \otimes \, \overline{\eps}_1$ is not indepedent: it can be obtained from $\Phi$ as
\begin{equation}
\label{eq:lambdaPhi}
\eps_2 \overline{\eps}_1 
= - (-)^{\text{deg} \Phi} \lambda ( \Phi) \, .
\end{equation}

From (\ref{eq:keps=0}), we see that  
\begin{equation}
\label{eq:annPhi1d}
K_1 \Phi = \Phi K_2 = 0 \, .
\end{equation}
If we define
\begin{equation}\label{eq:KKt-def}
K \equiv \frac{1}{2} (K_1 + K_2)^M\del_M \, , \qquad \widetilde{K} \equiv \frac{1}{2} (K_1 - K_2)_M d x^M \, ,
\end{equation}
we can rewrite \eqref{eq:annPhi1d} using \eqref{eq:gammadestrasinistra}:
\begin{equation}
( \iota_K+\widetilde{K} \wedge ) \Phi = 0 \, .
\end{equation}
In the same spirit we define
\begin{equation}
\label{eq:5formdef}
\Omega \equiv \frac{1}{2} (\Omega_1 \pm \Omega_2) \, , \qquad \widetilde{\Omega} \equiv \frac{1}{2} (\Omega_1 \mp \Omega_2) \qquad \text{for} \qquad \overset{\text{\tiny{IIA}}}{\text{\tiny{IIB}}} \, .
\end{equation}
Notice that $*\Omega=\widetilde\Omega$ in IIA while  $*\Omega=\Omega, *\widetilde\Omega=\widetilde\Omega$ in IIB. 

The vector $K$ will play a key role in our discussion and in particular it can be seen that 
\begin{equation}
	K^2\le 0\,.
\end{equation}
The case where $K^2=0$  is called the \emph{light-like case}; the case where $K^2<0$ is called the \emph{timelike case}, and will be the focus of this paper. 

We have used the metric and the spinors $\epsilon_1,\epsilon_2$ to construct the  forms $\Phi, \Omega_{1},\Omega_2$.\footnote{We recall that we don't have to consider also $K_1$ and $K_2$ since they are completely determined by $\Omega_{1},\Omega_2$.} 
Viceversa, in the spirit of \cite{10d}, we have to wonder if the geometric data encoded in  $\Phi, \Omega_{1},\Omega_2$ contain the complete information on metric and spinors, at least in the timelike case. We can use G-structures to address this question. 
As often happens dealing with G-structures, it is useful to enlarge the structure group of the tangent bundle $T$ to the one on the generalized tangent bundle $T \oplus T^*$, which is O$(10,10)$ \cite{gualtieri}. Furthermore, the generalized tangent bundle can be $B$-twisted in order to accomodate for a non-trivial $H$-field.
In this framework the common stabilizer of the metric and the $B$-field, i.e.\ the subgroup of O$(10,10)$ that does not transform $g$ and $B$, is known and it is given by $\text{O}(9,1) \times \text{O}(9,1)$. In generalized complex geometry differential forms are spinors with respect to the generalized tangent bundle metric, and the gamma matrices are given by contraction and wedge operators. We can then compute the infinitesimal action that leaves the appropriately $B$-twisted ``generalized spinors'' $\Phi, \Omega_{1},\Omega_2$ invariant. In appendix \ref{app:s-group} we prove that
\begin{equation}
\text{Stab} (\Phi, \Omega_1 , \Omega_2) \subset \text{O}(9,1) \times \text{O}(9,1) \, ,
\end{equation}
so that we have enough geometric data to define metric, $B$-field and spinors of our backgrounds. In the following we will in fact use the `untwisted' picture, in which the  $B$-field is treated as an external ingredient and $\Phi, \Omega_{1},\Omega_2$ determine only metric and spinors.


\subsection{Necessary and sufficient system} 
\label{app:sys}

We will now present a system of differential form equations which is equivalent to the supersymmetry conditions in the timelike case $K^2 < 0$.

For both IIA and IIB, the system is
\begin{subequations}
	\label{eq:systemII}
	\begin{align}
	&\d_H (\e^{-\phi} \Phi) = - (\iota_K+\widetilde{K} \wedge) F \,  , \label{eq:oldII} \\
	&\e^{ 2\phi} \d (\e^{- 2 \phi} \Omega) = - \iota_K  \ast H + \e^{\phi} ( \Phi , F )_6 \,  , \label{eq:dOmegaIIA}\\
	&\e^{ 2\phi} \d (\e^{- 2 \phi} \widetilde\Omega) =  - \ast (\widetilde{K}\wedge H) -\frac{1}{2}(-)^{|\Phi|} \,e^{\phi} \left( \Phi_M , F^M \right)_6 \,  , \label{eq:domegatII} \\
	& \mathcal{L}_K \phi= 0 \,  , \qquad   \d * \widetilde{K} = - \frac{1}{8}(-)^{|\Phi|}\, e^\phi  \left(\Phi , \g_M F \g^M \right) \label{eq:lastII} \, .
	\end{align}
\end{subequations}
As usual, $\d$ is the de Rham differential, $\d_H\equiv \d-H\wedge$ and $F$ is the polyform obtained by summing all RR field strength, which is $\d_H$-closed away from localized sources, and $|\Phi|$ is a short-hand for the form degree of $\Phi$.
 We have introduced the Chevalley--Mukai pairing between two forms $A$ and $B$:
\begin{equation}\label{eq:mukai}
(A,B) = \left( A \wedge \lambda(B) \right)_{10} \, \, ,
\end{equation}
where the subscript ${}_{10}$ denotes keeping the coefficient of the ten-form part only; and a similar novel six--form--valued pairing
\begin{equation}
	(A,B)_6\equiv \left( A \wedge \lambda(B)\right)_6\,\,.
\end{equation}

The first equation (\ref{eq:oldII}) of the system was derived in \cite{10d}, building upon results in \cite{koerber-martucci-ads} for a 1+9 splitting of spacetime. It was shown in \cite{10d} that it is enough to reproduce the pure spinor equations for Mink$_4$ or AdS$_4$ solutions \cite{gmpt2,gmpt3}. The second equation can again be found in \cite{koerber-martucci-ads} in the case of 1+9 splitting or in \cite{koerber-tsimpis} in the case of 4+6 splitting. The result $\mathcal{L}_K \phi= 0$ was also derived in \cite{10d}, and it is part of the more general result that $K$ is a symmetry of the solution (in other words, $\mathcal{L}_K= 0$ for all fields, not just the dilaton); in particular, $K$ is a Killing vector \cite{hackettjones-smith,figueroaofarrill-hackettjones-moutsopoulos,koerber-martucci-ads}. This follows from (\ref{eq:systemII}), rather than being part of the system as in \cite{10d}. The complete system (\ref{eq:systemII}) implies also the equation 
\begin{equation}
\label{eq:string_cal}
\d \widetilde{K} = \iota_K H
\end{equation}
which appears in the system of \cite{10d}.  

In later sections, we will interpret most of this system in terms of calibrations.  (\ref{eq:oldII}) and \eqref{eq:string_cal} already have a known interpretation in terms of D-branes and F1-string calibrations, which we will review. (\ref{eq:dOmegaIIA}) has a similar interpretation in terms of NS5 calibrations while (\ref{eq:domegatII}) seems to be related to the T-dual of an NS5, namely a KK5-monopole, although a calibration interpretation is  more subtle, for reasons we will see below.

In the next subsection, we will sketch how to derive the system (\ref{eq:systemII}) from the supersymmetry equations; in other words, we will show that those equations are necessary for supersymmetry. The proof of sufficiency is more technical, and we give it in appendix \ref{app:proof}.

\subsection{Derivation} 
\label{sub:der}
   
As we mentioned,  (\ref{eq:oldII}) was derived in \cite{10d}. So we will start with the second and the third equations of each system, which we will derive together. While this is in principle a straightforward application of gamma matrix identities and the supersymmetry equations, in practice it is simplified by some tricks; for this reason, we describe the computation below.       

We start by deriving an equation for the contraction with a three form, using \eqref{eq:gammadestrasinistra}: 
\begin{equation}
	\begin{split}
	\iota_H &= \frac{1}{8 \cdot 3!} H^{MNP} \left(  \overrightarrow{\g}_{MNP} -  \overleftarrow{\g}_{MNP} (-)^{\text{deg}} + 3  \overrightarrow{\g}_M  \overleftarrow{\g}_{NP} - 3  \overrightarrow{\g}_{NP}  \overleftarrow{\g}_M (-)^{\text{deg}} \right) \\
	&= \frac{1}{8} \left( \overrightarrow{H} - \overleftarrow{H} (-)^{\text{deg}} + \overrightarrow{\g}^M  \overleftarrow{H}_M -  \overrightarrow{H}_M  \overleftarrow{\g}^M (-)^{\text{deg}}  \right) \, .
	\end{split}
\end{equation}                                                           
This is the ``dual'' of the three-form wedge obtained in \cite[(A.10)]{gmpt3}.   

Now let us compute
\begin{align}
\label{eq:eps1noSUSY}
2 \e^{ 2\phi} \d (\e^{- 2 \phi} \eps_1 \overline{\eps}_1) + 2 \iota_H  \eps_1 \overline{\eps}_1 &= \left[ \g^M , D_M ( \eps_1 \overline{\eps}_1) - 2 \de_M \phi \eps_1 \overline{\eps}_1 \right] \\
&+ \frac{1}{4} \left( H \eps_1 \overline{\eps}_1 + \eps_1 \overline{\eps}_1 H + \g^M  \eps_1 \overline{\eps}_1 H_M + H_M \eps_1 \overline{\eps}_1 \g^M \right) = \nonumber \\
&= \left( D - \frac{1}{4} H - \de \phi \right) \eps_1 \overline\eps_1 + \g^M \eps_1 \left(  D_M \overline{\eps}_1 + \frac{1}{4} \overline{\eps}_1 H_M  \right) \nonumber \\
&-  \left( D_M - \frac{1}{4} H_M  \right) \eps_1 \overline\eps_1 \g^M - \eps_1 \left(  D_M \overline{\eps}_1 \g^M  + \frac{1}{4} \overline{\eps}_1 H - \overline\eps_1 \de \phi \right) \nonumber\\
&- \left( \de \phi - \frac{1}{2} H \right) \eps_1 \overline\eps_1 + \eps_1 \overline{\eps}_1 \left( \de \phi + \frac{1}{2} H \right) \, . \nonumber
\end{align}
If we now replace the supersymmetry equations \eqref{eq:SUSY1}, \eqref{eq:SUSY2} (and their transpose) and we use \eqref{eq:lambdaPhi} we get
\begin{equation}\label{eq:eps1epsbar1}
\begin{split}
\e^{ 2\phi} \d (\e^{- 2 \phi} \eps_1 \overline{\eps}_1)  = - \iota_H  \eps_1 \overline{\eps}_1 &+(-)^{|F|}  \frac{\text{e}^{\phi}}{32} \g^M \Phi \g_M  \lambda(F) -(-)^{|F|} \frac{\text{e}^{\phi}}{32} F \g^M \lambda(\Phi) \g_M    \\
&-(-)^{|F|} \frac{\text{e}^{\phi}}{32} \g^M F \g_M \lambda(\Phi) +(-)^{|F|} \frac{\text{e}^{\phi}}{32} \Phi \g^M \lambda(F) \g_M \, .
\end{split} 
\end{equation}
The same procedure can be applied for $\eps_2$:
\begin{align}\label{eq:eps2noSUSY}
2 \e^{ 2\phi} \d (\e^{- 2 \phi} \eps_2 \overline{\eps}_2) - 2 \iota_H  \eps_2 \overline{\eps}_2 &= 
\left( D + \frac{1}{4} H - \de \phi \right) \eps_2 \overline\eps_2 + \g^M \eps_2 \left(  D_M \overline{\eps}_2 - \frac{1}{4} \overline{\eps}_2 H_M  \right) \nonumber\\
&-  \left( D_M + \frac{1}{4} H_M  \right) \eps_2 \overline\eps_2 \g^M - \eps_2 \left(  D_M \overline{\eps}_2 \g^M  - \frac{1}{4} \overline{\eps}_2 H - \overline\eps_2 \de \phi \right) \nonumber\\
&- \left( \de \phi + \frac{1}{2} H \right) \eps_2 \overline\eps_2 + \eps_2 \overline{\eps}_2 \left( \de \phi - \frac{1}{2} H \right) \, \, , 
\end{align}
from which we obtain
\begin{equation}
\label{eq:eps2epsbar2}
\begin{split}
\e^{ 2\phi} \d (\e^{- 2 \phi} \eps_2 \overline{\eps}_2)  =  \iota_H  \eps_2 \overline{\eps}_2 &-(-)^{|F|}  \frac{\text{e}^{\phi}}{32} \g^M \lambda(\Phi) \g_M  F +(-)^{|F|} \frac{\text{e}^{\phi}}{32} \lambda(F) \g^M \Phi \g_M    \\
&+(-)^{|F|} \frac{\text{e}^{\phi}}{32} \g^M \lambda(F) \g_M \Phi -(-)^{|F|} \frac{\text{e}^{\phi}}{32} \lambda(\Phi) \g^M F \g_M \, \, .
\end{split} 
\end{equation}
From the difference between (\ref{eq:eps1epsbar1}) and (\ref{eq:eps2epsbar2}) we have
\begin{equation}
\label{epsepsbar_diff}
	\begin{split}
	\e^{ 2\phi} \d (\e^{- 2 \phi}( \eps_1 \overline{\eps}_1 - \eps_2 \overline{\eps}_2))  &= - \iota_H (\eps_1 \overline{\eps}_1 + \eps_2 \overline{\eps}_2) -(-)^{|F|} \frac{\text{e}^{\phi}}{32} \bigl( [\g^M F \g_M, \lambda(\Phi) ] \\
	&+ [F, \g^M \lambda(\Phi) \g_M ] + [\g^M \lambda(F) \g_M, \Phi] + [\lambda(F), \g^M \Phi \g_M]  \bigr) \, \, .
	\end{split}	
\end{equation}                        

Now, the calculation will be different depending on the theory. If we are in IIB  we obtain 
\begin{equation}
\label{eq:12f2}
\begin{split}
\e^{ 2\phi} \d (\e^{- 2 \phi}( \eps_1 \overline{\eps}_1 - \eps_2 \overline{\eps}_2))  &= - \iota_H (\eps_1 \overline{\eps}_1 + \eps_2 \overline{\eps}_2) - \text{e}^{\phi} \bigl\{ 2[F_1 , \Phi_1] \\
&+ [F_5 , \Phi_1]  + [F_1 , \Phi_5] - [F_3 , \Phi_3] \bigr\} \, \, ,
\end{split}
\end{equation}
from which, taking the six-form part: 
\begin{equation}
	\begin{split}
	\e^{ 2\phi} \d (\e^{- 2 \phi} \Omega) =& - \iota_H (\g K) - \frac{\e^{\phi}}{2} \left( F_5 \wedge \Phi_1 + F_1 \wedge \Phi_5 - F_3 \wedge \Phi_3 \right) \\
	=& - \iota_K  * H + \e^{\phi} \big( \Phi , F \big)_6 \, \, ,
	\end{split}	
\end{equation}                 
which is \eqref{eq:dOmegaIIA}. If we are in IIA, we have
\begin{equation}
\begin{split}
\e^{ 2\phi} \d (\e^{- 2 \phi}( \eps_1 \overline{\eps}_1 - \eps_2 \overline{\eps}_2))  &= - \iota_H (\eps_1 \overline{\eps}_1 + \eps_2 \overline{\eps}_2) + \frac{\text{e}^{\phi}}{2} \bigl\{ 3[F_2 , \Phi_2] \\
&+ [F_2 , \Phi_6]  + [F_6 , \Phi_2] - [F_4 , \Phi_4] \bigr\} \, \, ,
\end{split}
\end{equation}
where, using the relation
\begin{equation}
[F_4 , \Phi_4]_6 = -2 \Phi_4^M \wedge F_{4 \, M}\,, \qquad  [F_6 , \Phi_2] = -2 \Phi^M_2 \wedge F_{6 \, M}       	
\end{equation}
we get (\ref{eq:domegatII}).

We now turn to the sum of \eqref{eq:eps1epsbar1} and (\ref{eq:eps2epsbar2}):
\begin{equation}
\label{epsepsbar_sum}
	\begin{split}
	\e^{ 2\phi} \d (\e^{- 2 \phi}( \eps_1 \overline{\eps}_1 + \eps_2 \overline{\eps}_2))  &= - \iota_H (\eps_1 \overline{\eps}_1 - \eps_2 \overline{\eps}_2) - (-)^{|F|} \frac{\text{e}^{\phi}}{32} \bigl( \{ \g^M F \g_M, \lambda(\Phi) \} \\
	&+ \{ F, \g^M \lambda(\Phi) \g_M \} - \{\g^M \lambda(F) \g_M, \Phi \} - \{ \lambda(F), \g^M \Phi \g_M \}  \bigr) \, \, .
	\end{split}	
\end{equation}
Again, we distinguish the IIB from the IIA case. In IIB we have
\begin{equation}
\label{eq:22f2}
\begin{split}
\e^{ 2\phi} \d (\e^{- 2 \phi}( \eps_1 \overline{\eps}_1 + \eps_2 \overline{\eps}_2))  &= - \iota_H (\eps_1 \overline{\eps}_1 - \eps_2 \overline{\eps}_2) + \frac{\text{e}^{\phi}}{2} \bigl( 3 \{ F_1 , \Phi_3 \} - \{ F_9 , \Phi_3 \}  \\
&- 3 \{ F_3, \Phi_1 \} - \{ F_7, \Phi_1 \} - \{ F_3 , \Phi_5 \} + \{ F_5 , \Phi_3 \}  \bigr) \, \, .
\end{split} 
\end{equation}
Substituting in this equation the equalities
\begin{equation}
	\begin{split}
	\{ F_9 , \Phi_3 \}_6 &= 2 \, \iota_{\Phi_3} (\g F_1) = - 2 F_{1 \, M} \Phi_7^M\,, \qquad  \{ F_3 , \Phi_5 \}_6 = 2F_{3 \, M } \wedge \Phi_5^M \,,\\
	\{ F_7, \Phi_1 \}_6 &= 2 \, \iota_{\Phi_1} (\g F_3) =  2 \Phi_{1 \, M } F_7^M\,,   \qquad \, \, \, \,    \{ F_5 , \Phi_3 \}_6 =2 \Phi_{3 \, M} \wedge F_5^M\,,
	\end{split}	
\end{equation}
we get (\ref{eq:domegatII}).

On the other hand, for IIA we have
\begin{equation}
\label{IIAKOstarKteq}
\begin{split}
\e^{ 2\phi} \d (\e^{- 2 \phi}( \eps_1 \overline{\eps}_1 + \eps_2 \overline{\eps}_2))  &= - \iota_H (\eps_1 \overline{\eps}_1 - \eps_2 \overline{\eps}_2) - \text{e}^{\phi} \bigl( 2 \{ F_2 , \Phi_0 \} - 2 \{ F_0 , \Phi_2 \}  \\
&+ \{ F_2, \Phi_4 \} + \{ F_6, \Phi_0 \} - \{ F_4 , \Phi_2 \} - \{ F_0 , \Phi_6 \}  \bigr) \, \, ,
\end{split} 
\end{equation}
from which, taking the six-form part, \eqref{eq:dOmegaIIA} follows. 

We finally turn to the last line of  \eqref{eq:systemII}. The first equation, ${\cal L}_K \phi=0$, is obtained by multiplying from the left \eqref{eq:SUSY21} by $\overline{\eps}_1$, \eqref{eq:SUSY22} by $\overline{\eps}_2$, and subtracting the result. 
To derive the second, take the difference of
\begin{equation}
\begin{split}
D_M K_1^M =& - \frac{3}{4 \cdot 32} \, \overline{\eps}_1 H \eps_1 +(-)^{|F|} \frac{\e^\phi}{16 \cdot 32} \, \overline{\eps}_2 \g_M \lambda(F) \g^M \eps_1 \\
+& \frac{3}{4 \cdot 32} \, \overline{\eps}_1 H \eps_1 - \frac{\e^\phi}{16 \cdot 32} \, \overline{\eps}_1 \g_M F \g_N \eps_2 \\
=& - \frac{4\e^\phi}{32^2} \, \overline{\eps}_1 \g_M F \g^M \eps_2
\end{split}
\end{equation}
with
\begin{equation}
D_M K_2^M =  \frac{4\e^\phi}{32^2} \, \overline{\eps}_1 \g_M F \g^M \eps_2 \, ;
\end{equation}
then we get
\begin{equation}
D_M \widetilde{K}^M = - \frac{\e^\phi}{8 \cdot 32}  \overline\eps_1 \g^M F \g_M \eps_2 = (-)^{|\Phi|} \frac{\e^\phi}{8 \cdot 32} \text{Tr} (\lambda(\Phi) \g^M F \g_M) \, ,
\end{equation}
where we used \eqref{eq:lambdaPhi}. The left-hand-side is $* \d * \widetilde{K}$. For the right-hand-side, from (\ref{eq:gamma=}) we write $* \Phi = -(-)^{|\Phi|}\lambda(\Phi)$; using \eqref{eq:CMpair} we obtain the second equation in  (\ref{eq:lastII}).


\subsection{Supersymmetry and integrability}

We proved that the system \eqref{eq:systemII} is necessary and sufficient for supersymmetry for a configuration with timelike Killing vector $K$. However, in general \eqref{eq:systemII} guarantees only part of the equations of motion \cite{lust-tsimpis,gauntlett-martelli-sparks-waldram-ads5-IIB,lust-marchesano-martucci-tsimpis,giusto-martucci-petrini-russo, prins-tsimpis}.

First of all, we impose the Bianchi identity for the $B$-field and the RR fluxes:
\begin{equation} \label{BIid}
\d H=0\,,\qquad \d_H F=0\,.
\end{equation}  
(These equations must be appropriately corrected in presence of  localized sources.)
By using the results of \cite{lust-marchesano-martucci-tsimpis} as in appendix C of \cite{giusto-martucci-petrini-russo}, one can then prove that the supersymmetry implies the dilaton's equation of motion
and  the spinorial equations  
\begin{equation} \label{intcond}
(\cale_{MN}-\frac12\calh_{MN})\gamma^N\epsilon_1=0 \, ,\qquad
(\cale_{MN}+\frac12\calh_{MN})\gamma^N\epsilon_2=0\,,
\end{equation} 
where $\cale_{MN}=0$ gives the string-frame trace-reversed Einstein equations, while the vanishing $\calh_{MN}$ corresponds to the $B$-field equations of motion:
\begin{equation}
\calh\equiv \frac12 \calh_{MN}\d x^M\wedge\d x^N =e^{2\phi}*\Big[\d(e^{-2\phi}*H)-\frac12(F,F)_8\Big]=0\,.
\end{equation} 
As discussed in appendix \ref{app:proof}, we may choose a vielbein $e^a=(e^+,e^-,e^\alpha)$ such that $e^+$ and $e^-$ are proportional to the one-forms $K_1$ and  $K_2$ respectively. Remember that $\gamma^+\epsilon_1=\gamma^-\epsilon_2=0$, while $(\gamma^-\epsilon_1,\gamma^\alpha\epsilon_1)$ and   $(\gamma^+\epsilon_2,\gamma^\alpha\epsilon_2)$ give two sets of linearly independent spinors. Hence \eqref{intcond} implies the following components of the equations of motion:
\begin{equation}
\cale_{++}=\cale_{--}=\cale_{M\alpha}=\calh_{M\alpha}=0\,,
\end{equation}
together with 
\begin{equation}
\cale_{+-}=\frac12\calh_{+-}\,.
\end{equation}
Hence, once we have imposed \eqref{eq:systemII} and \eqref{BIid}, in order to solve the complete set of equations of motion, it remains to impose 
either $\cale_{+-}=0$ or $\calh_{+-}=0$. The latter condition may be written as
\begin{equation} \label{eq:Heq_integrability}
K\wedge \tilde K\wedge \Big[\d(e^{-2\phi}*H)-\frac12(F,F)_8\Big]=0\,,
\end{equation}
where the index of $K$ has been implicitly lowered by using the metric, while one can check that $\cale_{+-}=0$ is implied by
\begin{equation}
 \label{eq:EE_integrability}
\square e^{-2\phi} - e^{-2 \phi} H^2 -\frac{1}{4} \sum_{k} k F_k^2 = 0\,,
\end{equation}
which is a combination of the trace of the Einstein equation with the dilaton equation of motion.




\section{Dualities} 
\label{sec:dualities}

In this section we will discuss the action of T-duality and the type IIB SL($2,\mathbb{Z}$) duality on the geometric objects entering \eqref{eq:systemII}; moreover, we will also discuss the duality between M-theory and type IIA. The system \eqref{eq:systemII} is not manifestly invariant under the most general duality transformation. One can overcome this problem by combining  \eqref{eq:systemII}  with other (non-independent) equations which follow from it.  In particular,  we will see how \eqref{eq:systemII} for IIB backgrounds  can be replaced by a system which is manifestly  SL($2,\mathbb{Z}$) duality invariant.

\subsection{SL($2,\mathbb{Z}$) duality}
\label{sec:s}

We start by focusing on the IIB theory. As is well known, it enjoys an SL($2,\mathbb{Z}$) symmetry. The general element 
\begin{equation} \label{SL}
\left(\begin{array}{cc}  
\alpha & \beta \\
\gamma & \delta
\end{array}
\right)\in\text{SL($2,\mathbb{Z}$)} \qquad (\alpha\delta-\beta\gamma=1)
\end{equation}
acts on the axion-dilaton as 
\begin{equation}
\tau'= \frac{\alpha\tau+\beta}{\gamma\tau+\delta}\,,\qquad \tau\equiv C_0+\ii e^{-\phi} \, .
\end{equation}
The RR self-dual 5-forms $F_5$ is invariant  under  \eqref{SL}, while the  2-form potentials $(C_2,B_2)$ transform as a doublet. It is convenient to combine the corresponding field-strengths in the complex three-form
\begin{equation}
\calg_3\equiv e^{\frac12\phi}G_3\equiv e^{\frac12\phi}(F_3-\ii e^{-\phi}H)\,.
\end{equation}
One can then check that $\calg_3$ transforms by a phase under SL($2,\mathbb{Z}$):
\begin{equation}
\calg_3'= e^{-\ii\theta}\calg_3\,,\qquad \theta\equiv \arg(\gamma\tau+\delta) \, .
\end{equation}
We may think of $\calg_3$ as having charge $-1$ under the $\mathrm{U}(1)_{D}$ transformation defined by the phase $e^{\ii\theta}$; more generally, we will say that
a field has charge $q$ under $\mathrm{U}(1)_D$ if it transforms by a phase $e^{\ii q\theta}$. As another example, the one-form $e^\phi\d \tau$ has $\mathrm{U}(1)_D$-charge $q=-2$, that is
\begin{equation}
e^{\phi'}\d \tau'=e^{-2\ii\theta}(e^\phi\d \tau)\,.
\end{equation}

Notice that the $\mathrm{U}(1)_D$ transformations are typically point-dependent, since  $\tau$ is in general non-constant, and then they  do not commute with ordinary derivatives. One can however construct a composite compatible connection 
\begin{equation}
Q\equiv\frac12 e^\phi F_1
\end{equation}
and an associated covariant derivative $\del_M-\ii qQ_M$. In particular, we will need the covariant exterior derivative
\begin{equation} \label{d_Q}
\d_Q\equiv\d-\ii qQ \wedge \,.
\end{equation}

It is convenient to use the Einsten-frame metric
\begin{equation} \label{Eg}
g_{\rm E}\equiv e^{-\frac12\phi}g
\end{equation}
which is invariant under SL($2,\mathbb{Z}$) dualities. Finally,  the spinors $\epsilon_1,\epsilon_2$ transform in such a way that the complex combination $e^{-\frac12\phi}(\epsilon_1+\ii\epsilon_2)$ has $\mathrm{U}(1)_{D}$-charge $q=\frac12$ \cite{Schwarz:1983qr}.

By using the transformation rules of metric and spinors, we can  compute the transformation properties of the fields $K,\tilde K,\Phi,\Omega,\tilde\Omega$ appearing in  \eqref{eq:systemII}. It is  easy to check that the Killing vector $K$, the three-form $ e^{-\phi}\Phi_3 \equiv \Theta_{3}$ and the five-form $e^{-\frac32\phi}\widetilde\Omega \equiv \widetilde\Omega_{\rm E}$ are invariant under   SL($2,\mathbb{Z}$) duality, while the other forms  get mixed.
 It is then useful to express them in terms of the complex combinations 
\begin{equation}
\Theta_1\equiv e^{-\frac12\phi}(\widetilde K+\ii\Phi_1)\,,\qquad \Theta_5\equiv e^{-\frac32\phi}(\Omega + \ii \Phi_5)\,,
\end{equation}
and their Hodge-duals, which transform with definite $\mathrm{U}(1)_D$ charge $q=1$.  We have then reorganized all the relevant fields in combinations transforming  with definite  $\mathrm{U}(1)_D$-charges, summarized in table \ref{table:charges}.

\begin{table}[h!]
\centering
\begin{tabular}{ |c|c| } 
 \hline
 fields & $\mathrm{U}(1)_D$-charge  \\ 
  \hline
$g_{\rm E}$, $K$, $\Theta_{3}$, $\widetilde\Omega_E$, $F_5$ & $\ 0$  \\ 
$\Theta_1$, $\Theta_5$  & 1\\
$\calg_3$ & $-1$\\
 $e^\phi\d\tau$ & $-2$  \\ 
\hline
\end{tabular}\caption{$\mathrm{U}(1)_D$ charges of relevant fields.}
\label{table:charges}
\end{table}

Notice that the Einstein frame Hodge-operator $*_{\rm E}$ commutes with the duality transformation so that, for instance, $*_{\rm E}\,\calg_3$ has  $\mathrm{U}(1)_D$-charge $-1$.
These transformation rules will acquire a clear physical interpretation when we will identify the above differential forms in terms of calibrations for various extended objects that transform in a precise way under SL($2,\mathbb{Z}$) duality.

By manipulating  \eqref{eq:systemII} with the help of the (redundant) algebraic equations of appendix \ref{sec:algebraic} and \eqref{eq:string_cal}, we get the system of equations 
\begin{subequations}
	\label{Sl2Rsystem}
	\begin{align}
	&\mathcal{L}_K \tau =0\,,\quad \qquad e^\phi\d \tau \wedge *_{\rm E} \Theta_1+\frac\ii2\, \calg_3 \wedge *_{\rm E} \Theta_3 =0 \, ,\\
	&\d_Q\Theta_1-\frac\ii2e^\phi\d\overline\tau\wedge\overline\Theta_1 +\ii\, \iota_K\overline\calg_3 = 0 \, , \\
	&\d\Theta_3 + \iota_K F_5 + \text{Re}\big(\Theta_1 \wedge \calg_3\big)=0 \, ,\\
	& \d_Q \Theta_5+\frac\ii2 e^\phi \d \overline\tau\wedge\overline\Theta_5 + \Theta_3 \wedge \overline\calg_3 - \ii \iota_K (*_{\rm E}\, \overline\calg_3)+ \ii \Theta_1 \wedge F_5 =0 \, , \\
	&\d *_{\rm E} \Theta_3 +\frac12\text{Re} \left( \calg_3 \wedge \Theta_5- *_{\rm E}\, \calg_3 \wedge \Theta_1 \right) = 0\,, \\
	& \d_Q *_{\rm E} \Theta_1-\frac\ii2 e^{ \phi}\d\overline\tau\wedge  *_{\rm E} \overline{\Theta}_1 =0 \, , \label{Sinvariant9brane}\\
	&\d\widetilde\Omega_E + \frac14g^{MN}_{\rm E}\left[\text{Im}(\Theta_{5 \, M} \wedge \calg_{3\,N})-2 \Theta_{3 \, M} \wedge F_{5\,N}\right] - 3 *_{\rm E}\text{Im}(\Theta_1\wedge \calg_3) =0 \, .
	\end{align}
\end{subequations}
According to the general definition \eqref{d_Q} and the $\mathrm{U}(1)_D$-charges of table \ref{table:charges},  $\d_Q\equiv \d-\ii Q\equiv \d-\frac\ii2e^\phi F_1 \wedge$. 
From table \ref{table:charges} it is  also easy to see  that the system is manifestly  SL($2,\mathbb{Z}$) invariant.  

The system  \eqref{Sl2Rsystem} contains more equations than  \eqref{eq:systemII}. However, having used additional (redundant) equations,  the equivalence with supersymmetry may not be guaranteed anymore. Therefore, to be sure that none of the supersymmetry data is lost, one should check that the following algebraic constraints are satisfied:
\begin{equation}\label{auxalg}
\begin{split}
&g^{MN}_{\rm E}(\overline\calg_{3\,M} \wedge \Theta_{5 \, N}) - *_{\rm E}( \overline\calg_3\wedge \Theta_1)-2 e^{\phi}\d \overline\tau \wedge\widetilde\Omega_E + 2 \ii *_{\rm E}(e^{\phi}\d \overline\tau\wedge \Theta_3) =0 \, , \\
& \overline\calg_3 \wedge \Theta_5-\Theta_1\wedge *_{\rm E} \overline\calg_3+2 e^\phi \iota_K *_E\d \overline\tau + 2 \ii e^{\phi} \d \overline\tau \wedge *_{\rm E}\Theta_3 =0 \, ,
\end{split}
\end{equation}
which are complex combinations of  \eqref{eq:algebraicK_Omega}, \eqref{6Dalgebraiceq} and \eqref{8Dalgebraiceq}.
Again, by using  the $\mathrm{U}(1)_D$-charges of table \ref{table:charges} one can easily check that \eqref{auxalg} are manifestly invariant under SL($2,\mathbb{Z}$) dualities. 

While the system (\ref{Sl2Rsystem})--(\ref{auxalg}) we just presented might look alarmingly large, it lists separately each form degree, unlike for example (\ref{eq:systemII}).

\subsection{T-duality}
\label{sub:T-dual_main}

Type II theories with $d$ commuting isometries are characterized by an $\mathrm{O}(d,d;\mathbb{Z})$ group of T-dualities. Any  element of the $\mathrm{O}(d,d;\mathbb{Z})$ T-duality can be decomposed into a product of `simple' T-dualities along a given isometry, discrete diffeomorphisms and  shifts of the $B$-field.  We can then focus on the action of a simple T-duality along a certain Killing direction, parametrized by a coordinate  $y$.

Let us then split the coordinates as  $x^{M}=(x^m,y)$, with $m=0,\ldots,8$, and assume that all the geometric quantities do not depend on $y$. In order to describe the action of T-duality on bosonic fields it is convenient to define 
\begin{equation}
\begin{aligned}
	&E_{MN}\equiv g_{MN}-B_{MN}\,,\\
	&F_{\rm tw}\equiv e^{-B}\wedge F\,;
\end{aligned}
\end{equation}
the latter satisfies $\d F_{\rm tw}=0$ away from localized sources. Locally we can set $F_{\rm tw}=\d C_{\rm tw}$, with $C_{\rm tw}\equiv e^{-B}\wedge C$.
T-duality along $y$ gives \cite{buscher,buscher2}
\begin{equation} \label{TNS}
\begin{aligned}
	&g'_{yy}=\frac{1}{g_{yy}}\,,\qquad g'_{my}=-\frac{B_{my}}{g_{yy}}\,,\qquad B'_{my}=-\frac{g_{my}}{g_{yy}}\,,\\
	& E'_{mn}=E_{mn}-\frac{1}{g_{yy}}E_{ym}E_{ny}\,,\qquad e^{-2\phi'}= g_{yy}\,e^{-2\phi}
\end{aligned}
\end{equation}
on the type NS-NS bosonic sector, while the RR forms transform as, see e.g.\ \cite{hassan}:
\begin{equation} \label{TRRtw}
C'_{\rm tw}=\calt_y\cdot C_{\rm tw} \, , \qquad 
F'_{\rm tw}=\calt_y\cdot F_{\rm tw}\,.
\end{equation}
We have introduced the operator
\begin{equation} \label{genT}
\calt_y\cdot \equiv (\d y\wedge - \iota_{\del_y}) (-)^{\text{deg}} \, , \qquad \calt_y^2 = 1\,.
\end{equation}	
T-duality admits a natural formulation in terms of generalized complex geometry (see for instance \cite{gmpt3,grana-minasian-petrini-waldram}).
In particular, we can regard $C_{\rm tw}$ and $F_{\rm tw}$ as spinors associated with the $B$-twisted generalized tangent bundle. $\calt_y$ may be considered as a generalized vector of the  ($B$-twisted) generalized tangent bundle whose action \eqref{TRRtw} on $C_{\rm tw}$ and $F_{\rm tw}$ defines the spinorial $\mathrm{O}(10,10)$ representation. This observation can be immediately extended to a more general $\mathrm{O}(d,d,\mathbb{Z})$ T-duality group  \cite{hassan}, regarded as a  subgroup of $\mathrm{O}(10,10)$.

The action on $K,\tilde K, \Phi,\Omega,\tilde\Omega$ can be computed by using the spinorial T-duality rules derived by Hassan in \cite{hassan-RR}.\footnote{The relation with Hassan's conventions  in \cite{hassan-RR} is $\epsilon_1=\epsilon^{\rm H}_-\,,\quad \epsilon_2=\epsilon^{\rm H}_+, F=\lambda(F^{\rm H}),C=(-)^{\rm deg}\lambda(C^{\rm H}), H=-H^{\rm H}$. Furthermore  in \cite{hassan-RR} the T-duality transformation is defined up to an arbitrary  choice of sign, which corresponds to the choice of orientation 
	of the T-duality direction $y\equiv x^9$. We fix this ambiguity by choosing $a_{(A-B)}=-a_{(B-A)}=-1$ in Hassan's formulas.} In particular, the T-duality along $y$
transforms the spinors to $\epsilon'_1=\epsilon_1$ and $\epsilon'_2=T_y\epsilon_2$,   with $T_y\equiv - \frac{1}{\sqrt{g_{yy}}}\gamma_y$.
By using these formulas and the appropriate transformation rules for the vielbein  \cite{hassan-RR}, we obtain the simple T-duality rule\footnote{In components: $(e^{-\phi'}\Phi'_{\rm tw})_{m_1\ldots m_p}=-(e^{-\phi}\Phi_{\rm tw})_{ym_1\ldots m_p}$, $(e^{-\phi'}\Phi'_{\rm tw})_{ym_1\ldots m_p}=(e^{-\phi}\Phi_{\rm tw})_{m_1\ldots m_p}$.}
\begin{equation} \label{PhiT}
e^{-\phi'}\Phi'_{\rm tw}=\calt_y\cdot e^{-\phi}\Phi_{\rm tw} \, ,\qquad \Phi_{\rm tw}\equiv e^{-B}\wedge \Phi\,.
\end{equation}
We then see that $e^{-\phi}\Phi_{\rm tw}$, which can also be regarded as a spinor of the generalized tangent bundle, transforms exactly like $F_{\rm tw}$ under a simple T-duality. This correspondence clearly holds for discrete diffeomorphisms and shifts of $B$ too. 
Hence, a more general $\mathrm{O}(d,d,\mathbb{Z})$ T-duality group acts on $e^{-\phi}\Phi_{\rm tw}$ in the $\mathrm{O}(10,10)$ spinorial representation. 

The T-duality  action on $K$ and $\widetilde K$ can also be naturally described by using the language of generalized geometry. First of all, we may organize them in the generalized vector
\begin{equation}\label{calk}
\calk=K+\omega \, ,\qquad \omega\equiv \widetilde K+\iota_K B \, .
\end{equation} 	
By using Hassan's rules, one can check that $\calk$ transforms as an $\mathrm{O}(10,10)$ vector under a simple T-duality:
\begin{equation} \label{TT}
\calk'= \calk+2\cali(\calk,\calt_y)\calt_y 
\end{equation}
where $\cali$ denotes the canonical $(10,10)$ metric on the generalized tangent bundle, such that $\cali(v+\eta,v+\eta)=2\eta(v)$. In components, (\ref{TT}) reads $(K^{m})'=K^m$, $(K^{y})'=\omega_y$, $\omega'_{m}=\omega_m$, $\omega'_y=K^y$. One can check that \eqref{TT}
indeed defines an $\mathrm{O}(10,10)$ transformation, in the sense that $\cali(\calk',\calk')=\cali(\calk,\calk)$ since $\calk' =\calt_y\calk\calt_y$. As above, we can consider more general  $\mathrm{O}(d,d,\mathbb{Z})$ T-duality groups under which $\calk$ transforms in the fundamental  $\mathrm{O}(10,10)$ representation. 

Unfortunately  $\Omega$ and $\widetilde{\Omega}$ transform in a less nice form. The action of a simple T-duality in the $y$-direction can be written as: 
\begin{equation} \label{OmegaTT}
\begin{aligned}
	e^{-2\phi'}\Omega'_y&=e^{-2\phi}\Omega_y\,, \qquad	e^{-2\phi'}(\d y\wedge \Omega')_y=\,e^{-2\phi}(\eta\wedge\widetilde\Omega-\iota_yB\wedge\Omega)_y\,,\\
	e^{-2\phi'}\widetilde\Omega'_y&=e^{-2\phi}\widetilde\Omega_y\,, \qquad	e^{-2\phi'}(\d y\wedge \widetilde\Omega')_y= \,e^{-2\phi}(\eta\wedge\Omega-\iota_yB\wedge\widetilde\Omega)_y\,,
\end{aligned}
\end{equation} 
where $(\ )_y \equiv \iota_{\partial_y}$, and  we have introduced the one-form
\begin{equation}
\eta\equiv (\del_y)_M\d x^M=g_{yM}\d x^M\,.
\end{equation} 

We observe that the supersymmetry condition \eqref{eq:oldII} can be rewritten in the twisted form
\begin{equation}
\d (e^{-\phi}\Phi_{\rm tw})=-\calk\cdot F_{\rm tw}\,,
\end{equation} 
which is manifestly invariant by using the above T-duality rules and the fact that the Lie derivative along $y$ on all fields gives zero. The transformation of the remaining equations of  \eqref{eq:systemII} is less obvious and is discussed in appendix \ref{sub:t}.  The last line \eqref{eq:lastII}  and the $y$-longitudinal parts of  \eqref{eq:dOmegaIIA} and \eqref{eq:domegatII} remain invariant. On the other hand, the $y$-transversal parts of \eqref{eq:dOmegaIIA} and \eqref{eq:domegatII} transform in a more complicated way, as one might guess from \eqref{OmegaTT}. We will come back to this point when we will discuss the interpretation of   \eqref{eq:systemII} in terms of calibrations.


\subsection{Type IIA/M-theory duality and forms}
\label{sub:IIAM}

In this section we spell out the relations between the forms entering our system \eqref{eq:systemII} in the type IIA case  
and the forms that can be naturally be constructed from bispinors in M-theory.  These relations connect our results to those of \cite{gauntlett-pakis} and will be useful in the following. 

We adopt the same M-theory conventions of \cite{gauntlett-pakis} besides the use of a hat that will be useful to distinguish the eleven-dimensional objects from the ten-dimensional ones. One can use the Majorana supersymmetry generator $\hat\epsilon$ to construct the vector
\be
\begin{aligned}
\widehat K=\frac{1}{2^5} \,\overline{\hat{\eps}}\, \widehat{\Gamma}^M\,\hat\epsilon\,\del_M
\end{aligned}  
\ee
and the two- and five- forms
\begin{subequations}
\label{hatforms}
\begin{align}
\widehat{\Omega} &= \frac{1}{2^5\cdot 2!} \, \overline{\hat{\eps}} \, \widehat{\Gamma}_{M_1 M_2} \, \hat{\eps} \, \d x^{M_1}\wedge \d x^{M_2}  \,,\label{hatforms1}\\
\widehat{\Sigma} &= \frac{1}{2^5\cdot 5!} \, \overline{\hat{\eps}} \, \widehat{\Gamma}_{M_1 \dots M_5} \, \hat{\eps} \, \d x^{M_1} \wedge \dots \wedge \d x^{M_5}\,.\label{hatforms2} 
\end{align}  
\end{subequations}
By imposing that $\hat\epsilon$ is a Killing spinor, one can show that $\widehat K$ is a Killing vector and that $\widehat{\Omega}$ and 
$\widehat{\Sigma}$ satisfy a set of differential conditions \cite{gauntlett-pakis}. In particular, in the following we will need the equations:
\begin{subequations}\label{Meq}
\begin{align}
\d\widehat{\Omega} &= \iota_{\widehat K}\widehat F  \,,\label{Meq1}\\
\d\widehat{\Sigma} &=  \iota_{\widehat K}\hat*\,\widehat F- \widehat{\Omega}\wedge \widehat F \label{Meq2}\,,
\end{align}  
\end{subequations}
 where $\hat F=\d \hat A$ is the M-theory four-form field-strength.

As usual, in order to connect M-theory to type IIA,  we now perform a dimensional reduction following \cite{saffin} and \cite[Chap.~8]{becker-becker-schwarz}, which are consistent with our conventions except for $\eps_{1,2} \rightarrow \eps_{2,1}$ , $C_1 \rightarrow - C_1$ and $H \rightarrow -H$. In particular, the metric and the supersymmetry parameter split as follows 
\begin{subequations}
\label{11Dconv}
\begin{align}
&\d \hat s^2 = e^{- \frac{2}{3} \phi} \d s^2 + e^{\frac{4}{3} \phi} (\d x^{10} - C_1)^2\,, \\
&\hat{\eps} = \frac{1}{\sqrt{2}}e^{-\frac{1}{6} \phi} (\eps_1 + \eps_2)\, , \qquad \Gamma_{\underline{10}} \eps_1 = - \eps_1 \, ,
\end{align}
\end{subequations}
while for $\widehat{A}$ and the associated field-strength we have:
\begin{equation}
\begin{split}
&\widehat{A} = C_3 - B \wedge \d x^{10}\,, \\
&\widehat{F} =  F_4 - H \wedge (\d x^{10} - C_1)\, .
\end{split}
\end{equation}

By using \eqref{11Dconv}, one can then identify the relations between M-theory and IIA geometrical structures
\begin{subequations}
\label{11-10Drel}
\begin{align}
\widehat K&= K-e^{-\phi}\Phi_0\,\del_{10}\,, \label{11-10Drel-a}\\
\widehat{\Omega} &= -  e^{- \phi} \Phi_2- \widetilde{K} \wedge  (\d  x^{10} - C_1)\,,\label{11-10Drel-b}\\
\widehat{\Sigma} &=e^{-2 \phi} \Omega - e^{- \phi} \Phi_4 \wedge  (\d  x^{10} - C_1)\label{11-10Drel-c}\,.
\end{align}
\end{subequations}
We clearly see the compatibility of \eqref{11-10Drel-a} with the fact that $K$ describes a Killing vector in IIA. Furthermore, one can use \eqref{11-10Drel} into \eqref{Meq} to derive part of the supersymmetry conditions appearing in  \eqref{eq:systemII}. Notice also that the IIA forms which do not appear on the r.h.s.\ of \eqref{11-10Drel} can be obtained by dimensionally reducing the Hodge-duals of $\widehat{\omega}$ and $\widehat{\Sigma}$ in M-theory along the same lines. For example, in the following we will need
\be
\label{*hatSigma}
\hat*\widehat{\Sigma} =-e^{-2 \phi} \tilde\Omega\wedge (\d x^{10} - C_1) - e^{- \phi} \Phi_6 \,.
\ee


\section{Brane calibrations} 
\label{sec:d-cal}

It is known that \eqref{eq:oldII} has an interpretation in terms of D-brane calibrations \cite{martucci-electrified}. Let us give here a lightning review of this. In what follows, we will partially interpret the other equations in (\ref{eq:systemII}) in terms of calibrations.

In the original definition in Riemannian geometry \cite{harvey-lawson}, a calibration $\omega$ on a manifold $M$ is a $p$-form such that \emph{i)} $\omega|_N \le \mathrm{vol}_N$ for any $p$-dimensional subspace $N$ of the tangent space $T_x M$ at any point $x$, and \emph{ii)} $\d \omega = 0$. The idea of the definition is that if a calibrated submanifold $\Sigma$ exists, namely a $\Sigma$ such that the tangent space $T_x \Sigma\subset T_x M$ at any point obeys $\omega|_\Sigma = \mathrm{vol}_\Sigma$, then $\Sigma$ has minimal volume in its homology class. Indeed, given a ``deformed'' $\Sigma'$ in the same homology class of $\Sigma$, call $\Gamma$ a $p+1$ submanifold whose boundary $\del \Gamma= \Sigma-\Sigma'$; we can then write
\begin{equation}\label{eq:min-vol}
	 \mathrm{Vol}(\Sigma)- \mathrm{Vol}(\Sigma')= \int_\Sigma \mathrm{vol}- \int_{\Sigma'}\mathrm{vol} \le \int_\Sigma \omega - \int_{\Sigma'} \omega = \int_\Gamma \d \omega  = 0\,.
\end{equation}   

Calibrations show up naturally in string theory in various contexts, in particular dealing with solitonic objects that appear in the supergravity algebra; they are usually obtained as spinor bilinears, and calibrated submanifolds are wrapped by branes which obey BPS conditions. These are sometimes called ``generaralized'' calibrations \cite{gutowski-papadopoulos-townsend}, perhaps confusingly in the present context; a better name might be almost-calibrations since condition \emph{ii)} above is not met. In such a case, the failure of $\omega$ to be closed is related to the presence of a flux $F$. Schematically,
\begin{equation}\label{eq:gen-cal}
\d\omega=-\iota_K F	
\end{equation}
where  $F=\d C$ with $C$ a $(p+1)$-form potential and $K$ is the time-like Killing  associated to the supersymmetric configuration, as in (\ref{eq:KKt-def}). (Recall that in this paper we consider it to be timelike, even though the discussion applies also for general $K$, as argued in \cite{martucci-electrified}.)  For instance, \eqref{eq:string_cal} has exactly the form \eqref{eq:gen-cal}. The analogue of condition \emph{i)} above leads to minimizing the brane energy $-K^M \mathcal{P}_M$ (where $\calp_M$ is the brane momentum) rather than its volume, and (\ref{eq:min-vol}) is translated to the BPS bound:
\begin{equation}\label{BPSbound_intro}
-\int_\Sigma K^M \mathcal{P}_M \text{vol} \ge \int_\Sigma (\omega - \iota_K C)|_\Sigma \, .
\end{equation}
Choosing the gauge $\mathcal{L}_K C = 0$, we can write \eqref{eq:gen-cal} as 
\begin{equation}
\label{eq:cal_ex_intro}
\d\varphi = 0 \quad\text{with}\quad \varphi=\omega - \iota_K C \,;
\end{equation} 
 therefore the right-hand side  of \eqref{BPSbound_intro} is a topological quantity, which we interpret as the brane central charge. We refer to section \ref{sub:grav-cal} and to \cite{martucci-electrified} for further details about this brief discussion. Notice that, generically, $\varphi$ is not a globally defined differential form since the potential $C$ may be only locally defined. However this subtlety will not play a role in our discussion and in the following we will loosely  call calibration forms objects like $\varphi$ in \eqref{eq:cal_ex_intro}.

In \cite{martucci-smyth} it was shown that all the pure spinor equations for Mink$_4$ or AdS$_4$ compactifications \cite{gmpt2,gmpt3}, which are equivalent to the background supersymmetry, can be interpreted in terms of calibrations or generalized calibrations for D-branes extended along different numbers of spacetime dimensions. Since, as we mentioned, the pure spinor equations follow from (\ref{eq:oldII}), the latter also have a direct interpretation in terms of D-branes  \cite{martucci-electrified}. Indeed, with the help of \eqref{eq:string_cal}, one can check that (\ref{eq:oldII}) is equivalent to the closure of the forms
\be\label{Dpcal}
\varphi_{\text{D$p$}}=\left[e^{-\phi} \Phi_{\rm tw} - (\iota_{K} + \omega \wedge)C_{\rm tw}\right]_p \, ,
\ee
where we recall that $\Phi_{\rm tw}\equiv e^{-B} \wedge \Phi$, $C_{\rm tw}=e^{-B} \wedge C$ and $\omega=\tilde K+\iota_K B$, as in section \ref{sub:T-dual_main}, and we are assuming a gauge in which all form potentials like $C$ and $B$ are vanishing under $\call_K$. The forms \eqref{Dpcal}  are the D-brane counterpart  of $\varphi$ in \eqref{eq:cal_ex_intro}\footnote{The analogy between \eqref{Dpcal} and \eqref{eq:cal_ex_intro} is more explicit in terms of generalized complex geometry, i.e. considering $\iota_{K} + \omega \wedge$ as a generalized Killing vector (instead of $\iota_K$ alone) as in \cite{martucci-electrified}.}. 
For D-branes, the role of calibrations and the generalized calibration inequality \eqref{BPSbound_intro} can be argued rather directly using the D-brane action and kappa-symmetry as shown in \cite{martucci-electrified}. 
NS-branes are a little more subtle, since they do not have a direct definition in terms of open strings, but only as solitonic supergravity solutions. In this section, we will overcome this by using dualities. This issue will present itself again in the next section for KK5-monopoles; we will attack it there with a more general discussion of calibrations in a purely gravitational context.

Most of the present section will be now dedicated to provide an interpretation for (\ref{eq:dOmegaIIA}) in terms of calibrations for NS5-branes. Moreover, we will also briefly discuss the calibration condition for the more exotic NS9-brane. 

\subsection{Closure condition for NS5 calibration} 
\label{sub:ns5cal}

The purpose of this subsection is to rewrite (\ref{eq:dOmegaIIA}) as
\begin{equation}\label{dvarphi}
	\d\varphi_{\rm NS5}=0
\end{equation}
for a certain form $\varphi_{\rm NS5}$, which we will then interpret as NS5 calibration, in analogy with \eqref{eq:cal_ex_intro}. In the next two subsections we use the duality transformations discussed in section \ref{sec:dualities} to test this interpretation.

Let us first find a potential for the NS5 brane. Computations will be performed in type IIB and massless type IIA simultaneously. Consider the equation of motion for the NSNS three-form $H$: 
\begin{equation}
\label{H_EOM}
\d (e^{-2 \phi} * H) - \frac{1}{2} (F,F)_8 = 0\,,
\end{equation}
where recall $(A,B)_d \equiv (A\wedge \lambda(B))_d$.
Using also the property $(A,B)_d = (-)^{d(d-1)/2}(B,A)_d$, we can rewrite $(F,F)_8 =  - \d(F,C)_7$
and therefore \eqref{H_EOM} reads
\begin{equation}
\d \Big[( e^{-2 \phi} * H + \frac{1}{2} (F,C)_7  \Big]= 0 .
\end{equation}
This means that we can locally define the NS5 potential $\widetilde{B}$ such that:
\begin{equation}\label{eq:B}
\d \widetilde{B} = e^{-2 \phi} * H + \frac{1}{2} (F,C)_7. 
\end{equation}

Let us use (\ref{eq:B}) in (\ref{eq:dOmegaIIA}):
\begin{equation}
\d \left(e^{-2 \phi} \Omega \right) = -e^{-2 \phi} \iota_K * H + (e^{-\phi} \Phi , F)_6 = - \iota_K \d \widetilde{B} +  \iota_K  (F,C)_7  + (e^{-\phi} \Phi , F)_6\,.
\end{equation}
By recalling \eqref{eq:oldII}, we can manipulate $ (e^{-\phi} \Phi , F)_6$ a little bit 
so as to get
\begin{equation}
\label{NS5cal_intermediate}
\begin{aligned}
\dd \Big[\e^{-2 \phi} \Omega +  (\e^{-\phi} \Phi , C)_5  - \iota_K \widetilde{B} \Big] &=\frac{1}{2} (\iota_K F,C)_6 - \frac{1}{2} ( \iota_KC,F)_6  - ((\iota_K+\widetilde{K}\wedge)  F , C)_6 \\
&=\frac{1}{2} \dd ((\iota_K + \widetilde{K} \wedge)C , C)_5\,,
\end{aligned}
\end{equation}
In the first line we have chosen the gauge $\mathcal{L}_{K} \widetilde{B} = 0$, while in order to go to the second line  we have used the identity  $\{ \d_H ,  \iota_K  + \widetilde{K} \wedge \} = \mathcal{L}_K$ and chosen the gauge $\mathcal{L}_K C = 0$ for the RR potentials. 

Thus as promised (with the help of \eqref{eq:oldII}) we have rewritten (\ref{eq:dOmegaIIA}) in the form \eqref{dvarphi}, with
\begin{equation}
\label{NS5calibration}
\varphi_{\rm NS5} = e^{-2 \phi} \Omega +  (e^{-\phi} \Phi , C)_5 - \iota_K \widetilde{B}  - \frac{1}{2} \widetilde{K} \wedge (C,C)_4  - \frac{1}{2} (\iota_K C,C)_5 \, .
\end{equation}
In order to turn on the Romans mass in type IIA it is enough to repeat the same steps recalling that in this case $F = \dd_H C+ \e^{B} F_0$; the  outcome of this operation is
\begin{equation}
\label{NS5calibration_MassiveIIA}
\varphi_{\rm{NS5}_{\rm IIA}} = e^{-2 \phi} \Omega +  (e^{-\phi} \Phi , C)_5 - \iota_K \widetilde{B}  - \frac{1}{2} \widetilde{K} \wedge (C,C)_4  - \frac{1}{2} (\iota_K C,C)_5 + F_0 \sigma_5  \, ,
\end{equation}
where $\sigma_5$ is a five-form defined by $\varphi_{\rm D6} = \dd \sigma_5$ and $\varphi_{\rm D6}$ is the D6-brane calibration introduced in \eqref{Dpcal}. 
Given the appearance of the NS six-form potential $\widetilde{B}$, it is natural to interpret \eqref{NS5calibration} as calibration for a NS5-brane. In the same spirit, the last term in \eqref{NS5calibration_MassiveIIA} is in correspondence with the fact that  an isolated NS5-brane is anomalous in a Romans-mass background and requires the insertion of $F_0$ D6-branes ending it.
The interpretation of \eqref{NS5calibration} will be confirmed by exploiting the duality relations with other calibrations. 


\subsection{NS5 calibration in IIB from S-duality} 
\label{sub:ns5iib}

It is well known that type IIB S-duality relates  NS5-branes to D5-branes, which have well defined effective actions and calibrations.  We can then use S-duality to check our interpretation of  \eqref{NS5calibration} as calibration for IIB NS5-branes. S-duality is a   subgroup of the  $\mathrm{SL}(2,\mathbb{Z})$-duality group \eqref{SL}  generated by the element
\begin{equation}\label{Sduality}
\left(\begin{array}{rr}  
0 & -1 \\
1 & 0
\end{array}
\right)\, .
\end{equation} 
This S-duality transformation should then transform  the D5-brane calibration  to the NS5 one. 

We recall that the complete calibration for D-branes is actually given by the sum of  forms \eqref{Dpcal} of various (even/odd) degrees,  which allow to describe the energetics of  D-branes supporting non-trivial fluxes \cite{koerber,martucci-smyth,koerber-martucci-ads,martucci-electrified} and/or forming networks  \cite{evslin-martucci}. By duality, we expect the same to be true for NS5-branes as well but, for simplicity, 
in the following we will consider just isolated NS5-branes and D5-branes on which the world-volume flux can be consistently set to zero. In such a case, we can restrict our attention on the highest-rank contribution to the complete D5-brane calibration, which is given by $\varphi_{\rm D5}$ as defined in \eqref{Dpcal}.
By expanding $\Phi_{\rm tw}\equiv e^{-B} \wedge \Phi$ and $C_{\rm tw}\equiv e^{-B} \wedge C$ and straightforwardly applying the  transformation rules of section \eqref{sec:s} and appendix \ref{app:s} to the $\mathrm{SL}(2,\mathbb{Z})$ duality \eqref{Sduality}, one can explicitly check that $\varphi_{\rm D5}$ is mapped to  $\varphi_{\rm NS5}$ as defined in \eqref{NS5calibration}. Details are provided in appendix \ref{app:s}. 
This is a non-trivial consistency check for the interpretation of  (\ref{NS5calibration}) as NS5 calibration, at least in the IIB case. 

It would be interesting to derive \eqref{NS5calibration} directly from the $\mathrm{SL}(2,\mathbb{Z})$ covariant actions of \cite{bergshoeff-deroo-kerstan-ortin-riccioni,bergshoeff-kerstan-wulff}, but we will not try to do it in the present paper. In particular, by S-duality, we expect  \eqref{NS5calibration} to combine with the lower-rank calibrations for D3, D1 branes and F1-strings, as it happens for the complete  calibration  \cite{martucci-electrified} for D5-branes, whose (electric and/or magnetic) world-volume field-strength can induce lower-dimensional D3, D1 and F1 charges.

\subsection{NS5 and D4 calibrations from M-theory} 
\label{sub:ns5iia}

In order to check our interpretation of \eqref{NS5calibration} as NS5 calibration in type IIA too, we use the fact that it should uplift to a calibration for M5-branes in M-theory. In turn, by reducing back to IIA along a  direction    longitudinal to the M5-brane, the M5 calibration should give the calibration for D4-branes. 

The interpretation of the M-theory supersymmetry conditions \eqref{Meq} in terms of calibration conditions has been already considered in \cite{hackettjones-page-smith}. It is easy to rewrite \eqref{Meq1} as $\d\hat\varphi_{\rm M2}=0$ with
\be\label{M2cal}
\widehat\varphi_{\rm M2}\equiv \widehat\Omega+\iota_{\widehat K}\widehat A\,,
\ee
being the M2-brane calibration. Similarly, from \eqref{Meq2} one can identify the following M5 calibration 
\begin{equation}\label{eq:M5-cal}
\widehat{\varphi}_{\rm M5} \equiv \widehat{\Sigma} + \iota_{\widehat{K}} \widehat{C} +  \widehat{A} \wedge \widehat{\Omega} +  \frac{1}{2}\widehat{A}\wedge \iota_{\widehat{K}} \widehat{A}\,.
\end{equation}
Here $\widehat{C}$ is the `magnetic' potential associated with $\widehat{F}$. It can be defined starting from the eleven-dimensional equations of motion of $\widehat{F}$ 
\begin{equation}
\d \hat* \widehat{F} + \frac{1}{2} \widehat{F} \wedge \widehat{F} =\d\left(\hat* \widehat{F} + \frac{1}{2} \widehat{A} \wedge \widehat{F}\right)= 0 \, , 
\end{equation}
so that 
\begin{equation}
\label{Cdef}
 \d \widehat{C} = \hat* \widehat{F} + \frac{1}{2} \widehat{A} \wedge \widehat{F}\,.
\end{equation}

We can now reduce \eqref{eq:M5-cal} to IIA by applying the dimensional-reduction dictionary identified in section \ref{sub:IIAM}. The only necessary additional 
relation is 
\be
\hat * \widehat{F}  =   - e^{-2 \phi} * H - F_6 \wedge C_1 +  F_6 \wedge \d x^{10} \,,
\ee
which, combined with \eqref{Cdef} and \eqref{eq:B}, gives also
\begin{equation}
\widehat{C} =- \widetilde{B} - \frac{1}{2} C_5 \wedge C_1 + C_5 \wedge \d x^{10} - \frac{1}{2} B \wedge C_3 \wedge \d x^{10}  \, .
\end{equation}
After a straightforward computation, $\widehat{\varphi}_{\rm M5} $ splits as follows
\begin{equation}\label{M5calsplit}
\widehat{\varphi}_{\rm M5} =\varphi_{\rm NS5} -\varphi_{\rm D4}\wedge  \d x^{10}
\end{equation}
where $\varphi_{\rm NS5}$ is the (type IIA) NS5-calibration introduced in \eqref{NS5calibration_MassiveIIA} with $F_0=0$ and $\varphi_{\rm D4}$
is the D4 calibration, as defined in \eqref{Dpcal}.

The equation \eqref{M5calsplit} is indeed expected from the usual relation between M5-branes in M-theory and NS5- and D4-branes in IIA. Hence, it provides a non-trivial check of the mutual consistency between the corresponding calibrations. It would be interesting to motivate the M-theory and IIA NS5 calibrations from the world-volume effective actions \cite{pasti-sorokin-tonin,bandos-lechner-nurmagambetov-pasti-sorokin-tonin,bandos-nurmagambetov-sorokin}. 

As an additional consistency check, we prove in appendix \ref{sub:t} that also the longitudinal part of the NS5 calibration in IIA and IIB are T-dual to each other.

\subsection{NS9-branes}
\label{sec:NS9}

An analysis of the central charges of type II theories reveals the existence of a nine-brane called NS9 \cite[Sec.~6]{hull-charges}. In type IIB we can also think of it as the S-dual of a D9; in fact in type IIB we should then have a $(p,q)$ 9-branes.\footnote{\cite{bergshoeff-deroo-kerstan-ortin-riccioni} actually claims the existence of a SL$(2,\R)$ quadruplet of nine-branes, leading to $(p,q,r,s)$ bound states.} Just like a D9, the NS9 does not source any field strength, but it carries a ten-dimensional potential. 

A nine-brane is extended along all of spacetime; so a calibration would not tell us where it should sit. Nevertheless, in IIB we can extend formally the calibrations for D$p$-branes to $p=9$, and use S-duality to infer a similar nine-form for an NS9.  From \eqref{Sinvariant9brane} we can see that the S-dual of the D9 calibration condition is the first equation of \eqref{eq:10form_from_section2}
\begin{equation}
\d(e^{-2 \phi}* \widetilde{K}) =0 ,
\end{equation}
which we get in both IIA and IIB theories looking at the ten-form part of equations \eqref{eq:12f2} and \eqref{IIAKOstarKteq}. Therefore $e^{-2 \phi}* \widetilde{K}$ may be interpreted as the NS9 calibration for type IIB.

To check if this conclusion is valid also for IIA,  we perform a T-duality. Imposing a U(1) isometry and using the decomposition of (\ref{eq:T-dualdecomp}) we get 
\begin{equation}
\d(e^{-2 \phi}* \widetilde{K}) = e^{-C} \d(e^{-2 \phi + C}* \widetilde{k}_1) \wedge E^y + e^{-2 \phi + C}* \widetilde{k}_1 \wedge \d A_1-\d(e^{-2 \phi} \widetilde{k}_0) \wedge *_9 1 .
\end{equation}
We can notice that the last two terms are zero because they are ten-forms on a nine-dimensional subspace, so we have just
\begin{equation}
\d(e^{-2 \phi + C}* \widetilde{k}_1) = 0\,,
\end{equation}
which is invariant under T-duality as one can check from \eqref{eq:Flat_T-dual}. So $\d(e^{-2 \phi}* \widetilde{K}) =0$ in IIB transforms in the same equation in IIA and vice-versa. Therefore $e^{-2 \phi}* \widetilde{K}$ can indeed be interpreted as NS9 calibration for both IIA and IIB.



\section{KK5-monopoles} 
\label{sec:kk}

We will now turn to the third equation in the system (\ref{eq:domegatII}). We will argue that the form on the left-hand side can be interpreted as a kind of calibration form for the KK5-monopole. 

A KK5-monopole is a supersymmetric solitonic solution of type II theories \cite{hull-townsend}, obtained as $\mathbb{R}^6\times$ a four-dimensional Gibbons--Hawking space. It appears for example by T-dualizing a stack of $N$ NS5-branes along the Hopf isometry of its transverse $S^3$, or by lifting a stack of $N$ D6-branes to M-theory and reducing along another direction. Just like for NS5-branes, these solutions do not involve RR-fields and cannot be interpreted in terms of open strings; thus there is no simple way to derive a world-volume effective action. This issue is made even sharper for the KK5 by the fact that the $N=1$ case (a single monopole) is actually even completely smooth, and it is not even clear on which submanifold the putative world-volume action should be based. 
However, since the KK5-monopole charge appears in the superalgebra and in the BPS bound of every theory $d \ge 5$ \cite{hull-charges}, we expect that a concept of calibration should exists also for this object.

To address this, we will start by considering in section \ref{sub:grav-cal} a toy model in which central charges are defined using calibrations in purely gravitational terms. In section \ref{sub:m}, we will see that the speculations on our toy model are actually valid in M-theory and can be generalized also to the central charges given by KK5-monopoles. In this case it is less obvious which calibration condition corresponds to the conservation of the central charges. However,  the  M-theory/IIA dictionary of section \eqref{sub:IIAM} will allow us to guess at least the bispinorial part of the KK5 calibration.

\subsection{Gravitational calibrations} 
\label{sub:grav-cal}

At the beginning of section \ref{sec:d-cal} we introduced the concept of generalized calibration. In this section we will add some details to that discussion which aim at generalizing, at least schematically, the argument for the string calibration given in \cite[Sec.~3]{martucci-electrified} to the case of a general $p$-brane in a $d$-dimensional spacetime. Let us suppose that the action of the brane wrapping a $(p+1)$-dimensional surface ${\cal S}$ is the sum of a Nambu--Goto and a Wess--Zumino term
\begin{equation}
S_p =  - \mu_p \int_{\cal S}  \d^{p+1} \xi \sqrt{- \text{det} g|_S} + \mu_p \int_{\cal S} C \, ,
\end{equation}
where $\xi^\alpha=(\tau,\sigma^i)$ are the coordinates on $\cal S$; and that the supersymmetry of the background imposes the differential condition \eqref{eq:gen-cal}.
Now consider a space-like $(d-1)$-surface ${\cal M}$ and the space-like  $p$-surface  $\Sigma={\cal S}\cap {\cal M}$. Following \cite{martucci-electrified} the BPS bound can be algebraically derived from the $\kappa$-symmetry operator and it reads
\begin{equation}\label{BPSbound}
-K^MP_M\d^p\sigma\geq \omega|_\Sigma\,,
\end{equation}
where we have introduced the world-volume (or gauge-invariant) momentum conjugated to $x^M$:
\begin{equation}
P^M=-\sqrt{-h}\, h^{\tau\alpha}\del_\alpha X^M \, , \qquad h\equiv g|_{\cal S} \, .
\end{equation}

Now one can notice that the quantity
\begin{equation}
\label{eq:conserved_charge_toymodel}
-\int_\Sigma K^MP_M\d^n\sigma  -\int_\Sigma\omega \geq 0
\end{equation}
doesn't look like the BPS bound \eqref{BPSbound_intro}, even if it can be shown to be a conserved charge related to the $K$ isometry \cite[Sec.~3.2]{martucci-electrified}.
This is due to the fact that $P_M$ is not the canonical momentum, since it is obtained from the Legendre transformation of the Nambu--Goto part of the action only. Considering also the Wess--Zumino term, along the line of \cite[Sec.~3.4]{martucci-electrified}, we get that the canonical momentum is given by $\mathcal{P}_M = P_M + \iota_M C |_\Sigma$ and therefore \eqref{eq:conserved_charge_toymodel} becomes exactly \eqref{BPSbound_intro}. This
allows us to interpret the right-hand side of \eqref{BPSbound_intro} as a central charge:
\begin{equation}\label{firstZ}
	Z=\int_\Sigma(\omega-\iota_K C)=\int_{{\cal M}}(\omega-\iota_K C)\wedge \delta_{\cal S} \, ,
\end{equation} 
where $\delta_{\cal S}$ is a delta-like $d-(p+1)$ form localized on ${\cal S}$. 
 
Up to now the $p$-brane was regarded as a probe, meaning that we were considering a regime where the back-reaction can be neglected. Now, let us take $\delta_{\cal S}$ as a source for the flux $F$, that we suppose to satisfy the usual equation of motion
\begin{equation}
\d*F=(-)^p\delta_{\cal S} \, .
\end{equation}
Then, by using the bulk equation \eqref{eq:gen-cal}, we can write the r.h.s.\ of \eqref{firstZ} as:
\begin{equation}
\int_{{\cal M}}(\omega-\iota_K C)\wedge \d *F=(-)^p\int_{\cal B} (\omega-\iota_K C)\wedge *F
\end{equation}
where ${\cal B}=\del{\cal M}$.
Therefore we are left with the following identification of the brane central charge
\begin{equation}\label{eq:CC}
Z= (-)^p \int_{\cal B} \varphi\wedge *F 
\end{equation}
with $\varphi$ as defined in \eqref{eq:cal_ex_intro}.
Notice that the integral is evaluated not on the brane world-volume but on the space boundary. This has been possible by the promotion of the brane from probe to back-reacting. 
Assuming that ${\cal S}\cap {\cal B}=0$, the integrated quantity is invariant under deformations of the boundary ${\cal B}$. More generally, we may assume fixed boundary conditions such that this continues to hold.

The idea is to take \eqref{eq:CC} as definition of central charge, carried by the flux $ F$, that is associated  with the brane charge. We can then consider more general backgrounds with somehow fixed boundary conditions so that \eqref{eq:CC} makes sense more generally, even in absence of branes, and does not change under deformations preserving the boundary conditions. The above argument starts from a back-reacting brane. In the following subsections, we will re-derive our conclusions from purely gravitational arguments in M-theory.


\subsection{Gravitational BPS bound in M-theory and central charges} 
\label{sub:m}

Let us focus on a family of backgrounds in M-theory with certain boundary conditions fixed as in \cite{hull-charges} admitting a spinor $\epsilon$ and defining an asymptotic Killing vector $K^M=\bar\epsilon\Gamma^M\epsilon$. We are using the conventions of section \ref{sub:IIAM}, but for simplicity we omit all hats, since in the present section we just consider M-theory quantities and no confusion should arise. Furthermore, one should select some asymptotic  configuration $g^{(0)}, A^{(0)}$ such that 
\begin{equation}
{\cal D}^{(0)}_M\epsilon|_{\cal B}=0 \, ;
\end{equation}
we refer to \cite{hull-charges} for all the details concerning how this asymptotic configuration must be chosen and we restrict ourselves to a more formal discussion. For more general configurations, we have
\begin{equation}\label{eq:TM}
{\cal D}_M\epsilon|_{\cal B}= ({\cal D}_M- {\cal D}^{(0)}_M)\epsilon|_{\cal B}\equiv T_M\epsilon|_{\cal B} \, ,
\end{equation}
where $T_M\equiv {\cal D}_M- {\cal D}^{(0)}_M$ is some linear combination of  tensors contracted with gamma matrices. 

By following \cite{hull-positivity,hull-charges}, the supercharge associated with a Killing spinor $\epsilon$ takes the form
\begin{equation}
Q[\epsilon]=\int_{\cal B} \d x^M\wedge \bar\epsilon\, \Gamma_{(8)}\psi_M
\end{equation}
and, up to normalization,  we can write
\begin{equation}
\label{eq:QQ_bracket}
\{Q(\epsilon),Q(\epsilon)\}=\int_{\cal B} \d x^M\wedge \bar\epsilon\, \Gamma_{(8)}{\cal D}_M\epsilon=\int_{\cal B} \d x^M\wedge \bar\epsilon\, \Gamma_{(8)}T_M\epsilon = 
\int_{\cal B} \d x^M\wedge (T_M \cdot \epsilon \overline{\epsilon})_{8}\,.
\end{equation}
The quantum mechanical realization of the l.h.s.\ of \eqref{eq:QQ_bracket} implies, as usual,  the positivity condition 
$ \{Q(\epsilon),Q(\epsilon)\}\geq 0$. The idea is now to manipulate \eqref{eq:QQ_bracket} and try to deduce from $ \{Q(\epsilon),Q(\epsilon)\}\geq 0$ a BPS bound of the form
\begin{equation}\label{gravBPSbound}
\calp[K]-\sum_aZ_a \geq 0
\end{equation}
where $Z_a$ are central charges as defined in the previous section, with corresponding calibrations $\varphi_a$ and ``fluxes" $F_a$, obtained by expanding $T_M$.

Let's check this. By using the eleven-dimensional gamma-matrices properties, we can write the bulk supersymmetry condition in the form
\begin{equation}
\nabla_M \epsilon - \frac{1}{12} \iota_M\left( * F+ 2  F \right) \epsilon =0 \,.
\end{equation}
Therefore the $T_M$ defined in (\ref{eq:TM}) reads
\begin{equation}\label{eq:TMexp}
T_M = \frac{1}{4} \Delta \omega_M -\frac{1}{12}  \iota_M\left(  \Delta *  F+ 2 \Delta F \right)\,,
\end{equation}
where $\Delta F \equiv F - F^{(0)}$, and $\Delta \omega_M \equiv (\omega_M^{AB} - \omega_M^{AB\,(0)})\gamma_{AB}$ (a difference of connections, and hence a tensor at the boundary).  Notice that $\Delta * F$ in \eqref{eq:TMexp}  is not a closed form: indeed it satisfies
\begin{equation}
\d \Delta * F + F \wedge \Delta F = 0 \, ,
\end{equation}
where we considered just the leading order at the boundary (in other words, the first order in $\Delta F$).
We could instead introduce a closed (gauge-dependent) field-strength
\be\label{defDeltaG}
\Delta G\equiv \Delta * F + A \wedge \Delta F
\ee
whose flux along the boundary measures   the M2 Page charge.

Plugging (\ref{eq:TMexp}) into (\ref{eq:QQ_bracket}) and using (\ref{eq:gammadestrasinistra}) repeatedly together with \eqref{defDeltaG}, we obtain
\begin{equation}
\label{central_charges}
\{Q(\epsilon),Q(\epsilon)\} = \frac{1}{4} \int_{\cal B} * (\d x^{AB} \wedge {K} ) \wedge \Delta \omega_{AB} +\frac{1}{4} \int_{\cal B} \big[ {\Omega} \wedge \Delta G - ({\Sigma}+\Omega \wedge A) \wedge \Delta F- * {\Sigma} \wedge \Delta \omega\big]\,.
\end{equation}
${\Sigma}$ and ${\Omega}$ were defined in (\ref{hatforms}) (recall that in this section we are omitting hats), and we have introduced the three-form $\Delta \omega\equiv \frac12 \Delta \omega_{MAB}\d x^M \wedge E^A \wedge E^B$.\footnote{This is a three-form; it does depend on the choice of vielbein, but ultimately when we go back to (\ref{central_charges}), where only differences appear, this does not matter.} Here, all the Hodge-duals are meant  with respect to the eleven-dimensional metric.

The first term is exactly the ADM momentum $P[K]$ as defined in \cite[(3.2)]{hull-charges}, which we interpret as the equivalent of our gauge-invariant momentum (as one can see from the absence of form potentials). The BPS bound obtained by imposing $\{Q(\epsilon),Q(\epsilon)\}\geq 0$ in \eqref{central_charges} is then similar to \eqref{eq:conserved_charge_toymodel}.  
This bound can also be partially rewritten  in the form \eqref{gravBPSbound}: we can clearly isolate the contribution of the M2 and M5 central charges, 
\be\label{ZM2M5}
Z_{\rm M2}=\int_{\cal B} \varphi_{\rm M2} \wedge \Delta G \, ,\qquad Z_{\rm M5}=-\int_{\cal B} \varphi_{\rm M5} \wedge  \Delta F\,,
\ee
defined in analogy with \eqref{eq:CC}, where $\varphi_{\rm M2}$ and $\varphi_{\rm M5}$ are the (closed) M2 and M5 calibrations introduced in \eqref{M2cal} and \eqref{eq:M5-cal} (with the addition of the hats). 

The last term on the r.h.s.~should be associated with  KK6-monopoles. Indeed, following \cite{hull-charges}, we can interpret $\Delta \omega$ as `geometric' flux sourced by the KK6-monopoles,  since  its integral corresponds to the NUT charge in the case of a Taub--NUT solution. We are then led to identify $*\Sigma$ with (part of) the calibration for KK6-monopoles in M-theory. However, we have not been able to write down a clean corresponding topological central charge of the form \eqref{ZM2M5}. This is due to the purely gravitational nature of the KK6-monopole, which appears to be mixed with the ADM momentum in the BPS bound (\ref{central_charges}). So, taking into account these subtleties, we will refer to  $*\Sigma$ as a KK6 {\em calibrating form}, in order to distinguish it from the more standard calibrations. 
In the following subsection we will see that this interpretation is consistent with the reduction to IIA, which relates M-theory KK6-monopoles to IIA KK5-monopoles  and  D6-branes. This will allow us to identify the analogous IIA KK5 calibrating forms.

\subsection{Type II KK calibrating forms from dualities} 
\label{sub:moniiared}

We now revisit the above conclusions from the IIA point of view, by using  the general relations discussed in section \ref{sub:IIAM}.   Let us reintroduce the hat to distinguish M-theory quantities, as in that section. Then, the M-theory KK6 calibrating form $\hat * \widehat{\Sigma}$ decomposes as in \eqref{*hatSigma},  while  the associated geometric flux $ \Delta\widehat{\omega}$ reduces to 
\begin{equation}
 \Delta\widehat{\omega} =e^{-\frac{2}{3} \phi} \Delta\omega_{10} -\frac12e^{\frac{4}{3} \phi} \Delta F_2 \wedge (\d x^{10}-C_1)
\end{equation}
By identifying $x^{10}\simeq x^{10}+1$, we conclude that the last term in \eqref{central_charges} reads
\begin{equation}\label{eq:KK6-red}
 \int_{\widehat{\cal B}} \hat* \widehat{\Sigma} \wedge  \Delta \widehat\omega_{11} = \int_{{\cal B}} \Big(e^{-2 \phi}  \widetilde{\Omega}  \wedge \Delta \omega_{10}+ \frac{1}{2} e^{-\phi} \Phi_6 \wedge \Delta F_2 \Big)\,.
\end{equation}
where we have integrated of the $S^1$ of the M-theory nine-dimensional boundary $\widehat{\cal B}\simeq S^1\times {\cal B}$.  Both terms on the right-hand side of (\ref{eq:KK6-red}) are expected. In the last term in (\ref{eq:KK6-red}), $e^{-\phi}\Phi_6$ is gauge-invariant contribution to the D6 calibration $\varphi_{\rm D6}$, see \eqref{Dpcal}. This can be completed to give an associated topological central charge $Z_{\rm D6}$ as we did in the previous section. On the other hand, in analogy with the M-theory case, we are led to identify $e^{-2 \phi}\widetilde{\Omega}$
with the  type IIA KK5 calibrating form.

The analogous KK5 calibrating form  for IIB can be obtained from T-duality. A transverse T-duality maps a KK5-monopole into a NS5-brane and vice-versa, while under a longitudinal one both KK5 and NS5 remain invariant. Using \eqref{OmegaTT}, we  see that the longitudinal part of  $e^{-2\phi} \Omega$ becomes proportional to $e^{-2\phi} \widetilde{\Omega}$ after a T-duality transformation. Since we saw that $e^{-2\phi} \Omega$  is part of the NS5-calibration, it is natural to identify  $e^{-2 \phi}\widetilde{\Omega}$ with  the KK5 calibration in IIB as well.

\section{Ansatz for AdS$_2$ horizons} 
\label{sec:ads2}

Supersymmetric solutions with a timelike Killing vector are suitable to describe static space-times, and in particular can be used to study black holes. We present here an application of our system to this problem. Instead of looking at a full black-hole solution, we restrict ourselves to the near horizon geometry which can be viewed as an AdS$_2 \times M_8$ vacuum solution where $M_8$ is typically a fibration of a compact manifold $M_6$ over $S^2$. Classifications of black hole horizons in a similar spirit were given in \cite{gran-gutowski-papadopoulos-IIB-hor1,gran-gutowski-papadopoulos-IIB-hor2}. 

\subsection{Spinor Ansatz}
\label{app:spinor_Ansatz}

In general, a solution describing an AdS$_2$ vacuum is not the near-horizon of a black hole, so we have to be particularly careful about finding a proper Ansatz for the supersymmetry parameters. Moreover, not every Ansatz leads to the timelike case; for example, if we start from spinors like the ones in \cite{rosa} we will get a solution which is light-like and to which therefore we can not apply our system. 

For these reasons we will derive our Ansatz starting from the black-hole background in \cite{katmadas-t}, which is an uplift of the Cacciatori--Klemm solution \cite{cacciatori-klemm} to M-theory with a regular Sasakian internal manifold S$_7$. In the near-horizon limit we have an AdS$_2$ with coordinates $t$, $r$ and a compact space $M_9$, which can be seen as a U(1) fibration over an eight-dimensional K\"ahler manifold K$_8$ which contains both the horizon $S^2$ and the K\"ahler K$_6 \subset$ S$_7$ (even if this distinction is not important in our discussion). The vertical vector $\del_\psi$ is called Reeb vector.
The eleven-dimensional supersymmetry parameter $\hat \epsilon$ defines, via its bilinears, an SU(5) structure on every constant-time surface \cite{gauntlett-pakis}, which is determined by the differential forms
\begin{equation}
\label{SE_forms}
\widehat{K} = \partial_t \, , \qquad \Delta (E^r \wedge E^\psi + J_{\rm{K}_8} ) \, , \qquad  \Delta^{\frac{5}{2}} (E^r + \ii E^\psi)\wedge \Omega_{\rm{K}_8} \,,
\end{equation}
already decomposed according to \cite{katmadas-t}. (\ref{hatforms1}) and (\ref{hatforms2}) are respectively the two-form and the real part of the five-form in (\ref{SE_forms}).
$E^\psi$ and $E^r$ are the vielbein one-forms associated to the coordinate $\psi,r$ respectively; $\Delta$ is a warping function; $J_{{K}_8}$ and $\Omega_{{K}_8}$ are the real two-form and holomorphic four-form of the K$_8$. Since $\widehat{K}$ points in the time direction, when we perform the dimensional reduction to recover type IIA the Killing vector $K$ will be timelike, and then our system will apply.

Now let us see which eleven-dimensional spinor defines the bilinears \eqref{SE_forms} via Clifford map. This spinor has a particularly simple form with the following choice of eleven-dimensional gamma matrices
\begin{equation}\label{eq:clifford-11}
\begin{split}
&\Gamma_{\underline{1}} = \Gamma_{\underline{r}} = \sigma_1 \otimes \bbone_2\otimes \bbone_2\otimes \bbone_2\otimes \bbone_2 \, ,\qquad
\Gamma_{\underline{2}} = \Gamma_{\underline{\psi}} = \sigma_2 \otimes \bbone_2\otimes \bbone_2\otimes \bbone_2\otimes \bbone_2 \\
&\Gamma_{\underline{3}} = \sigma_3\otimes \sigma_1 \otimes \bbone_2 \otimes \bbone_2\otimes\bbone_2  \, ,\qquad \ldots ,\qquad
 \Gamma_{\underline{0}} =\Gamma_{\underline{t}}= \ii \sigma_3\otimes\sigma_3\otimes\sigma_3\otimes\sigma_3\otimes\sigma_3 \, .
\end{split}
\end{equation}
In this basis, we reproduce \eqref{SE_forms} by taking 
\begin{equation}
\hat \epsilon = \frac{1}{\sqrt{2}} \big(|+++++\rangle+|-----\rangle\big) \, .
\end{equation}

The black hole solution we are considering has a $\mathrm{U}(1)$ R-symmetry which is generated by the Reeb vector; it produces a fibration of the internal manifold over the horizon $S^2$. Since the spinors that live on $S^2$ are twisted with respect to this connection, the R-symmetry action on one spinor produces another one and thus the solution has a multiple of two supercharges. However, since the R-symmetry does not involve the time and the radial direction, from the perspective of an AdS$_2$ vacuum we just have $\mathcal{N}=1$. 

When we perform the dimensional reduction to IIA we have to be careful to preserve the Reeb direction in order not to break supersymmetry. We decide to reduce along the coordinate $x^{10}$, which  parametrizes a (generic) direction of K$_8$. In ten dimensions, it is convenient to choose a representation of gamma matrices such that the chiral operator  is given by
\be\label{GammaM}
\Gamma=\sigma_3\otimes\sigma_3\otimes\sigma_3\otimes\sigma_3\otimes\sigma_3\,.
\ee
We then see that we cannot identify $\Gamma$ with $\Gamma_{\underline{10}}$ appearing in \eqref{eq:clifford-11}, as usual in dimensional reduction. Rather, we have $\Gamma=-\ii\Gamma_{\underline{0}}$.\footnote{This choice was important in \cite{katmadas-t} because the SU(5) structure defined by $J$ and $\Omega$ in (\ref{SE_forms}) lives in the space directions.} 
Thus we have to change spin representation  by finding an operator $O$ which defines a new set of gamma matrices $\Gamma_{\underline{M}}^{\text{new}} = O \Gamma_{\underline{M}}^{\text{old}} O^{-1}$ such that the new $\Gamma_{\underline{10}}$ coincides with $\Gamma$ in \eqref{GammaM}. A useful choice is 
\begin{equation}
O = \frac12(\bbone_{32}- \ii Q_2)(\bbone_{32}- \ii Q_1)
\end{equation}
where
\begin{equation}
Q_1 = \sigma_1\otimes\sigma_3\otimes\sigma_3\otimes\sigma_3\otimes\sigma_3 \, , \qquad Q_2 =\bbone_2 \otimes\bbone_2 \otimes\bbone_2 \otimes\bbone_2 \otimes \sigma_1 \, ,
\end{equation}
so that
\begin{equation}
\Gamma_{\underline{0}}^{\text{new}} = \ii \Gamma_{\underline{2}}^{\text{old}}  \, , \quad \Gamma_{\underline{2}}^{\text{new}} = \Gamma_{\underline{10}}^{\text{old}}  \, , \quad \Gamma_{\underline{10}}^{\text{new}} = - \ii \Gamma_{\underline{0}}^{\text{old}}  \, , \quad \Gamma_{\underline{i}}^{\text{new}} = \Gamma_{\underline{i}}^{\text{old}} \quad i \neq 0,2,10 \, .\\
\end{equation}
Notice that with such a choice the new $\Gamma_{\underline{0}}$ and $\Gamma_{\underline{1}}$ act only on the first spin-$\frac12$ factor of the  usual tensorial decomposition of the ten-dimensional spinors.  
Splitting the rotated spinor $O \hat \eps$ in two chiralities we get the supersymmetry parameters for IIA:
\begin{equation}
\label{spinor_Ansatz_11D}
\begin{split}
\epsilon_1 &= \frac{1}{2} \big(|+++++\rangle- \ii|----+\rangle \big) + \text{Maj.\,conj.} \,,\\
\epsilon_2 &= \frac{1}{2} \big(- \ii|++++-\rangle+|-----\rangle \big) + \text{Maj.\,conj.} \, 
\end{split}
\end{equation}

This decomposition suggest the following, more general, spinor Ansatz for near-horizon geometries
\begin{equation}
\label{eq:ads2_spinor_Ansatz}
\begin{split}
&\eps_1 = \alpha_+ \otimes \eta_{1 \, +} + \alpha_- \otimes \eta_{1 \, -} = P_+(\alpha \otimes \eta_1)\\
&\eps_2 = \alpha_+ \otimes \eta_{2 \, \mp} + \alpha_- \otimes \eta_{2 \, \pm} = P_\mp(\alpha \otimes \eta_2)
\end{split}
\end{equation}  
where $\alpha = \alpha_+ + \alpha_-$ is a real Killing spinor on AdS$_2$, $P_{\pm}$ are the chiral projectors and $\eta_{i} = \eta_{i \, +}+ \eta_{i \, -} $ are Majorana spinors on $M_8$, that we can take to be real.\footnote{In the particular case we were considering in \eqref{spinor_Ansatz_11D} we have $\eta_{1 +} = (|++++\rangle - \ii |----\rangle)$, $\eta_{1 -} = - \ii |---+\rangle - |+++-\rangle$, $\eta_{2 +} = |----\rangle - \ii |++++\rangle$ and $\eta_{2 -} =- \ii |+++-\rangle - |---+\rangle$.}
 Notice that the presence of just one Killing spinor ensures the correct amount of supersymmetry. 

\subsection{Bispinors on AdS$_2 \times M_8$}
\label{sub:ads2-bisp}

Having determined the spinor Ansatz, in this subsection we will compute the ten-dimensional bilinears to plug in \eqref{eq:systemII}. 

To preserve the isometry of AdS$_2$, we will split the metric as usual as
\begin{equation}
\d s^2_{10} = e^{2A} \d s^2_{{\rm AdS}_2} + \d s^2_{M_8} \, .
\end{equation}
This suggests the gamma-matrix decomposition
\begin{equation}
\begin{split}
&\Gamma_\mu = e^A \sigma_\mu \otimes \bbone_{16} \qquad \mu = 0,1 \, , \\
&\Gamma_m = \sigma_3 \otimes \gamma_m \qquad m = 2,\dots,9 \, ,
\end{split}
\end{equation}
where $\sigma_0 = \ii \sigma_2$ and $A$ is a function of $M_8$.

All possible choices of the AdS$_2$ Killing spinor $\alpha$ are equivalent, since our solution must be invariant under $\mathrm{SO}(2,1)$. For definiteness we will take 
\begin{equation}
\alpha = e^{r/2}\left( |+ \rangle + |- \rangle \right) \, , \qquad \d s^2_{\mathrm{AdS}_2} = e^{2r} \d t^2 + \d r^2 \, .
\end{equation}
Its bilinears read
\begin{equation}
\label{eq:alpha-fierz}
\alpha \otimes  \overline \alpha = -e^{2r} \d t + e^{2r} \d t \wedge \d r \, , \qquad \sigma_3 \alpha \otimes  \overline \alpha = - e^r + e^r \d r \, .
\end{equation}
Thus $\alpha$ by itself already defines a vielbein (i.e.~an identity structure) on AdS$_2$.

On the other hand, the spinors on $M_8$ are not enough to define an identity structure so we will simply rename their bilinears as:
\begin{equation}
\begin{split}
&\omega_1 = \eta_1 \eta_1^t \, \qquad \quad \omega_2 = \eta_2 \eta_2^t \, \qquad \quad  \omega = (\omega_1+\omega_2)/2  \qquad \quad  \widetilde\omega = (\omega_1-\omega_2)/2 \\
&\omega_1^{\g} = \g \eta_1 \eta_1^t \, \qquad \omega_2^{\g} = \g \eta_2 \eta_2^t \, \qquad  \omega_{\g} = (\omega_1^{\g}+\omega_2^{\g})/2 \, \qquad  \widetilde\omega_{\g} = (\omega^{\g}_1-\omega^{\g}_2)/2 \\
&\psi = \eta_1 \eta_2^t \, \quad \qquad \psi_{\g}= \g \eta_1 \eta_2^t 
\end{split}
\end{equation}
where $\gamma$ is the chiral operator on $M_8$.
In particular, to keep the analogy with the ten-dimensional notation, we will give special names to the zero and one-form part:
\begin{equation}
\begin{split}
& a = (\omega)_0 \, \qquad  \widetilde{a} = (\widetilde{\omega})_0 \, \qquad a_{\g} = (\omega_{\g})_0 \, \qquad  \widetilde{a}_{\g} = (\widetilde{\omega}_{\g})_0 \\
&k_1 = (\omega_1)_1 \, \qquad k_2 = (\omega_2)_1 \, \qquad k = (\omega)_1 \, \qquad \widetilde{k} = (\widetilde{\omega})_1
\end{split}
\end{equation}
where the subscript $0$ and $1$ indicates to take the zero and one-form part only. 

Now we are ready to explicitly write the ten-dimensional bilinears of section \ref{sub:geo} in terms of two and eight-dimensional ones. 
We will report here the calculation for $K_1$ in all the details, while for the other bilinears one can proceed by analogy:
\begin{equation}
\begin{split}
32 K_1 =& (\alpha \otimes \eta_1)^t P_+^t \Gamma_{\underline{0}} \Gamma_M P_+ (\alpha \otimes \eta_1) E^M =  (\alpha \otimes \eta_1)^t \Gamma_0 \Gamma_M P_+ (\alpha \otimes \eta_1) E^M  \\
=& (\alpha \otimes \eta_1)^t \Gamma_{\underline{0}} \Gamma_M  (\alpha \otimes \eta_1) E^M + (\alpha \otimes \eta_1)^t \Gamma_{\underline{0}} \Gamma_M \Gamma (\alpha \otimes \eta_1) E^M \\
=& \frac{e^A}{2} \left(  \eta_1^t \eta_1 \, \overline{\alpha} \sigma_\mu \alpha \, e^\mu +  \eta_1^t \g \eta_1 \, \overline{\alpha} \sigma_\mu \sigma_3 \alpha \, e^\mu \right) + \frac{1}{2}  \overline{\alpha} \sigma_3 \alpha (16k_1) \, ,
\end{split}
\end{equation}
where $e^\mu$ is the vielbein on AdS$_2$. Using \eqref{eq:alpha-fierz} we get:
\begin{equation}
32 K_1 = - e^{r+A} (\eta_1^t \eta_1 e^0-\eta_1^t \g \eta_1 e^1) - e^r (16k_1) \, .
\end{equation}
We can perform the same steps for $K_2$
\begin{equation}
32 K_2 = - e^{r+A} (\eta_2^t \eta_2 e^0+\eta_2^t \g \eta_2 e^1) - e^r (16k_2) 
\end{equation}
and from these expressions we can calculate $K$ (considered as a one form) and $\widetilde{K}$:
\begin{equation}
K = -\frac{e^{r+A}}{2} \left( a e^0 - \widetilde{a}_{\g} e^1 \right)-\frac{e^{r}}{2}k \, ,  \qquad \widetilde{K} =-\frac{e^{r+A}}{2} \left( \widetilde{a} e^0 - a_{\g} e^1 \right)- \frac{e^{r}}{2}\widetilde{k} \, .
\end{equation}

Following a similar logic for $\Omega$ and $\widetilde{\Omega}$ we get:
\begin{equation}
\begin{split}
&\Omega = -\frac{e^r}{2} \left((\omega)_5 + e^A e^0 \wedge (\omega)_4 - e^A e^1 \wedge (\widetilde{\omega}_{\g})_4  - e^{2A} e^0 \wedge e^1 \wedge (\widetilde{\omega}_{\g})_3 \right) \, , \\
&\widetilde{\Omega} = -\frac{e^r}{2} \left((\widetilde{\omega})_5 + e^A e^0 \wedge (\widetilde{\omega})_4 - e^A e^1 \wedge (\omega_{\g})_4  - e^{2A} e^0 \wedge e^1 \wedge (\omega_{\g})_3 \right) \, , 
\end{split}
\end{equation}
and, finally, $\Phi$ reads:
\begin{equation}
\begin{split}
\Phi =& P_+ (\alpha \otimes \eta) \overline{(\alpha \otimes \eta)} P_+ = \left[ P_+ (\alpha \otimes \eta) \overline{(\alpha \otimes \eta)} \right]_+ \\
=& -\frac{e^r}{2} \left[ (\psi_{\g})_+ - e^{A} e^0 \wedge (\psi_{\g})_- + e^{A} e^1 \wedge (\psi)_- - e^{2A} e^0 \wedge e^1 \wedge  (\psi)_+  \right] \, ,
\end{split}
\end{equation}
where the subscripts $+$ and $-$ indicate to take the even or the odd forms degree respectively.

\subsection{Supersymmetry conditions}

In the previous subsection we showed how bilinears decompose under our Ansatz of section \ref{app:spinor_Ansatz}. The fluxes must also be decomposed and if we want to preserve AdS$_2$ they have to be singlet under the action of its isometry group:
\begin{equation}\label{eq:ads2-fluxes}
\begin{split}
&H = H_3 + e^{2A} e^0 \wedge e^1 \wedge H_1 \, , \qquad * H = *_8 H_1 + e^{2A}  e^0 \wedge e^1 \wedge *_8 H_3 \, , \\
&F = f + e^{2A}  e^0 \wedge e^1 \wedge *_8 \lambda(f) \, .
\end{split}
\end{equation}

Now we can derive the supersymmetry conditions just by plugging the expressions for the fluxes (\ref{eq:ads2-fluxes}) and the ones for the bispinors in section \ref{sub:ads2-bisp} in \eqref{eq:systemII}. The first three equations in \eqref{eq:systemII} will split in four equations each, one for every independent form on AdS$_2$. This abundance is in part due to the fact that our system contains some redundancy, unlike the system in \cite{10d}. However we expect that specializing the Ansatz on $M_8$ a little more, some equations turn out to be dependent.

We will now show the AdS$_2 \times M_8$ supersymmetry conditions. From \eqref{eq:oldII} we get:
\begin{subequations}
\begin{align}
&\d_{H_3}(e^{-\phi}\psi_{\g})_+ = -  (\widetilde{k}+\iota_k)f \, , \\
&\d_{H_3}(e^{A-\phi}\psi_{\g})_- =  e^A (\widetilde{a}f + \widetilde{a}_{\g}*_8 \lambda(f) ) \, , \\
&\d_{H_3}(e^{A-\phi}\psi)_- = - e^A (a_{\g}f - a*_8 \lambda(f)) - (e^{-\phi}\psi_{\g})_+ \, , \\
&\d_{H_3}(e^{2A-\phi}\psi)_+ =  e^{2A} (\widetilde{k}+\iota_k)*_8 \lambda(f) + 2 (e^{A-\phi}\psi_{\g})_- - H_1 \wedge (e^{2A-\phi}\psi_{\g})_+ \, ,
\end{align}
\end{subequations}
from \eqref{eq:dOmegaIIA}
\begin{subequations}
\begin{align}
&e^{2\phi} \d(e^{-2\phi}\omega)_5 = -  \iota_k *_8 H_1 + (e^\phi \psi_\gamma ,f)_6 \, ,\\
&e^{2\phi-A} \d(e^{-2\phi+A}\omega)_4 = -\widetilde{a}_\gamma *_8 H_3-e^{\phi}( \psi_\gamma ,f)_5 \, ,\\
& e^{2\phi-A} \d(e^{-2\phi+A}\widetilde{\omega}_{\g})_4 = -a *_8 H_3 - e^{\phi} ( \psi ,f)_5 + e^{-A} (\omega)_5 \, , \\
& e^{2\phi-2A} \d(e^{-2\phi+2A}\widetilde{\omega}_{\g})_3 = \iota_k *_8 H_3 + e^{\phi} [(\psi,f)_4 + (\psi_{\gamma},*_8 \lambda(f))_4] - 2 e^{-A} (\omega)_4 \, ,
\end{align}
\end{subequations}
and in the end from \eqref{eq:domegatII}
\begin{subequations}
\begin{align}
&e^{2\phi} \d(e^{-2\phi}\widetilde{\omega})_5 = - \iota_{\widetilde{k}} *_8 H_1 - \frac{e^\phi}{2} ( \psi_\gamma^m  ,f_m)_6 \, , \\
&e^{2\phi-A} \d(e^{-2\phi+A}\widetilde{\omega})_4 = - a_\gamma *_8 H_3- \frac{e^{\phi}}{2} [( \psi_\gamma^m ,f_m)_5- (\psi,*_8 \lambda(f))_5] \, , \\
&e^{2\phi-A} \d(e^{-2\phi+A}\omega_{\g})_4 = - \widetilde{a} *_8 H_3
-\frac{e^{\phi}}{2} [(\psi^m ,f_m)_5 + (\psi_{\g} ,*_8 \lambda(f))_5] + e^{-A} (\widetilde{\omega})_5 \, , \\
&e^{2\phi-2A} \d(e^{-2\phi+2A}\omega_{\g})_3 = \iota_{\widetilde{k}} *_8 H_3 - \frac{e^{\phi}}{2} [(\psi^m,f_m)_4 + (\psi_{\gamma}^m,*_8 \lambda(f)_m)_4] - 2 e^{-A} (\widetilde{\omega})_4 \, .
\end{align}
\end{subequations}
Now we are left to deal with the last line of \eqref{eq:systemII}. In this case we have a scalar and a ten-form equation, so we will get just one equation on $M_8$ for each of them
\begin{equation}
\mathcal{L}_k \phi = 0 \, , \quad e^{-2A}\d (e^{2A}*_8 \widetilde{k}) = 2 e^{-A} a_\gamma \text{Vol}_8- \frac{e^\phi}{4} \left[(\psi_{\gamma}, (3 - \text{deg})  *_8 \lambda (f)) -(\psi, (5 - \text{deg})f) \right] \,.
\end{equation}
As usual, these equations are equivalent to setting the supersymmetric variations of the fields to zero. Recall that for a solution one also has to impose the Bianchi identities \eqref{BIid} and one equation between \eqref{eq:Heq_integrability} and \eqref{eq:EE_integrability}.



\section{4d type IIB vacua and $\mathrm{SL}(2,\mathbb{Z})$-duality}
\label{sec:4d-sl2z}

In section \ref{sec:s} we have derived a system of necessary and sufficient conditions for supersymmetry in the timelike case which is invariant under the $\mathrm{SL}(2,\mathbb{Z})$ symmetry of type IIB supergravity. We can easily extend this result to the light-like case for AdS$_d\times M_{10-d}$ or Mink$_d\times M_{10-d}$ solutions, for $d=$4, 5, 6 or 7. 

Indeed in \cite{10d} it is shown  that, in these cases,  equation \eqref{eq:oldII} together with the string calibration condition $\d \widetilde{K} = \iota_K H$ and the Killing spinor equation $\mathcal{L}_K g =0$ are enough to impose the supersymmetry of the solution. One can then just look at the $\mathrm{SL}(2,\R)$-covariantization of (\ref{eq:oldII}) alone.
In the language of section \ref{sec:s}, this  is composed of the equations
\begin{subequations}
	\label{eq:Sl2Rsystem_vacua}
	\begin{align}
	&\mathcal{L}_K g_{\rm E} =0 \, , \\
	&\d_Q\Theta_1-\frac\ii2e^\phi\d\overline\tau\wedge\overline\Theta_1 +\ii\, \iota_K\overline\calg_3 = 0 \, , \\
	&\d\Theta_3 + \iota_K F_5 + \text{Re}\big(\Theta_1 \wedge \calg_3\big)=0 \, ,\\
	& \d_Q \Theta_5+\frac\ii2 e^\phi \d \overline\tau\wedge\overline\Theta_5 + \Theta_3 \wedge \overline\calg_3 - \ii \iota_K (*_{\rm E}\, \overline\calg_3)+ \ii \Theta_1 \wedge F_5 =0 \, , \\
	&\d *_{\rm E} \Theta_3 +\frac12\text{Re} \left( \calg_3 \wedge \Theta_5- *_{\rm E}\, \calg_3 \wedge \Theta_1 \right) = 0 
	\end{align}
\end{subequations}
which must be supplemented with the algebraic constraint:
\begin{equation}
\label{eq:sl2R_vacua_algebraic}
\overline\calg_3 \wedge \Theta_5-\Theta_1\wedge *_{\rm E} \overline\calg_3+2 e^\phi \iota_K *_{\rm E}\d \overline\tau + 2 \ii e^{\phi} \d \overline\tau \wedge *_{\rm E}\Theta_3 =0 \, .
\end{equation}

\subsection{Application to four dimensions}

Now we will provide an application of \eqref{eq:Sl2Rsystem_vacua} to four-dimensional $\mathcal{N}=1$ vacuum solutions, where as usual the metric is decomposed as $\dd s^2_{10}= e^{2A}\dd s^2_4 + \dd s^2_{M_6}$, the RR flux decomposes as $F = f + e^{4A} \mathrm{vol}_4 \wedge *_6 \lambda (f)$ with $f$ an internal form, and $H$ is a three-form on $M_6$ only. The spinor Ansatz for this case is
\begin{equation}
\eps_i = \zeta_+ \otimes \eta^i_+ + \zeta_- \otimes \eta^i_- \, , \quad \qquad \zeta_+ = \zeta_-^* \, , \quad \eta^i_+ = \eta_-^{i \, *} \, .
\end{equation}
Using standard notation of \cite[Sec.~4.1.1]{10d}, we have the four-dimensional bispinors
\begin{equation}
\zeta_+ \otimes \overline{\zeta}_+ = v + \ii *_4 v \, , \qquad \zeta_+ \otimes \overline{\zeta}_+ = v \wedge w
\end{equation}
where $v$ is a real null vector while $w = w_1 + i w_2$ is complex, and the six-dimensional bispinors:
\begin{equation}
\begin{split}
& \eta^1_+ \eta^2_+ = \phi_+ \, , \qquad \qquad \qquad \qquad \, \, \eta^1_+ \eta^2_- = \phi_- \, , \\
& \eta^i_+ \eta^{i \, \dag}_+ = (1-i *_6 \lambda) (\omega^i_0+ \ii \omega^i_2) \, , \qquad \eta^i_+ \eta^{i \, \dag}_- = \omega^i_3 + \ii *_6 \omega^i_3  \\
\end{split}
\end{equation}
where $\phi_{\pm}$ are complex self-dual forms while $\omega_k^i$ are real $k$-forms. We again give names to the sum and the difference of the forms generated by the same spinor:
\begin{equation}
\omega_k = \frac{\omega^1_k + \omega^2_k}{2} \, , \qquad \qquad \widetilde{\omega}_k = \frac{\omega^1_k - \omega^2_k}{2} \, .
\end{equation}

Given the four and the six-dimensional bilinears, we can calculate, following \cite{10d}, how the ten-dimensional ones decompose:
\begin{equation}
\begin{split}
&\Phi = 2 \text{Re} \left((e^A v + \ii e^{3A} *_4 v) \wedge \phi_+ + e^{2A} v \wedge w \wedge \phi_- \right) \, , \\
&K = 2 e^{-A} \omega_0 \partial_{v} \, , \qquad \widetilde{K} = 2 e^A \widetilde{\omega}_0 v \, ,\\
&\widetilde{\Omega} = 2 \text{Re} \left(-e^A v \wedge *_6 \widetilde{\omega}_2 - e^{3A} *_4 v \wedge \widetilde{\omega}_2  + e^{2A} v \wedge w \wedge \omega_3 \right) \, .
\end{split}
\end{equation} 

As we have seen in section \ref{sec:s}, to make more explicit the $\mathrm{SL}(2,\R)$-invariant structure of the supersymmetry conditions it is convenient to express all the bilinears in terms of the Einstein metric
\begin{equation}
g_{\rm E} = e^{2 A_{\rm E}} g_4 + g_{6 \, {\rm E}} \, , \qquad g_{\rm E} = e^{- \frac{\phi}{2}} g =  e^{- \frac{\phi}{2}} \left(e^{2 A} g_4 + g_{6} \right)\,,
\end{equation}
and to organize the components of the six-dimensional bilinears in terms of their U(1)$_D$ charges. In particular we have five real neutral forms
\begin{equation}
\alpha_0 = e^{-\frac{1}{4}\phi} \text{Im} (\phi_{+})_0 \, , \quad \alpha_2 = e^{-\frac{3}{4}\phi} \text{Re} (\phi_{+})_2 \, , \quad k_0 = e^{-\frac{1}{4}\phi}\omega_0 \, , \quad \alpha_1 = e^{-\frac{1}{2}\phi} \phi_-  \,,
\end{equation}
and three complex forms
\begin{equation}
\theta_0 = e^{-\frac{1}{4}\phi}(\widetilde{\omega}_0 + \ii \text{Re}(\phi_+)_0) \, , \quad \theta_2 = e^{-\frac{3}{4}\phi}(\widetilde{\omega}_2 + \ii \text{Im}(\phi_+)_2) \, , \quad \theta_3 = e^{-\phi}(\omega_3 + \ii \text{Re}(\phi_-)_3) \, ,
\end{equation}
whose of U(1)$_D$ charge equal to $+1$. 
The ten-dimensional multiplets in terms of the six-dimensional ones read:
\begin{equation}
\label{10to6multiplet}
\begin{split}
&\Theta_1 = 2 e^{A_{\rm E}} \theta_0 v \, ,\\
&\Theta_3 = 2 \left( e^{A_{\rm E}} v \wedge \alpha_2 - e^{3A_{\rm E}} *_4 v  \alpha_0 + e^{2A_{\rm E}} v \wedge w_1 \wedge \text{Re}\alpha_1-e^{2A_{\rm E}} v \wedge w_2 \wedge \text{Im}\alpha_1  \right) \, ,\\
&\Theta_5= 2 \left(- e^{A_{\rm E}} v \wedge *_{\rm E} \theta_2 - e^{3A_{\rm E}} *_4 v \wedge \theta_2 + e^{2A_{\rm E}} v \wedge w_1 \wedge \theta_3 - e^{2A_{\rm E}} v \wedge w_2 \wedge *_{\rm E} \theta_3 \right)
\end{split}
\end{equation}
where $*_{\rm E}$ is a shorthand for $*_{6 \, , \, E}$.
We apply the same logic also to redefine fluxes:
\begin{equation}
\mathcal{G} = f_3 - \ii e^{-\phi} H \, , \qquad \tau = C_0 + \ii e^{- \phi} \, , \qquad F_5 = f_5 + e^{4A_{\rm E}} \text{Vol}_4 \wedge *_{\rm E} f_5 \, . 
\end{equation}
The $\mathrm{U}(1)_D$ charges for the new forms are given in table \ref{table:charges_4D}.

\begin{table}[h!]
	\centering
	\begin{tabular}{ |c|c| } 
		\hline
		fields & $\mathrm{U}(1)_D$-charge  \\ 
		\hline
		$\alpha_0$,	$\alpha_1$,	$\alpha_2$, $k_0$, $f_5$ & $\ 0$  \\ 
		$\theta_0$, $\theta_2$ , $\theta_3$  & 1\\ 
		\hline
	\end{tabular}\caption{$\mathrm{U}(1)_D$ charges of relevant fields on the internal manifold.}
	\label{table:charges_4D}
\end{table}

Now it is enough to substitute \eqref{10to6multiplet} in \eqref{eq:Sl2Rsystem_vacua} to get the $\mathrm{SL}(2,\R)$ invariant conditions for four-dimensional vacua. Since the four-dimensional spinor must be a Killing spinor in order to preserve the isometry of the vacuum, its behavior under the action of the external derivative can be computed (see for example \cite[(2.13)]{passias-prins-t}):
\begin{equation}
\d v = 2 \mu \, \, v \wedge w_1 \, , \qquad \d(v \wedge w) = -3 \ii \mu *_4 v \, , \qquad \d *_4 v = 0 \,,
\end{equation}
where $\mu$ is related to the cosmological constant by $\Lambda=-3 |\mu|^2$. 

Now we can show the supersymmetry conditions, which are the generalization of \cite{heidenreich} to the case $\theta_0 \neq 0$ is allowed\footnote{Since $\text{Re} \theta_0$ is the difference of the norms of the two spinors, $\theta_0 \neq 0$ corresponds to allowing spinors with not-equal norm.}. For an AdS$_4\times M_6$ solution ($\mu \neq 0$) they read
\begin{subequations}
	\label{eq:S4AdS}
	\begin{align}
	&\theta_0 = 0 \, , \qquad k_0 = c_+ e^{A_{\rm E}} \, , \\
	& \d(e^{2A_{\rm E}} \text{Re} \alpha_1) + 2 \mu e^{A_{\rm E}} \alpha_2 = 0 \, , \\
	& \d(e^{3A_{\rm E}} \alpha_0) - k_0 e^{3A_{\rm E}} *_{\rm E} f_5+ 3 \mu e^{2A_{\rm E}} \text{Im} \alpha_1=0 \, ,\\
	& \d_Q(e^{A_{\rm E}} *_{\rm E} \theta_2) +\frac{\ii}{2} e^{\phi+A_{\rm E}} \d \overline\tau\wedge*_{\rm E} \overline\theta_2+ e^{A_{\rm E}} \alpha_2 \wedge \overline{\mathcal{G}}=0 \, ,\\
	& \d_Q(e^{2A_{\rm E}} \theta_3)+\frac{\ii}{2}e^{\phi + 2 A_{\rm E}} \d \overline \tau \wedge \overline \theta_3 + e^{2A_{\rm E}} \text{Re} \alpha_1 \wedge \overline{\mathcal{G}}- 2 \mu e^{A_{\rm E}}*_{\rm E} \theta_2 = 0 \, ,\\
	&\d_Q(e^{2A_{\rm E}} *_{\rm E} \theta_3) +\frac{\ii}{2}e^{\phi + 2 A_{\rm E}} \d \overline \tau \wedge *_{\rm E} \overline  \theta_3+ e^{2A_{\rm E}} \text{Im} \alpha_1 \wedge \overline{\mathcal{G}}=0 \, ,\\
	& \d_Q(e^{3A_{\rm E}} \theta_2)+\frac{\ii}{2}e^{\phi + 3 A_{\rm E}} \d \overline \tau \wedge \overline \theta_2 -e^{3A_{\rm E}}\alpha_0 \overline{\mathcal{G}} + \ii e^{3A_{\rm E}} k_0 *_{\rm E} \overline{\mathcal{G}} +3 \mu e^{2A_{\rm E}}*_{\rm E} \theta_3 = 0 \, , \\
	&\d(e^{2A_{\rm E}} *_{\rm E} \text{Im}\alpha_1) + \frac{e^{2A_{\rm E}}}{2} \text{Re}(\theta_{\rm E} \wedge \mathcal{G})+2 \mu e^{A_{\rm E}} *_{\rm E} \alpha_0=0 \, , \\
	&\d(e^{2A_{\rm E}} *_{\rm E} \text{Re}\alpha_1) - \frac{e^{2A_{\rm E}}}{2} \text{Re}(*_{\rm E} \theta_3 \wedge \mathcal{G})=0 \, ,\\
	&\d(e^{3A_{\rm E}} *_{\rm E} \alpha_2) -\frac{e^{3A_{\rm E}}}{2} \text{Re}(\theta_2 \wedge \mathcal{G})- 3 \mu e^{2A_{\rm E}} *_{\rm E} \text{Re} \alpha_1=0 \, ,
	\end{align}
\end{subequations} 
while for Mink$_4\times M_6$  solutions ($\mu=0$)
\begin{subequations}
	\label{eq:S4Mink}
	\begin{align}
	&\d_Q(e^{A_{\rm E}} \theta_0)- \frac{\ii}{2} e^{\phi + A_{\rm E}} \overline{\theta}_0 \d \overline \tau = 0 \, , \qquad k_0 = c_+ e^{A_{\rm E}} \, , \\
	&\d(e^{A_{\rm E}} \alpha_2) - e^{A_{\rm E}} \text{Re} (\theta_0 \mathcal{G})=0 \, , \qquad \d(e^{2A_{\rm E}} \alpha_1) = 0 \, , \\
	& \d(e^{3A_{\rm E}} \alpha_0) - k_0 e^{3A_{\rm E}} *_{\rm E} f_5=0 \, ,\\
	& \d_Q(e^{A_{\rm E}} *_{\rm E} \theta_2) +\frac{\ii}{2} e^{\phi+A_{\rm E}} \d \overline\tau\wedge*_{\rm E} \overline\theta_2+ e^{A_{\rm E}} \alpha_2 \wedge \overline{\mathcal{G}} + \ii e^{A_{\rm E}} \theta_0 f_5 =0 \, , \\
	& \d_Q(e^{2A_{\rm E}} \theta_3)+\frac{\ii}{2}e^{\phi + 2 A_{\rm E}} \d \overline \tau \wedge \overline \theta_3 + e^{2A_{\rm E}} \text{Re} \alpha_1 \wedge \overline{\mathcal{G}} = 0 \, ,\\
	&\d_Q(e^{2A_{\rm E}} *_{\rm E} \theta_3) +\frac{\ii}{2}e^{\phi + 2 A_{\rm E}} \d \overline \tau \wedge *_{\rm E} \overline  \theta_3+ e^{2A_{\rm E}} \text{Im} \alpha_1 \wedge \overline{\mathcal{G}}=0 \, ,\\
	& \d_Q(e^{3A_{\rm E}} \theta_2)+\frac{\ii}{2}e^{\phi + 3 A_{\rm E}} \d \overline \tau \wedge \overline \theta_2 -e^{3A_{\rm E}}\alpha_0 \overline{\mathcal{G}} + \ii e^{3A_{\rm E}} k_0 *_{\rm E} \overline{\mathcal{G}} = 0 \, ,\\
	&\d(e^{2A_{\rm E}} *_{\rm E} \text{Im}\alpha_1) + \frac{e^{2A_{\rm E}}}{2} \text{Re}(\theta_{\rm E} \wedge \mathcal{G})=0 \, ,\\
	&\d(e^{2A_{\rm E}} *_{\rm E} \text{Re}\alpha_1) - \frac{e^{2A_{\rm E}}}{2} \text{Re}(*_{\rm E} \theta_3 \wedge \mathcal{G})=0 \, ,\\
	&\d(e^{3A_{\rm E}} *_{\rm E} \alpha_2) -\frac{e^{3A_{\rm E}}}{2} \text{Re}(\theta_2 \wedge \mathcal{G})=0 \, .
	\end{align}
\end{subequations} 
As said before, this equations must be supplemented with the algebraic constraint \eqref{eq:sl2R_vacua_algebraic} which, in terms of the internal-space forms, reads:
\begin{equation}
\begin{split}
& \theta_3 \wedge \overline{\mathcal{G}} + 2 \ii e^{\phi} \d \overline{\tau} \wedge *_{\rm E} \text{Im}\alpha_1 =0 \, ,\\
&*_{\rm E} \theta_3 \wedge \overline{\mathcal{G}} - 2 \ii e^{\phi} \d \overline{\tau} \wedge *_{\rm E} \text{Re}\alpha_1 =0 \, , \\
&\theta_2 \wedge \overline{\mathcal{G}} - 2 \ii e^{\phi} \d \overline{\tau} \wedge *_{\rm E} \alpha_2- 2 e^{\phi}k_0 *_{\rm E} \d \overline{\tau} =0 \, .
\end{split}
\end{equation}

At first sight the systems \eqref{eq:S4AdS} and \eqref{eq:S4Mink} seem to contain a huge amount of equations compared to pure spinor equations of \cite{gmpt2}, which are just three. However this is due to the fact that to write the system in $\mathrm{SL}(2,\mathbb{Z})$-invariant form we had to write all form degrees separately. In fact the total number of equations is actually the same as in \cite{gmpt2}.


\section*{Acknowledgements}
We would like to thanks N. Macpherson and A. Zaffaroni for useful discussions. We are supported in part by INFN.

\appendix


\section{Some properties of spinors} 
\label{app:tech}

In this section we will review some properties of ten-dimensional spinors.

The Clifford multiplication of a single gamma matrix $\gamma^M$ with $C_k$, defined as in \eqref{eq:cl}, can be rewritten in terms of some familiar operations on the corresponding forms:
\begin{equation}
\label{eq:gammadestrasinistra}
\overrightarrow{\g}^M C_k = \g^M C_k = (\dd x^M \wedge+\iota^M) C_k \, , \qquad \overleftarrow{\g}^M C_k = C_k \g^M = (-)^k(\dd x^M \wedge -\iota^M ) C_k \, 
\end{equation}
where $\iota_M$ indicates contraction along direction $M$.

We can also combine these operators with the action of $\gamma$ in (\ref{eq:gamma=}): 
\begin{equation}
	\overrightarrow{\g}^M \overrightarrow{\g} = - \overrightarrow{\g} \overrightarrow{\g}^M \, ,\qquad \overleftarrow{\g}^M \overrightarrow{\g} =\overrightarrow{\g} \overleftarrow{\g}^M  \,.
\end{equation}
It follows that
\begin{equation}
\label{eq:gammaiwedge}
\d x^M \wedge \overrightarrow{\g} = - \overrightarrow{\g} \iota^M\,, \qquad \iota^M  \overrightarrow{\g} = -  \overrightarrow{\g} \d x^M \, .
\end{equation}
From the definition of $\lambda$ just below \eqref{eq:gamma=} we also get 
\begin{equation}\label{eq:lambdagammaprop}
\lambda(\d x^M \wedge C_k) = (-)^k \d x^M \wedge \lambda( C_k)\, ,\quad \lambda(\iota^M C_k) = - (-)^k \iota^M \lambda( C_k) \, ,\quad\lambda(\g^M C_k) = \lambda(C_k) \g^M.
\end{equation}

The generalization of (\ref{eq:fierzeps}) to any bispinor $C$ reads:
\begin{equation}
C = \sum_{k = 0}^{10} \frac{1}{32 k!} \mathrm{tr}(C \g_{M_k \dots M_1}) \g^{M_1 \dots M_k} \,  .
\end{equation}
In particular, for $C = \eps \otimes \overline{\eta}$, by cyclicity of the trace one gets
\begin{equation}
\label{eq:fierzid2}
\eps \otimes \, \overline{\eta} = \sum_{k = 0}^{10} \frac{1}{32 k!} \, (\overline{\eta}\g_{M_k \dots M_1}\eps) \, \g^{M_1 \dots M_k} \,  ,
\end{equation}
and, by imposing $\eta=\epsilon$, we get back to (\ref{eq:fierzeps}).

The Chevalley--Mukai pairing between two forms $A$ and $B$ defined in (\ref{eq:mukai}) is also related to a bispinor trace by
\begin{equation}
\label{eq:CMpair}
(A,B) = - \frac{(-1)^{\text{deg}(A)}}{2^{5}} \mathrm{tr} (* A B) \,  .
\end{equation}



\section{Sufficiency of the main system} 
\label{app:proof}

In this appendix we will describe the proof of the sufficiency of (\ref{eq:systemII}) for supersymmetry. We will work only with IIB theory; the discussion for IIA is analogous. 

In the notation of \cite{10d}, the supersymmetry conditions read
\begin{subequations}
	\label{eq:SUSY1}
	\begin{align}
	\left( D_M - \frac{1}{4} H_M  \right) \eps_1 & + \frac{\text{e}^{\phi}}{16}  F \g_M \eps_2 = 0 \, , \qquad \qquad \qquad 
	\left( D - \frac{1}{4} H - \de \phi \right)\eps_1 = 0 \, , \\
	\left( D_M + \frac{1}{4} H_M  \right) \eps_2 &+(-)^{|F|} \frac{\text{e}^{\phi}}{16} \lambda(F) \g_M \eps_1 = 0 \, , \qquad
	\left( D + \frac{1}{4} H - \de \phi \right)\eps_2 = 0 \, ,
	\end{align}
\end{subequations}
where the sign $(-)^{|F|} = (-)^{\text{deg}(F)}$ is the only difference between IIA and IIB. 
Acting with $\overrightarrow{\g}^{M}$ on the two equations on the left and subtracting the ones on the right side we also obtain the original dilatino equations:
\begin{subequations}
	\label{eq:SUSY2}
	\begin{align}
	\left( \de \phi - \frac{1}{2} H \right) \eps_1 &+ \frac{\text{e}^{\phi}}{16} \g^M F \g_M \eps_2 = 0 \label{eq:SUSY21} \\
	\left( \de \phi + \frac{1}{2} H \right) \eps_2 &+ (-)^{|F|} \frac{\text{e}^{\phi}}{16} \g^M \lambda(F) \g_M \eps_1 = 0 \label{eq:SUSY22} \,.
	\end{align}
\end{subequations}
During calculations we will also need the transposed version of these equations.

\subsection{Structure groups and intrinsic torsion}
\label{app:s-group}

We first review some spinorial geometry in ten dimensions, following \cite{10d} (especially Sec.~2 and App.~B there), to which we refer for details.

A spinor leads to a reduction of the structure group of the tangent bundle to its stabilizer. Let us determine the structure group defined by a spinor $\eps$ of chirality $+$. For convenience, we choose a frame in which $K=e_-$ is part of the vielbein:
\begin{equation}
\label{eq:vielbein}
e_+ \cdot e_- = \frac{1}{2} \, \, , \quad  e_{\pm} \cdot e_{\pm} = 0 \,  , \quad e_{\pm} \cdot e_{\alpha} = 0 \, \, , \quad e_{\alpha} \cdot e_{\alpha} = 1 \,  , 
\end{equation}
with $\alpha = 1, \dots , 8$. This choice of indices suggests to decompose the Clifford algebra as $\mathrm{Cl}(1,9) \simeq \mathrm{Cl}(1,1) \otimes \mathrm{Cl}(0,8)$, so that we can rewrite
\begin{equation}
\label{eq:decepseta}
\eps = | \uparrow \, \rangle \otimes \eta \, ,
\end{equation}
in terms of a two-dimensional $| \uparrow \rangle$ and of an eight-dimensional Majorana--Weyl spinor $\eta$. We can now look at the infinitesimal action of a Lorentz transformation on $\eps$ to compute its stabilizer:
\begin{equation}
	\delta \eps = \omega_{MN} \g^{MN} \eps \, , \qquad \omega_{MN} \g^{MN} \in \mathrm{spin}(1,9)\,.
\end{equation}
Since $K \eps = \g_- \eps = \g^+ \eps = 0$, we have that $\g^{+ \alpha}$ annihilates $\eps$. Moreover the eight-dimensional spinor $\eta$ is annihilated by 21 out of 28 of the eight-dimensional gamma matrices $\g^{\alpha \beta}$; so we can write:
\begin{equation}
\label{stab:1eps}
\mathrm{stab}(\eps) = \text{span}\{ \omega^{21}_{\alpha \beta} \g^{\alpha \beta} , \g^{+ \alpha} \} \, \, .
\end{equation}
The elements $\omega^{21}_{\alpha \beta} \g^{\alpha \beta}$ are in the adjoint representation of $\mathrm{Spin}(7)$. Moreover because $[\g_{\alpha \beta} , \g^{+ \delta}] = 2 \delta^\delta_{[\alpha} \g_{\beta]}^+$ we have that  
\begin{equation}
\mathrm{Stab}(\eps) =\mathrm{Spin}(7) \ltimes \R^8 = \mathrm{ISpin}(7) \, ,
\end{equation}
where ISpin is the inhomogeneous spin group, in analogy with $\mathrm{ISO}(n)$ for  inhomogeneous $\mathrm{SO}(n)$ groups.

We expect that the same structure group can be deduced also by using the forms generated by $\eps$. Let's start from the stabilizer of $K$; since $K$ is null
\begin{equation}
\text{Stab}(K) = \mathrm{ISO}(8) =\mathrm{SO}(8) \ltimes \R^8 \, .
\end{equation}
\eqref{eq:ikk^omega}--(\ref{eq:OKPsi}) tell us that the four-form $\Psi$ contains only components which are orthogonal to $K$ different from $K$ itself; i.e., in the basis \eqref{eq:vielbein}, only $\alpha$ components. If we restrict our original spinor $\eps$ to this eight-dimensional subspace we obtain the eight-dimensional Majorana--Weyl spinor $\eta$ defined in (\ref{eq:decepseta}), which is known to give rise to a $\mathrm{Spin}(7)$ structure. In fact $\Psi$ is nothing but the four-form that describes this $\mathrm{Spin}(7)$ structure which we can find in the eight-dimensional bispinor $\eta \otimes \eta^t$. The local Lorentz transformations that leaves $\Psi$ invariant reduce $\mathrm{SO}(8)$ to $\mathrm{Spin}(7)$ and then we have again 
\begin{equation}
\text{Stab}(K , \Omega) =  \mathrm{ISpin}(7) \, .
\end{equation}

The stabilizer \eqref{stab:1eps} of the infinitesimal action of a Lorentz transformation on $\eps$ is 29 dimensional. The orbit of the Lorentz group action, which is given by all spinors that can be written as $\g^{MN} \eps$, has the dimension of $\mathrm{Spin}(9, 1)$ minus the dimension of the isotropy group, which is $45-29=16$. Since the space of Majorana--Weyl spinors with the same chirality is 16-dimensional,  
\begin{equation}\label{eq:basis+}
	\{ \g^{MN} \eps \}
\end{equation}
is a basis for the space of spinors  with the same chirality as $\epsilon$. For Majorana--Weyl spinors with opposite chirality, we can find a basis by picking a particular spinor with negative chirality and acting on it with $\g^{MN}$. A natural choice for this spinor is $\g_+ \eps$, and hence our basis for spinors with chirality opposite to $\eps$ is 
\begin{equation}\label{eq:basis-}
	\{ \g^{MN} \g_+ \eps \} \, \, .
\end{equation}

Type II theories actually contain two fermionic parameters $\eps_{1,2}$. Each one of them defines an $\mathrm{ISpin}(7)$ structure. To evaluate the stabilizer of $\eps_{1,2}$ in $\mathrm{SO}(1,9)$ we have to look at the intersection of the two copies of $\mathrm{ISpin}(7)$. Various possibilities exist for this intersection and for the $G$-structure on $M_{10}$; for details see \cite[Sec.~2.2]{10d}. In fact the common stabilizer of $\epsilon_{1,2}$ may change from a point to another even for a single solution.

However, in the spirit of generalized complex geometry, we can try to define the common stabilizer on the generalized tangent bundle $T M_{10} \oplus T^*M_{10}$. Considering the bilinears defined by $\eps_{1,2}$ as spinors on this generalized tangent bundle, we are able to define a structure group as a subgroup of $\mathrm{O}(10,10)$. The action of $\mathrm{Cl}(10,10)$ can be decomposed as two copies of ordinary $\mathrm{Cl}(9,1)$ gamma matrices acting from the left and from the right of a bispinor as in \eqref{eq:gammadestrasinistra}. For example, the presence of a metric and a $B$ field on $M_{10}$ restricts the structure group to $\mathrm{o}(9,1) \times \mathrm{o}(9,1) = \text{span}\{ \overleftarrow{\g}^{MN} , \overrightarrow{\g}^{MN} \}$. 
If moreover we add as geometric data also the two spinors $\eps_1$ and $\eps_2$ we have a basis of the type (\ref{eq:basis+})--(\ref{eq:basis-}) associated to both, we will the use a subscript $1$ or $2$ to distinguish index relative to $\eps_1$ from the index relative to $\eps_2$. The common stabilizer therefore reads:
\begin{equation}
\text{stab}(g,B,\eps_1 , \overline{\epsilon}_2) = \text{span} \{ \omega_{21}^{\alpha_1 \beta_1} \overrightarrow{\g}_{\alpha_1 \beta_1} , \overrightarrow{\g}_{-_1 \alpha_1},\omega_{21}^{\alpha_2 \beta_2} \overleftarrow{\g}_{\alpha_2 \beta_2} , \overleftarrow{\g}_{-_2 \alpha_2}  \} = \mathrm{ispin}(7) \times \mathrm{ispin}(7) \, .
\end{equation}

Now let's evaluate the stabilizer of our bilinears, it is easy to see that:
\begin{equation}
\begin{split}
&\text{stab}(\Phi) = \text{span} \begin{Bmatrix}
\omega_{21}^{\alpha_1 \beta_1} \overrightarrow{\g}_{\alpha_1 \beta_1} , \overrightarrow{\g}_{-_1 \alpha_1},\omega_{21}^{\alpha_2 \beta_2} \overleftarrow{\g}_{\alpha_2 \beta_2} , \overleftarrow{\g}_{-_2 \alpha_2} , \overrightarrow{\g}_{-_1 +_1} + \overleftarrow{\g}_{-_2 +_2}   \\
\overrightarrow{\g}_{-_1}\overleftarrow{\g}_{\alpha_2} , \overrightarrow{\g}_{-_1}\overleftarrow{\g}_{+_2} , \overrightarrow{\g}_{\alpha_1}\overleftarrow{\g}_{-_2} , \overrightarrow{\g}_{+_1}\overleftarrow{\g}_{-_2} , \overrightarrow{\g}_{-_1}\overleftarrow{\g}_{-_2}
\end{Bmatrix} \, , \\
&\text{stab}(\eps_i \overline{\eps}_i) = \text{span} \begin{Bmatrix}
\omega_{21}^{\alpha_i \beta_i} \overrightarrow{\g}_{\alpha_i \beta_i} , \overrightarrow{\g}_{-_i \alpha_i},\omega_{21}^{\alpha_i \beta_i} \overleftarrow{\g}_{\alpha_i \beta_i} , \overleftarrow{\g}_{-_i \alpha_i} , \overrightarrow{\g}_{-_i +_i} + \overleftarrow{\g}_{-_i +_i}   \\
\overrightarrow{\g}_{-_i}\overleftarrow{\g}_{\alpha_i} , \overrightarrow{\g}_{-_i}\overleftarrow{\g}_{+_i} , \overrightarrow{\g}_{\alpha_i}\overleftarrow{\g}_{-_i} , \overrightarrow{\g}_{+_i}\overleftarrow{\g}_{-_i} , \overrightarrow{\g}_{-_i}\overleftarrow{\g}_{-_i} 
\end{Bmatrix} \, .
\end{split}
\end{equation}
In the timelike case, since $K^2 = \frac{1}{2} K_1 \cdot K_2  \neq 0$, we are allowed to choose $e_{+_1} \sim K_2$ and $e_{+_2} \sim K_1$ and therefore the common stabilizer reads:
\begin{equation}
\text{stab}(\Phi,\Omega_1,\Omega_2) \subseteq \text{span} \{ \omega_{21}^{\alpha_1 \beta_1} \overrightarrow{\g}_{\alpha_1 \beta_1},\omega_{21}^{\alpha_2 \beta_2} \overleftarrow{\g}_{\alpha_2 \beta_2} \} = \mathrm{spin}(7) \times \mathrm{spin}(7)
\end{equation}
where we considered $\Omega_i$ instead of $\eps_i \overline{\eps}_i$ since in the timelike case $\Phi$ alone is enough to determine $K_1$ and $K_2$. 
So we discovered that in the timelike case the differential forms we have in \eqref{eq:systemII} contain enough information to define the metric, the B field and two spinors.
However, to prove the sufficiency we must find a way to count the independent components of the supersymmetry equations. Along the lines of  \cite[Sec.~A.4]{gmpt3} we can define:
\begin{equation}
	\label{eq:intr1}
	\begin{split}
	\left( D_M - \frac{1}{4} H_M  \right) \eps_1 &= Q^1_{MNP} \g^{NP} \eps_1  \,  ,  \quad 
	\left( D - \frac{1}{4} H - \de \phi \right)\eps_1 = T^1_{MN} \g^{MN} \g_{+_1} \eps_1 \, , \\
	\left( D_M + \frac{1}{4} H_M  \right) \eps_2 &= Q^2_{MNP} \g^{NP} \eps_2  \, , \quad 
	\left( D + \frac{1}{4} H - \de \phi \right)\eps_2 =  T^2_{MN} \g^{MN} \g_{+_2} \eps_2 \, .
	\end{split}
\end{equation}
There is no assumption so far: the left hand sides are spinors that can be expanded on our basis and the $Q$'s and $T$'s are coefficients of this expansion. They can be viewed as intrinsic torsion coefficients. Notice that some elements of the expansion are trivially zero because belong to the stabilizers of $\eps_1$ and $\eps_2$, so we can put to zero the corresponding intrinsic torsion components:
\begin{equation}
\label{eq:torsintescl}
Q^a_{M \alpha_a +_a} = 0 \, \, , \quad  \omega_{21}^{\alpha_a \beta_a} Q^a_{M \alpha_a \beta_a} = 0 \, \, , \quad T^a_{\alpha_a -_a} = 0 \, \, , \quad  \omega_{21}^{\alpha_a \beta_a} T^a_{\alpha_a \beta_a} \qquad a = 1,2 \, .
\end{equation}
For the same reason, we can assume that the $Q$'s are antisymmetric in their last two indices while the $T$'s are totally antisymmetric.
Now, tensoring two copies of (\ref{eq:basis+})--(\ref{eq:basis-}), we can produce a basis for the space of bispinors from the spinors one:
\begin{equation}
	\g_{MN} \eps_1 \overline{\eps}_2 \g_{PQ} \, , \quad \g_{MN} \g_{+_1} \eps_1 \overline{\eps}_2 \g_{+_2} \g_{PQ}  \, , \quad \g_{MN} \g_{+_1} \eps_1 \overline{\eps}_2 \g_{PQ}  \, , \quad \g_{MN}  \eps_1 \overline{\eps}_2 \g_{+_2} \g_{PQ} \, .
\end{equation}
In IIB, the first two sets of generators are a formal sum of odd forms while the second two are a sum of even ones. $F$ is sum of odd forms and furthermore it is self dual: $\g F = F$. This tells us that it will be a linear combination of the first set of generators:
\begin{equation}
\label{eq:Fintrtro}
F = R_{MNPQ} \g^{MN} \eps_1 \overline{\eps}_2 \g^{PQ} \, \, ,
\end{equation}
and moreover we also have that
\begin{equation}
\label{eq:lambdaFintrtro}
\lambda(F) = R_{MNPQ} \g^{QP} \eps_2 \overline{\eps}_1 \g^{MN} \, \, .
\end{equation}

Replacing (\ref{eq:Fintrtro}) and (\ref{eq:lambdaFintrtro}) in \eqref{eq:SUSY1} and, comparing them with \eqref{eq:intr1}, we get:
\begin{equation}
\label{eq:susyintorsint}
\begin{split}
Q^1_{MNP} =& 4 \e^\phi R_{NPM -_2} \, \, , \qquad T^1_{MN} = 0 \, \, , \\
Q^2_{MNP} =& 4 \e^\phi R_{-_1MNP} \, \,  , \qquad T^2_{MN} = 0 \, \, .
\end{split}
\end{equation}
These are the supersymmetry equations rewritten in terms of intrinsic torsion components.

For what follows, it is also useful to rewrite \eqref{eq:SUSY2}: combining  equations \eqref{eq:intr1} we get
\begin{subequations}
	\label{eq:intr2}
	\begin{align}
	\left( \de \phi - \frac{1}{2} H \right) \eps_1 &= Q^1_{MNP} \g^M \g^{NP} \eps_1 -  T^1_{MN} \g^{MN} \g_{+_1} \eps_1 \, \, ,  \label{eq:intr21} \\
	\left( \de \phi + \frac{1}{2} H \right) \eps_2 &=  Q^2_{MNP} \g^M \g^{NP} \eps_2 -  T^2_{MN} \g^{MN} \g_{+_2} \eps_2 \,\, .  \label{eq:intr22}
	\end{align}
\end{subequations}

\subsection{Intrinsic torsion for form equations}

We will now rewrite all the equations derived in the previous subsection in the language of the intrinsic torsion components $Q$ and $T$. 
The intrinsic torsion components for equation \eqref{eq:oldII} were derived in \cite{10d}: 
\begin{subequations}\label{eq:Phiintr}
\begin{align}
Q^1_{MN \alpha_1} =& 4 \e^{\phi} R_{N \alpha_1 M -_2} \quad (M \neq +_2) \,  , \qquad T^1_{\alpha_1 \beta_1} = 0 \,  , \label{eq:Phiintr-Q1MNa} \\
Q^2_{MN \alpha_2} =& 4 \e^{\phi} R_{-_1M N \alpha_2} \quad (M \neq +_1) \,  , \qquad T^2_{\alpha_2 \beta_2} = 0 \,  ,  \label{eq:Phiintr-Q2MNa}\\
Q^1_{\alpha_2 +_1 -_1} +& T^2_{\alpha_2 +_2} =  4 \e^{\phi} R_{+_1 -_1 \alpha_2 -_2} \,  , \qquad T^2_{+_2 -_2} = -2 Q^1_{-_2 +_1-_1}\,  , \label{eq:Phiintr-Q1a+-}\\
Q^2_{\alpha_1 +_2 -_2} +& T^1_{\alpha_1 +_1} =  4 \e^{\phi} R_{-_1  \alpha_1 +_2 -_2} \,  , \qquad T^1_{+_1 -_1} = -2 Q^2_{-_1 +_2-_2}\, . \label{eq:Phiintr-Q2a+-}
\end{align}
\end{subequations}
In order to prove sufficiency  of \eqref{eq:systemII} for supersymmetry, we have to show that the last three equations of the system contain the intrinsic torsion equations that appear in (\ref{eq:susyintorsint}) and do not appear in (\ref{eq:Phiintr}).

Let's start from the six-form equations first, (\ref{eq:dOmegaIIA}), (\ref{eq:domegatII}) (recall that we are looking at IIB). Since the intrinsic torsion components inside the equations for $d \Omega$ and $d \widetilde{\Omega}$ is the same as the one we will find evaluating  $d \Omega_{1}$ and $d \Omega_2$ separately, we can directly start from the latter. Replacing \eqref{eq:intr1} and \eqref{eq:intr2} inside \eqref{eq:eps1noSUSY} we get:
\begin{align}\label{eq:dOmega1intr}
	2 \e^{ 2\phi} \d (\e^{- 2 \phi} \eps_1 \bar{\eps}_1) &+ 2 \iota_H  \eps_1 \bar{\eps}_1 = 2 T^1_{MN} \left( \g^{MN} \g_{+_1} \eps_1 \bar\eps_1 + \eps_1 \bar\eps_1 \g_{+_1}  \g^{MN} \right) \\
	&- Q^1_{MNP} \left( \g^M \eps_1 \bar\eps_1 \g^{NP} + \g^{NP} \eps_1 \bar\eps_1 \g^M + \g^M \g^{NP} \eps_1 \bar\eps_1 + \eps_1 \bar\eps_1 \g^{NP} \g^M   \right) . \nonumber
\end{align}
Notice that the terms in the first bracket are independent, because they are tensor products of spinors with different chirality.

The situation is a little more complicated for the second bracket in (\ref{eq:dOmega1intr}). First of all, since $\g^{+_1}$ annihilates $\eps_1$ and commutes with all the $\g^{\alpha_1}$, the components of $Q^1_{MNP}$ with
\begin{equation}
	Q^1_{+_1 \alpha_1 \beta_1 }\, \, 
\end{equation}
are absent from it. Moreover, the bracket contains tensor products of spinors with the same chirality which can therefore add up to zero: indeed
\begin{equation}
\begin{sistema}
\g^M \g^{NP} \eps_1 = c \g^M \eps_1 \\ 
\g^{NP} \eps_1 = c \eps_1 
\end{sistema} \quad \, \, 
\end{equation}
has solution for $\g^{NP} = \g^{+_1-_1}$ and $c = 2$. So in fact also the components
\begin{equation}
	Q^1_{M +_1 -_1}
\end{equation}
are absent. 

(\ref{eq:dOmega1intr}) gives independent equations for the remaining components of  $Q^1_{MNP}$  and for $T^1_{MN}$. In other words, no choice of indices in $Q^1_{MNP}$ and in $T^1_{MN}$ multiply the same bispinor. If we focus on tensor products of spinors with $-+$ chirality, we need to compare the first  $T^1_{MN}$  term ($\gamma^{MN} \gamma_{+_1} \epsilon_1 \overline{\epsilon_1}$) with the first and third $Q^1_{MNP}$ terms ($\gamma^M \epsilon_1 \overline{\epsilon_1}\gamma^{NP}$ and $\gamma^M \gamma^{NP} \epsilon_1 \overline{\epsilon_1}$). If we want to sum the first  $T^1_{MN}$  term  with the first and  third $Q^1_{MNP}$ terms we have to choose $NP$ to be such that $\g^{NP} \eps_1 = \eps_1$, which has solution for $NP=+_1 -_1$; but in this case the first and third $Q^1_{MNP}$ terms  actually cancel. We can have that the first $T^1_{MN}$ term can sum with the second $Q^1_{MNP}$ term, but in this case the first $Q^1_{MNP}$ term multiplies a different bispinor and implies a separate equation.

If we now replace \eqref{eq:Fintrtro} and \eqref{eq:lambdaFintrtro} in \eqref{eq:eps1epsbar1} and use $\bar\eps_2 \g^{PQ} \g_S \eps_2 = 64 K_2^{[P} \delta^{Q]}_S$ we have
\begin{equation}\label{eq:de1e1R}
	\begin{split}
	2 \e^{ 2\phi} \d (\e^{- 2 \phi} \eps_1 \bar{\eps}_1) + 2 \iota_H  \eps_1 \bar{\eps}_1 &= 4 \e^\phi R_{MN-_2Q} \big(  \g^Q \eps_1 \bar\eps_1 \g^{MN} +  \g^{MN} \eps_1 \bar\eps_1 \g^Q \\ &+ \g^Q \g^{MN} \eps_1 \bar\eps_1 + \eps_1 \bar\eps_1 \g^{MN} \g^Q \big) \, \, .
	\end{split}	
\end{equation}

Comparing (\ref{eq:dOmega1intr}) and (\ref{eq:de1e1R}) we can extract the content of (\ref{eq:eps1epsbar1}) in terms of intrinsic torsion:
\begin{equation}
\label{eq:eps1comp}
\begin{split}
&Q^1_{MNP} = \, 4 \e^\phi R_{NPM-_2} \quad \text{with} \quad MNP \neq +_1 \alpha_1 \beta_1 \, , \, M +_1 -_1 \\
&T^1_{MN} = \, 0 \, \, .
\end{split}
\end{equation}
A similar procedure can be applied to (\ref{eq:eps2epsbar2}) and leads to 
\begin{equation}
\label{eq:eps2comp}
\begin{split}
&Q^2_{MNP} = \, 4 \e^\phi R_{-_1MNP} \quad \text{with} \quad MNP \neq +_2 \alpha_2 \beta_2 \, , \, M +_2 -_2 \\
&T^2_{MN} = \, 0 \, \, .
\end{split}
\end{equation}

Notice that \eqref{eq:eps1comp}--\eqref{eq:eps2comp} are not yet the intrinsic torsion component for the equations for $\d \Omega_{1,2}$ because we still have to project (\ref{eq:eps1epsbar1} and (\ref{eq:eps2epsbar2}) on the six-form part (they include also a two-form and a ten-form part). It may happen that projecting on the six-form part makes one lose some  intrinsic torsion equations in (\ref{eq:eps1comp})--(\ref{eq:eps2comp}). In other words, at this point some of those equations might be actually due to the two-form and ten-form, rather than to the six-form part we're interested in.

Moreover, (\ref{eq:dOmegaIIA})--(\ref{eq:domegatII}) does not help us find anyway all the missing component of \eqref{eq:Phiintr}. First of all some components are missing from both systems; for example we never get $Q^{1}_{+_2 +_1 -_1}$. More troubling still, in the new intrinsic equations (\ref{eq:eps1comp})--(\ref{eq:eps2comp}) the missing $Q^1_{MNP}$ components are written with the first index being of 1 type, whereas in all the equations in (\ref{eq:Phiintr}) the missing components of $Q^1_{MNP}$ are written with the first index being of 2 type. In the most general case, we cannot even compare the two types of indices. In the light-like case, for example, $+_1$ is equal to $+_2$; in the timelike case, this is no longer true, and $+_1$ might even be taken to be $-_2$.

All this suggests that we simplify the problem by restricting ourselves to one case; for reasons that will become clear, we have found it easier to work with the timelike case, which as we explained in the introduction is actually generic.

\subsection{Intrinsic torsion in the timelike case}

In this case, $K_1$ and $K_2$ are two different null vectors: we can choose them to define the $+$ and $-$ indices respectively, and we can define the remaining eight indices $\alpha$ to be the directions orthogonal to them. In other words:
\begin{equation}
\label{eq:basetimelike}
e_{+_1} = e_{-_2} \, ,\qquad
e_{-_1} = e_{+_2} \, ,\qquad e_{\alpha_1}= e_{\alpha_2}\equiv e_\alpha \, .
\end{equation}
We stress that $K_1$ and $K_2$ are not equal but just proportional to the vielbein vectors since in general $K_1 \cdot K_2 \neq \frac{1}{2}$. However, since all the computations in this section are algebraic, this difference doesn't really matter and we are free to ignore this normalization.

We can now go back to the problem of taking the six-form part of \eqref{eq:eps1comp}--\eqref{eq:eps2comp}. Starting from 
\begin{equation}\label{eq:dO1}
	\begin{split}
	2 \e^{ 2\phi} \d (\e^{- 2 \phi} \Omega_1) &+ 2 \iota_{K_1} * H  = 2 T^1_{MN} \left( \g^{MN} \g_{+1} \eps_1 \bar\eps_1 + \eps_1 \bar\eps_1 \g_{+1}  \g^{MN} \right)_6 \\
	&- Q^1_{MNP} \left\{ \g^M ,\left\{ \g^{NP} , \eps_1 \bar\eps_1 \right\} \right\}_6 \, ;
	\end{split}
\end{equation}
we will try to see if it contains equations for the components of $Q$ and $T$ that appear in (\ref{eq:susyintorsint}) and not in (\ref{eq:Phiintr}).

We will start from the $T$ components for which we would like to have an equation, namely  $T^1_{\alpha +_1}$ and $T^1_{+_1 -_1}$. The first, $T^1_{\alpha +_1}$, appears in (\ref{eq:dO1}) multiplying
\begin{equation}\label{eq:T1a+1}
	\begin{split}
	\left( \g^{\alpha +_1} \g_{+_1} \eps_1 \bar\eps_1 + \eps_1 \bar\eps_1 \g_{+_1}  \g^{\alpha +_1} \right)_6 &= 2 \, \left( \g^{\alpha}\eps_1 \bar\eps_1 - \eps_1 \bar{\eps_1 } \g^{\alpha}    \right)_6 \\
	= 4 \, \d x^{\alpha}\wedge \Omega_1 &=4 \, \,\d x^{\alpha}\wedge\Psi_1 \wedge K_1   \,.
	\end{split}
\end{equation}
$\Psi_1$ is the Spin(7) four-form associated to the spinor $\epsilon_1$ via (\ref{eq:OKPsi}); it is not annihilated by the wedge product with any one-form.
So the components $T^1_{\alpha +_1}$ do appear in (\ref{eq:dO1}). As for  $T^1_{+_1 -_1}$, it appears in (\ref{eq:dO1}) multiplied by
\begin{equation}\label{eq:T1+1-1}
	\begin{split}
	\left( \g^{-_1 +_1} \g_{+_1} \eps_1 \bar\eps_1 + \eps_1 \bar\eps_1 \g_{+_1}  \g^{-_1 +_1} \right)_6 =& \left( \g^{+_1}\eps_1 \bar\eps_1 + \eps_1 \overline{\g^{+_1}  \eps_1 }  \right)_6  \\
	=& 2 \, K_2\wedge \Omega_1 \neq 0 \, \, ,
	\end{split}
\end{equation}
so also $T^1_{+_1-_1}$ does appear in (\ref{eq:dO1}).

Now let us consider the $Q$ terms. 
This time we are interested in the components $Q^1_{-_1 \alpha \beta}$, $Q^1_{-_1 -_1 \alpha}$. For the first, $Q^1_{-_1 \alpha \beta}$, we see that it appears in (\ref{eq:dO1}) with
\begin{equation}\label{eq:Q1-1ab}
	\begin{split}
	\left\{ \g^{-_1} ,\left\{ \g^{\alpha \beta} , \eps_1 \bar\eps_1 \right\} \right\}_6&= 2\iota_{K_2} \left\{	\gamma^{\alpha \beta}, \epsilon_1 \bar \epsilon_1\right\}_7= 2\gamma \left[ K_2 \wedge  \left\{	\gamma^{\alpha \beta}, \epsilon_1 \bar \epsilon_1\right\}_3 \right]\\
  &=2\gamma \left[K_2 \wedge K_1 \wedge \left\{	\gamma^{\alpha \beta}, \eta_1 \eta_1^t\right\}_2 \right]
	\, ,
	\end{split}
\end{equation}
where $\eta_1$ is the eight-dimensional part of $\eps_1$ (as in \eqref{eq:decepseta}). Now, the bispinor $\left\{	\gamma^{\alpha \beta}, \eta_1 \eta_1^t\right\}$ only has a two-form and a six-form part. (This can be seen for example by writing it as $\tilde \eta_1 \eta^t- \eta_1 \tilde{\eta}^t$.) Moreover, it is invariant under left multiplication by the eight-dimensional chiral matrix, so it is self-dual; thus the six-form part is the eight-dimensional Hodge dual of the two-form part. We thus conclude that (\ref{eq:Q1-1ab}) is zero if and only if $ \left\{ \g^{\alpha \beta} , \eta_1 \eta_1^t \right\} = 0$. This could only happen if
\begin{equation}
	 \g^{\alpha \beta} \eta_1 = c \eta_1 	
\end{equation}
for some real $c$. Contracting this from the left by $\eta_1^t$ however gives $c=0$. Another way of seeing this is that, as seen above (\ref{stab:1eps}), 
 $\g^{\alpha \beta} \eta$ is non-zero only for seven linear combinations; and those are independent from $\eta$, since they form with it a basis for eight-dimensional spinors.  Since those seven combinations are the only ones that matter in $Q_{M \alpha \beta}$, see (\ref{eq:torsintescl}), we conclude that the components $Q^1_{-_1 \alpha \beta}$ are all present in (\ref{eq:dO1}). For $Q^1_{-_1 -_1 \alpha}$ we can proceed similarly: it multiplies
\begin{equation}\label{eq:Q1-1-1a}
	\left\{ \g^{-_1} ,\left\{ \g^{-_1 \alpha} , \eps_1 \bar\eps_1 \right\} \right\}_6= 
	2\gamma \left[K_2 \wedge  \left\{	\gamma^{-_1 \alpha}, \epsilon_1 \bar \epsilon_1\right\}_3\right]=
	  4(K_1 \cdot K_2) \g \left( \iota^{\alpha}\Psi_1 \wedge K_2   \right)\,,
\end{equation}
where in the last step we have used that $\{ \gamma^{MN}, \cdot \}= 2 (\dd x^M\wedge \dd x^N + \iota^M \iota^N)$ on even forms (which follows from (\ref{eq:gammadestrasinistra})).
(\ref{eq:Q1-1-1a}) is always nonzero because the Spin(7) four-form $\Psi$ is never annihilated by any single contraction. 

We have shown that $T^1_{\alpha +_1}$, $T^1_{+_1 -_1}$, $Q^1_{-_1 \alpha \beta}$, $Q^1_{-_1 -_1 \alpha}$ are present in (\ref{eq:dO1}); they occur multiplying (\ref{eq:T1a+1})--(\ref{eq:Q1-1-1a}). Now we also observe that those forms are all independent; to see this, it is enough to look at the way the $K_i$ appear. Thus, (\ref{eq:dO1}) gives equations for all of these components. Moreover, once we assume (\ref{eq:Phiintr-Q1MNa}), in fact $T^1_{\alpha +_1}$, $T^1_{+_1 -_1}$, $Q^1_{-_1 \alpha \beta}$, $Q^1_{-_1 -_1 \alpha}$ are the only intrinsic torsions appearing in (\ref{eq:dO1}). So there is also no danger of them mixing with anything else.\footnote{Indeed we also observed that the equations in \eqref{eq:Phiintr-Q1MNa} which mixed $Q$ and $T$ always contains a term like $Q^i_{M +_i -_i}$, for which we proved that it cannot appear in the 6-form equations. Therefore we will never have the same mixed component, which means that the $Q$ and $T$ intrinsic torsion is always decoupled.} In other words, after assuming (\ref{eq:Phiintr-Q1MNa}), the left-hand side of (\ref{eq:dO1}) consists of the four terms  (\ref{eq:T1a+1})--(\ref{eq:Q1-1-1a}), multiplied by $T^1_{\alpha +_1}$, $T^1_{+_1 -_1}$, $Q^1_{-_1 \alpha \beta}$, $Q^1_{-_1 -_1 \alpha}$ respectively. Since (\ref{eq:dO1}) follows from supersymmetry, we conclude that it gives us the equations (\ref{eq:susyintorsint}) for these components. 

Going back to (\ref{eq:Phiintr-Q1a+-}), we now also obtain (\ref{eq:susyintorsint}) for $Q^1_{\alpha_2+_1-_1}$, $Q^1_{-_2+_1-_1}$.

The same steps can be used also for $Q^2$, $T^2$. We then conclude that the five-form equations (\ref{eq:dOmegaIIA}), (\ref{eq:domegatII}) have all the missing components of \eqref{eq:Phiintr} except $Q^1_{+_2 +_1-_1}=Q^1_{-_1 +_1 -_1}$ and $Q^2_{+_1 +_2-_2} = Q^2_{+_1-_1+_1}$.

These however can be found in (\ref{eq:lastII}). Multiplying \eqref{eq:intr21} on the left by $\overline{\eps}_1$ and \eqref{eq:intr22} by  $\overline{\eps}_2$ we get from the sum:
\begin{equation}
	\frac{1}{64} (\bar\eps_1 \de \phi \eps_1 +\bar\eps_2 \de \phi \eps_2 ) = Q^{1 \, M}_{\quad M -_1} - T^1_{-_1 +_1}+Q^{2 \, M}_{\quad M -_2} -T^2_{-_2 +_2} = 0 \, .
\end{equation}
The left-hand side is nothing but ${\cal L}_K \phi$. Since we have proven that all the $T$ component are zero, we have
\begin{equation}
\label{eq:intrtorslastline}
Q^{1 \, M}_{\quad M -_1} = - Q^{2 \, M}_{\quad M -_2}\,.
\end{equation}

Let us now see the intrinsic torsion of the second in (\ref{eq:lastII}). From \eqref{eq:intr1} we get
\begin{equation}
	D_M \widetilde{K}^M = - \frac{1}{64} \left( Q^1_{MPQ} \, \overline{\eps}_1 \left[ \g^{PQ} , \g^M \right] \eps_1-Q^2_{MPQ} \, \overline{\eps}_2 \left[ \g^{PQ} , \g^M \right] \eps_2  \right) =-2  Q^{1 \, M}_{\quad M -_1} + 2 Q^{2 \, M}_{\quad M -_2}\,;
\end{equation}
on the other hand, using \eqref{eq:Fintrtro},
\begin{equation}
\begin{split}
	D_M \widetilde{K}^M =& - \frac{4\e^\phi}{32^2} \, \overline{\eps}_1 \g_M F \g^M \eps_2 = - \frac{4\e^\phi}{32^2} R_{RSPQ} \, \left(  \overline{\eps}_1 \g_M \g^{RS} \eps_1 \overline{\eps}_2 \g^{PQ} \g^M \eps_2 \right) \\
	=& - 16 \e^\phi R_{-_1 M N -_2 } \, .
\end{split}
\end{equation}

Therefore, (\ref{eq:lastII}) overall gives the two conditions
\begin{equation}
Q^{1 \, M}_{\quad M -_1} = - Q^{2 \, M}_{\quad M-_2}  = \, 4 \e^\phi R^{\quad \, M}_{-_1 \quad M -_2} \,.
\end{equation}
Since in fact we have obtained equations for all the $Q$ components except $Q^1_{+_2 +_1-_1}=Q^1_{-_1 +_1 -_1}$ and $Q^2_{+_1 +_2-_2} = Q^2_{+_1-_1+_1}$, only these two components survive in the sum, and then we get the desired missing equations.

This completes the proof that (\ref{eq:systemII}) is sufficient for supersymmetry.



\section{Dualities} 
\label{sec:dual}

In this appendix we collect some details about S- and T-duality. 

\subsection{S-duality} 
\label{app:s}

First of all let us summarize the $\mathrm{SL}(2,\R)$ formalism of type IIB supergravity. Given a general element 
\begin{equation}
\Lambda = \begin{pmatrix}
\alpha & \beta \\
\gamma & \delta
\end{pmatrix} \in \mathrm{SL}(2,\R)\,,
\end{equation}
the following transformation is a symmetry of the action:
\begin{equation}
\tau^\prime  = \frac{\alpha \tau + \beta}{\gamma \tau + \delta} , \qquad F^\prime _5 = F_5 , \qquad g^\prime  = |\gamma \tau + \delta| g \, ,\qquad\begin{pmatrix} C_2^\prime  \\ B^\prime  \end{pmatrix}
= \begin{pmatrix}
\alpha & \beta \\
\gamma & \delta
\end{pmatrix} \begin{pmatrix} C_2 \\ B \end{pmatrix} \, ,
\end{equation}
where $\tau = C_0 + \ii e^{- \phi}$.
From these rules we can derive how potentials transform. From $F_5^\prime  = F_5$ we get
\begin{equation}
\begin{split}
\dd C_4^\prime =&\dd C_4 - \dd B \wedge C_2 +  \dd B^\prime  \wedge C_2^\prime \\
=&\dd C_4 + \beta \delta B \wedge \dd B + \beta \gamma (B\wedge\dd C_2+ \dd B \wedge C_2)+ \alpha \gamma C_2 \wedge\dd C_2  
\end{split}
\end{equation}
and thus
\begin{equation}
C_4^\prime  = C_4 + \beta \gamma B \wedge C_2 + \frac{1}{2} (\alpha \gamma C_2 \wedge C_2 + \beta \delta B_2 \wedge B_2)\, .
\end{equation}
Moreover, performing an S-duality on the equation $F_7 = - * F_3$ we get:
\begin{equation}
\label{C6_12}
\dd (C_6^\prime ) = - *^\prime \dd C_2^\prime  + C_0^\prime  *^\prime  H^\prime  + H^\prime  \wedge C_4^\prime  \,.
\end{equation}
Under conformal transformation $g \to \alpha^2 g$ the Hodge dual of a $k$-form $\Omega_k$ transforms as $*  \Omega_k \to \alpha^{D-2k} * \Omega_k$. Using this, we can explicitly evaluate \eqref{C6_12}:
\begin{equation}
\begin{split}
&\dd (C_6^\prime ) = \gamma e^{-2 \phi} * H + (C_0 \gamma+ \delta)\dd C_6 - \gamma C_0 C_4 \wedge d B + \gamma C_4 \wedge\dd C_2 + \frac{1}{2}(\beta \delta^2 B^2 \wedge \dd B\\ 
&+ \beta \gamma \delta B^2 \wedge \dd C_2 + \beta \gamma \delta B \wedge C_2 \wedge \dd B+ \beta \gamma^2 B \wedge C_2 \wedge\dd C_2 + \alpha \gamma\delta C_2^2 \wedge \dd B + \alpha \gamma^2 C_2^2 \wedge\dd C_2 ) \\
&=\gamma \dd \widetilde{B} + \delta\dd C_6 + \frac{1}{2}\big( \gamma  (C_0\dd C_6+\dd C_0 \wedge C_6 + C_4 \wedge\dd C_2 +\dd C_4 \wedge C_2) + \beta \delta^2 B^2 \wedge \dd B\\ 
&+ \beta \gamma \delta B^2 \wedge\dd C_2 + \beta \gamma \delta B \wedge C_2 \wedge \dd B+ \beta \gamma^2 B \wedge C_2 \wedge\dd C_2 + \beta \gamma^2 C_2^2 \wedge \dd B + \alpha \gamma^2 C_2^2 \wedge\dd C_2 \big) 
\end{split}
\end{equation}
From the last line one can check that the correct transformation rule for $C_6$ is
\begin{equation}
C_6^\prime  = \gamma \widetilde{B}+\delta C_6 + \frac{\gamma}{2} \left(  C_0 C_6 +  C_4 \wedge C_2 + \beta B \wedge C_2 \wedge (\delta B + \gamma C_2)\right) + \frac{1}{3} \left(\beta \delta^2 B^3 + \alpha \gamma^2 C_2^3 \right)\,.
\end{equation} 

Another important ingredient we need is how the bilinears transform under $\mathrm{SL}(2,\R)$. The spinors transform under a $\mathrm{U}(1)_D$ subgroup of the original $\mathrm{SL}(2,\R)$ symmetry, indeed the transformation rule reads 
\begin{equation}
\begin{pmatrix} \eps_1^\prime  \\ \eps_2^\prime  \end{pmatrix}
= |\gamma \tau + \delta|^{\frac{1}{4}} \begin{pmatrix}
\ \cos (\theta /2)  & -\sin (\theta /2) \\
\sin (\theta /2) & \cos (\theta /2)
\end{pmatrix} \begin{pmatrix} \eps_1 \\ \eps_2 \end{pmatrix}
\end{equation}
where $\theta = \text{arg}(\gamma \tau + \delta)$. Therefore we have that $K, \Phi_3, \widetilde{\Omega}$ are singlets:
\begin{equation}
\label{sl2singlet}
K' = |\gamma \tau + \delta| K\,, \qquad \Phi_3' = |\gamma \tau + \delta|^2 \Phi_3\,, \qquad \widetilde{\Omega}' = |\gamma \tau + \delta|^3 \widetilde{\Omega}\,. \\
\end{equation}
The other bilinears are components of a doublet:
\begin{equation}
\label{sl2doublet}
\begin{split}
&(\widetilde{K} + \ii \Phi_1)^\prime= |\gamma \tau + \delta|e^{\ii \theta} (\widetilde{K} + \ii \Phi_1) = (\gamma \tau + \delta) (\widetilde{K} + i \Phi_1) \,,\\
&(\Omega + \ii \Phi_5)^\prime= |\gamma \tau + \delta|^3e^{\ii \theta} (\Omega + i \Phi_5) =  |\gamma \tau + \delta|^2 (\gamma \tau + \delta) (\Omega + \ii \Phi_5)\,.
\end{split}
\end{equation}

For completeness, we also give the transformation rule for the other fluxes:
\begin{equation}
\begin{split}
&F_1^\prime = \frac{|\gamma \tau + \delta|^2\dd C_0- 2 \gamma^2 e^{-\phi} (e^{-\phi}\dd C_0 - (C_0+ \delta/\gamma) \dd e^{-\phi})}{|\gamma \tau + \delta|^4} \,,\\
&F_3^\prime = \frac{(\gamma C_0 + \delta) F_3 - \gamma e^{-2 \phi} H}{|\gamma \tau + \delta|^2} \,, \qquad \quad H^\prime = (C_0 \gamma + \delta)H +\gamma F_3 \, ,\\
&F_7^\prime = (\gamma C_0+\delta) F_7 + \gamma e^{-2 \phi} * H\,, \qquad (e^{-2 \phi} * H)^\prime = e^{-2 \phi}  \frac{(\gamma C_0+\delta)* H- \gamma F_7}{|\gamma \tau + \delta|^2}\,.
\end{split}
\end{equation}

As an application of what we have just seen, we can check that the D5 calibration \eqref{Dpcal} gives the NS5 one \eqref{NS5calibration} after a simple S-duality \eqref{Sduality}. In this particular case the transformation rule reads:
\begin{equation}
\begin{split}
&C_0' = - \frac{C_0}{|\tau|^2} \,, \qquad e^{-\phi '} = \frac{e^{-\phi}}{|\tau|^2} \,\\
&C_2' = -B\, ,  \qquad B' =  C_2\,, \qquad  C_4' = C_4 - B \wedge C_2 \,, \\
&C_6' = \widetilde{B} + \frac{1}{2} ( C_0 C_6 + C_4 \wedge C_2 - C_2 \wedge C_2 \wedge B) \,.
\end{split}
\end{equation}
The D5 calibration is more explicitly given by
\begin{equation}
\varphi_{\rm D5}=\left( e^{-B} \wedge (e^{-\phi} \Phi - (\iota_{K} + \widetilde{K} \wedge)C) \right)_5 .
\end{equation}
We expand $e^{-B}$ and transform the resulting terms one by one. We begin with the $B^2$ term:
\begin{equation}
\left( \frac{1}{2} B^2 \wedge \left( e^{-\phi} \Phi_1 - \iota_{K} C_2 - \widetilde{K} C_0 \right) \right)' = \frac{1}{2} C_2^2 \wedge (\widetilde{K} + \iota_K B)\,.
\end{equation}
Notice that $e^{-\phi} \Phi_1 - \iota_{K} C_2 - \widetilde{K} C_0$ is the D1 calibration, while $\widetilde{K} + \iota_K B$ is the calibration for a fundamental string, in agreement with the S-duality. Next we transform the term linear in $B$:
\begin{equation}
\Big( -B \wedge \big( e^{-\phi} \Phi_3 - \iota_{K} C_4 - \widetilde{K} \wedge C_2 \big) \Big)' = - C_2 \wedge \big( e^{-\phi} \Phi_3 - \iota_{K} (C_4 - B \wedge C_2) + (C_0 \widetilde{K} - e^{- \phi} \Phi_1) \wedge B \big),
\end{equation}
and finally
\begin{equation}
\begin{split}
&\left( e^{-\phi} \Phi_5 - \iota_{K} C_6 - \widetilde{K} \wedge C_4 \right)' = e^{- 2 \phi} \widetilde{\Omega} + e^{-\phi} C_0 \Phi_5 - \iota_{K} \widetilde{B} \\ 
&- \frac{1}{2} \iota_K  ( C_0 C_6 + C_4 \wedge C_2 - C_2^2 \wedge B) -  (C_0 \widetilde{K} - e^{- \phi} \Phi_1) \wedge C_4 +  (C_0 \widetilde{K} - e^{- \phi} \Phi_1) \wedge C_2 \wedge B\,  .
\end{split}
\end{equation}
Summing up we find
\begin{equation}
\begin{split}
&\left( e^{-B} \wedge (e^{-\phi} \Phi - (\iota_{K} + \widetilde{K} \wedge)C) \right)_5' = e^{- 2 \phi} \widetilde{\Omega}+ e^{-\phi} C_0 \Phi_5-e^{-\phi} C_2 \wedge \Phi_3+ e^{- \phi} \Phi_1 \wedge C_4\\
&- C_0 \widetilde{K} \wedge C_4 + \frac{1}{2} C_2 \wedge C_2 \wedge \widetilde{K} - \iota_{K} \widetilde{B} - \frac{1}{2} (C_0 \iota_K C_6+C_4 \wedge \iota_K C_2-\iota_K C_4 \wedge C_2) = \varphi^{\rm IIB}_{\rm NS5}\, ,
\end{split}
\end{equation}
which is exactly what we were looking for.


\subsection{Flat-index T-duality} 
\label{sub:t}

In this section, following \cite{kelekci-lozano-macpherson-ocolgain}, we will revisit the T-duality formalism of section \ref{sub:T-dual_main} using flat-index notation. The benefits of this formulation are that the transformation rules of fields and bilinears assume a simpler form, especially for the five-form \eqref{OmegaTT}. This makes particularly easy to check that the longitudinal part of \eqref{eq:dOmegaIIA} and \eqref{eq:domegatII} is invariant; however, we will partially lose the explicit $\mathrm{O}(d,d)$ interpretation we had in section \ref{sub:T-dual_main}.

We again assume that we can define a compact and isometric direction $\partial_y$. We decompose the fields as:
\begin{equation}
\label{eq:T-dualdecomp}
\begin{split}
&\dd s_{10}^2 = \dd s_{9,A}^{2}+ e^{2 C} (\dd y + A_1)^2  ,\quad 
B = B_2 + B_1 \wedge \dd y \, ,\quad F = F_\perp +   F_\parallel \wedge E^y \, , \\
&\Phi = \Phi_\perp + \Phi_\parallel \wedge E^y \, , \qquad K = k_1+k_0 E^y \, , \qquad \widetilde{K} = \widetilde{k}_1+\widetilde{k}_0 E^y \, , \\
&\Omega = \omega_5+\omega_4 \wedge E^y \, , \qquad \widetilde{\Omega} = \widetilde{\omega}_5+ \widetilde{\omega}_4 \wedge E^y \, , \qquad E^y = e^C (\dd y + A_1) \, .
\end{split}
\end{equation}
Using section \ref{sub:T-dual_main}, one can perform a T-duality from IIB to IIA, which in this case transforms the components of fields and bilinears as:
\begin{equation}
\label{eq:Flat_T-dual}
\begin{split}
&\dd s_{9,B}^2 = \dd s_{9,A}^2 \, ,\qquad \phi^B = \phi^A - C^A \, ,\qquad   C^B = - C^A \,,\\
 &B_2^B = B_2^A+ A_1^A \wedge B_1^A \, ,\qquad  A^B_1 = -B_1^A \, ,\qquad B^B_1 = -A_1^A \,,\\
& F_\perp^B= e^{C^A} F_\parallel^A \, ,\qquad F_\parallel^B = e^{C^A} F_\perp^A \, ,\qquad \Phi_\perp^B= \Phi_\parallel^A \, ,\qquad  \Phi_\parallel^B = \Phi_\perp^A \,,\\
& k_1^B = k_1^A \, ,\qquad k_0^B = k_0^A \, ,\qquad  \widetilde{k}_1^B = \widetilde{k}_1^A  \, ,\qquad  \widetilde{k}_0^B = \widetilde{k}_0^A \,\\ 
&\omega_5^B = \widetilde{\omega}_5^A  \, ,\qquad \omega_4^B = \omega_4^A \, ,\qquad	 \widetilde{\omega}_5^B = \omega_5^A \, ,\qquad \widetilde{\omega}_4^B = \widetilde{\omega}_4^A\,,
\end{split}
\end{equation}
where superscripts $A,B$ denote in which theory the field is sitting. For what it follows, it is convenient to rewrite $H$ as
\begin{equation}
H = \dd B_2-\dd B_1 \wedge A_1+ e^{-C} \dd B_1 \wedge E^y \, ,
\end{equation}
where we can notice that $\dd B_2-\dd B_1 \wedge A_1$ is a T-duality invariant.

Let us now decompose \eqref{eq:dOmegaIIA} and \eqref{eq:domegatII} according to \eqref{eq:T-dualdecomp}. The longitudinal parts read:
\begin{equation}
\label{eq:T-dual-longitudinal_match}
\begin{split}
&\dd(e^{-2 \phi+C} \omega_4) = e^{-2 \phi+C} \iota_{k_1} *_9 (\dd B_2 - \dd B_1 \wedge A_1)+ e^{-\phi+C}((\Phi_\parallel,F_\perp)_6-(\Phi_\perp, F_\parallel)_6)   \,,\\
&\dd(e^{-2 \phi+C} \widetilde{\omega}_4) = e^{-2 \phi+C}\iota_{\widetilde{k}_1} *_9 (\dd B_2 - \dd B_1 \wedge A_1)- \frac{e^{-\phi+C}}{2}((\Phi_\parallel^m,F_{\perp \, m})_6-(\Phi_\perp^m, F_{\parallel \, m})_6)\,,
\end{split}
\end{equation}
where $m$ runs from $0$ to $8$. Using \eqref{eq:Flat_T-dual} it is immediate to see that these equations are invariant, which is what we should expect from the calibration conditions of the NS5-brane and the KK5-monopole, since the corresponding objects are invariant when they lie along the T-duality direction.
However, it is also easy to see that the transverse part of the KK5-monopole equation (\ref{eq:domegatII}) does not transform in the transverse part of the NS5-brane equation (\ref{eq:dOmegaIIA}), as it is required from the duality properties of these objects. This leads us to think that, to make the T-duality working, the KK5-monopole equation requires some improvement or a better interpretation in the future.

\section{More differential-form equations}
\label{sec:algebraic}

In this appendix we will derive some form equations that do not appear in \eqref{eq:systemII}. We will see that in most cases we will manage to arrange all the components in compact expressions. For this purpose, we define the following bracket which is an analogous to the Chevalley--Mukai pairing 
\begin{equation}
\{A,B\}_d = \big(A \wedge \lambda  [(d - 2 \text{deg}) B]\big)_d
\end{equation}
with opposite symmetry: $\{A,B\}_d = -(-)^{d(d-1)/2}\{B,A\}_d$.
For example, we can see that the 2-form part of \eqref{epsepsbar_diff} and \eqref{epsepsbar_sum} for both IIA and IIB reads:
\begin{equation}
\label{eq:dK}
\begin{split}
&e^{2 \phi} \d(e^{-2 \phi} K) = * \left( H \wedge \Omega+ \frac{e^\phi}{4} \{\Phi, F\}_8 \right) \, , \\ 
&e^{2 \phi} \d(e^{-2 \phi} \widetilde{K}) = * \left( H \wedge \widetilde{\Omega}+ \frac{e^\phi}{8} \{\Phi_M, F^M\}_8 \right) \, ,
\end{split}
\end{equation}
while from the 10-form part we have:
\begin{equation}
\label{eq:10form_from_section2}
\d\left( e^{-2 \phi} * \widetilde{K} \right) =0  \qquad \d\left( e^{-2 \phi} * K \right) =0 \, .
\end{equation}

Some new equations can be obtained by using the two supersymmetry conditions on the left of \eqref{eq:SUSY1}:
\begin{equation}
\begin{split}
&\d(\eps_1 \overline{\eps}_1) = \frac{1}{2} \left[ \gamma^M , D_M (\eps_1 \overline{\eps}_1) \right] = \frac{1}{2} \left[ \gamma^M , \frac{1}{4} [ H_M ,\eps_1 \overline{\eps}_1] + (-)^{|F|} \frac{e^{\phi}}{16} (F \gamma_M \lambda(\Phi) + \Phi \gamma_M \lambda(F)) \right] \\
&= H_M \wedge \iota_M \eps_1 \overline{\eps}_1+(-)^{|F|} \frac{e^{\phi}}{32} ( \gamma^M F \gamma_M \lambda(\Phi) + \gamma^M \Phi \gamma_M \lambda(F)-F \gamma_M \lambda(\Phi) \gamma^M - \Phi \gamma_M \lambda(F) \gamma^M )
\end{split}
\end{equation} 
and 
\begin{equation}
\begin{split}
\d(\eps_2 \overline{\eps}_2) =& -H_M \wedge \iota_M \eps_2 \overline{\eps}_2 +(-)^{|F|} \frac{e^{\phi}}{32} \big( -\gamma^M \lambda(F) \gamma_M \Phi \\
&- \gamma^M \lambda(\Phi) \gamma_M F+\lambda(F) \gamma_M \Phi \gamma^M + \lambda(\Phi) \gamma_M F \gamma^M \big) \, .
\end{split}
\end{equation} 
Taking the sum and the difference of these last two equations we get:
\begin{equation}
\begin{split}
\d \left( \frac{\eps_1 \overline{\eps}_1 \pm \eps_2 \overline{\eps}_2}{2}\right) &= H_M \wedge \iota_M  \left( \frac{\eps_1 \overline{\eps}_1 \mp \eps_2 \overline{\eps}_2}{2}\right) + (-)^{|F|} \frac{e^{\phi}}{64} \Big( [\gamma_M F \gamma^M, \lambda(\Phi)]_{\pm} \\
&+[\gamma_M \Phi \gamma^M, \lambda(F)]_{\pm}-[F, \gamma_M \lambda(\Phi) \gamma^M]_\pm -[\Phi, \gamma_M \lambda(F) \gamma^M]_\pm \Big)
\end{split}
\end{equation}
where $[\, \, , \,]_-$ indicates the usual commutator while $[\, \, , \,]_+$ is the anticommutator. Analogously to what we have seen in section \ref{sub:der}, the result is different depending on whether we are considering IIA or IIB. From the sum in IIB we get
\begin{equation}
\begin{split}
\d \left( \frac{\eps_1 \overline{\eps}_1 + \eps_2 \overline{\eps}_2}{2}\right) =& H_M \wedge \iota_M  \left( \frac{\eps_1 \overline{\eps}_1 - \eps_2 \overline{\eps}_2}{2}\right) - \frac{e^{\phi}}{4} \big( \{ \Phi_1,F_3 \}+ \{ \Phi_3,F_1 \} \\
&- \{ \Phi_3 ,F_5 \} - \{\Phi_5,F_3\} + \{\Phi_7, F_1\} + \{\Phi_1,F_7\}  \big) 
\end{split}
\end{equation}
and from the difference
\begin{equation}
\begin{split}
\d \left( \frac{\eps_1 \overline{\eps}_1 - \eps_2 \overline{\eps}_2}{2}\right) =& H_M \wedge \iota_M  \left( \frac{\eps_1 \overline{\eps}_1 + \eps_2 \overline{\eps}_2}{2}\right) + \frac{e^{\phi}}{4} \Big( 4 [\Phi_1 , F_1]+2[\Phi_1,F_5] \\
-&2[\Phi_5,F_1]+ 2[\Phi_9,F_1]-2 [\Phi_3,F_3]+ [\Phi_7,F_3]- [\Phi_3,F_7] \Big) \, ,
\end{split}
\end{equation}
while for IIA:
\begin{equation}
\begin{split}
&\d \left( \frac{\eps_1 \overline{\eps}_1 + \eps_2 \overline{\eps}_2}{2}\right) = H_M \wedge \iota_M  \left( \frac{\eps_1 \overline{\eps}_1 - \eps_2 \overline{\eps}_2}{2}\right) + \frac{e^{\phi}}{4} \Big(-\{\Phi_2 , F_0\} + \{\Phi_2 , F_4\}+3\{\Phi_2,F_8\} \\
& \qquad \qquad - 3\{\Phi_6,F_0 \} - \{\Phi_6, F_4\} + \{\Phi_4,F_2\}-5 \{\Phi_{10},F_0\}-3\{\Phi_0,F_6 \} - \{\Phi_0,F_2 \} \Big) \, , \\
&\d \left( \frac{\eps_1 \overline{\eps}_1 - \eps_2 \overline{\eps}_2}{2}\right) = H_M \wedge \iota_M  \left( \frac{\eps_1 \overline{\eps}_1 + \eps_2 \overline{\eps}_2}{2}\right) + \frac{e^{\phi}}{2} \Big( [\Phi_6,F_2]-[\Phi_2,F_6] \Big) \, .
\end{split}
\end{equation}
Splitting the various degrees and massaging a little bit the expressions we get the 2-form conditions:
\begin{equation}
\label{eq:dKtilde_dK}
\d\widetilde{K} = \iota_K H \,,\qquad \d K = \iota_{\widetilde{K}}H- \frac{e^{\phi}}{2} * (\Phi,F)_8 \, ,\\
\end{equation}
where the first one is nothing but the F1 calibration condition. From the 6-form part we get 
\begin{equation}
\label{eq:Omega_extra}
\d\widetilde{\Omega} = H_M \wedge \Omega^M - \frac{e^{\phi}}{4} \left\{ \Phi_M, F^M\right\}_6 \, , \quad
\d \Omega = H_M \wedge \widetilde{\Omega}^M - \frac{e^{\phi}}{4} \Big( \{\Phi, F\}+ \frac{1}{4} \{\Phi_{MN}, F^{MN}\} \Big)_6 . 
\end{equation}
We need not write the ten-form part, since it was derived in section \ref{sub:der}.

Notice that these equations can be used as a different definition for the exterior derivative of the bilinears, or they can be combined with the old definitions to get algebraic constraints. For example combining the first of \eqref{eq:dK} with the second of \eqref{eq:dKtilde_dK}, and \eqref{eq:Omega_extra} with the equation for the KK5-monopole in IIB \eqref{eq:domegatII}, we get the following relations that turn out to be useful in section \ref{sec:s}:
\begin{equation}
\label{eq:algebraicK_Omega}
\begin{split}
&2 F_1 \wedge \Phi_7 - F_3 \wedge \Phi_5 - e^{-\phi} H \wedge \Omega + e^{-\phi} \widetilde{K} \wedge * H+ \Phi_1 \wedge F_7+ 2 \iota_K * \dd e^{-\phi} = 0 \,,\\
&e^{-\phi} H_M \wedge \Omega^M +F_{3 \, M} \wedge \Phi^M_5 - \iota_{\Phi_1} F_7 + e^{-\phi} \iota_{\widetilde{K}} * H + 2 \dd e^{-\phi} 
\wedge \widetilde{\Omega}- 2 F_{1 \, M} \Phi_7^M=0\,.
\end{split}
\end{equation}                                                                                                                                      
We can provide some others algebraic equations that can be obtained from \eqref{eq:SUSY2}. Since we will mainly use this result in section \ref{sec:s}, we will perform all the computations for type IIB. However pure algebraic equations impose strict constraints, so we hope these equations to be useful also for the classification of complicated cases (e.g.~vacuum solutions with high supersymmetry). 

We first take the tensor product of \eqref{eq:SUSY21} with $\overline{\eps}_2$ and of the transpose of \eqref{eq:SUSY22} with $\eps_1$:
\begin{equation}
\begin{split}
&\left( \de \phi - \frac{1}{2} H \right) \Phi + \frac{\text{e}^{\phi}}{2} (2F_1 + F_3) \eps_2\overline{\eps}_2 = 0 \,,\\
&\Phi \left( \de \phi - \frac{1}{2} H \right)  - \frac{\text{e}^{\phi}}{2} \eps_1 \overline{\eps}_1 (2 F_1 + F_3)  = 0\,.
\end{split}
\end{equation}
Taking the difference of these two expressions:
\begin{equation}
\begin{split}
0&=-2 \Phi \wedge \dd e^{-\phi} + e^{- \phi} H \wedge \Phi- e^{- \phi} \g (H_M \wedge \Phi^M) + 2 \iota_{F_1} \frac{\eps_1 \overline{\eps}_2+\eps_2 \overline{\eps}_2}{2}- 2 F_1 \wedge \frac{\eps_1 \overline{\eps}_2-\eps_2 \overline{\eps}_2}{2} \\
&-F_3 \wedge \frac{\eps_1 \overline{\eps}_2-\eps_2 \overline{\eps}_2}{2}+F_3^M \wedge \iota_M \frac{\eps_1 \overline{\eps}_2+\eps_2 \overline{\eps}_2}{2}+ \g \left( F_3^M \wedge \iota_M \frac{\eps_1 \overline{\eps}_2-\eps_2 \overline{\eps}_2}{2} \right)+ \iota_{F_3} \frac{\eps_1 \overline{\eps}_2+\eps_2 \overline{\eps}_2}{2}\,.
\end{split}
\end{equation}
(We recover the sum by taking $\g$ times the difference.) The most interesting equations come from the ten-, eight-, four- and zero-form components. They read:
\begin{subequations}
	\begin{align}
	&2 \dd e^{-\phi} \wedge \Phi_9+ e^{-\phi} H \wedge \Phi_7 - 2 \widetilde{K} \wedge F_9 =0 \label{eq_2Bonly} \,,\\
	&2 \dd e^{-\phi} \wedge \Phi_7+ e^{-\phi} H \wedge \Phi_5- e^{-\phi} \Phi_1 \wedge * H - 2 \iota_K F_9 - F_3 \wedge \Omega - \widetilde{K} \wedge F_7 =0 \,,\label{8Dalgebraiceq} \\
	&2 \iota_{ \dd e^{-\phi}} \Phi_7+ e^{- \phi} \iota_{\Phi_1} * H + e^{- \phi} H_M \wedge \Phi_5^M + 2 F_1 \wedge \widetilde{\Omega}+\iota_{\widetilde{k}} F_7- F_3^M \wedge \Omega_M =0 \,,\label{6Dalgebraiceq} \\
	& \iota_K F_1 =0\,.
	\end{align}
\end{subequations}
where we have taken the Hodge dual of the four-form part.


\bibliography{at}
\bibliographystyle{at}

\end{document}